\documentclass[aps,prl,twocolumn,showpacs,superscriptaddress]{revtex4}

\usepackage{graphicx}
\usepackage{color}
\usepackage{natbib}
\usepackage{epsfig}

\usepackage{bibunits}

\usepackage{amsmath,amssymb,amsfonts}

\newcommand{\apjl}{Astrophys. J. Lett.}%
\newcommand{\apjs}{Astrophys. J. Supp.}%
\newcommand{\aap}{Astron. Astrophys.}%
\newcommand{\mnras}{Mon. Not. Roy. Astron. Soc.}%
\newcommand{\lrr}{Living Reviews in Relativity}%
\newcommand{\pasa}{Publications of the Astronomical Society of Australia}

\begin{document}

\title{Identifying a first-order phase transition in neutron star mergers through gravitational waves}

\author{Andreas Bauswein}
\affiliation{GSI Helmholtzzentrum f\"ur Schwerionenforschung, Planckstra{\ss}e 1, 64291 Darmstadt, Germany}
\affiliation{Heidelberg Institute for Theoretical Studies, Schloss-Wolfsbrunnenweg 35, 69118 Heidelberg, Germany}

\author{Niels-Uwe F. Bastian}
\affiliation{Institute of Theoretical Physics, University of Wroclaw, 50-205 Wroclaw, Poland}

\author{David B. Blaschke}
\affiliation{Institute of Theoretical Physics, University of Wroclaw, 50-205 Wroclaw, Poland}
\affiliation{National Research Nuclear University (MEPhI), 115409 Moscow, Russia}
\affiliation{Bogoliubov Laboratory for Theoretical Physics, Joint Institute for Nuclear Research, 141980 Dubna, Russia}

\author{Katerina Chatziioannou}
\affiliation{Canadian Institute for Theoretical Astrophysics, 60 St. George Street, University of Toronto, Toronto, ON M5S 3H8, Canada}

\author{James A. Clark}
\affiliation{Center for Relativistic Astrophysics, School of Physics, Georgia Institute of Technology, Atlanta, Georgia 30332, USA}

\author{Tobias Fischer}
\affiliation{Institute of Theoretical Physics, University of Wroclaw, 50-205 Wroclaw, Poland}

\author{Micaela Oertel}
\affiliation{LUTH, Observatoire de Paris, PSL Research University, CNRS, Université Paris Diderot, Sorbonne Paris Cité, 5 place Jules Janssen, 92195 Meudon, France}

\date{\today} 

\begin{abstract}
We identify an observable imprint of a first-order hadron-quark phase transition at supranuclear densities on the gravitational-wave (GW) emission of neutron star mergers. Specifically, we show that the dominant postmerger GW frequency $f_\mathrm{peak}$ may exhibit a significant deviation from an empirical relation between $f_\mathrm{peak}$ and the tidal deformability if a strong first-order phase transition leads to the formation of a gravitationally stable extended quark matter core in the postmerger remnant. A comparison of the GW signatures from a large, representative sample of microphysical, purely hadronic equations of state indicates that this imprint is only observed in those systems which undergo a strong first-order phase transition. Such a shift of the dominant postmerger GW frequency can be revealed by future GW observations, which would provide evidence for the existence of a strong first-order phase transition in the interior of neutron stars.

\end{abstract}

   \pacs{04.30.Tv,26.60.Kp,26.60Dd,97.60.Jd}

   \maketitle
   
\begin{bibunit}

{\it Introduction:} The theory of strong interactions, quantum chromodynamics (QCD), with quarks and gluons as fundamental degrees of freedom predicts a transition from nuclear matter to quark matter. At vanishing baryonic chemical potential, numerical solutions of QCD are available, which state a smooth crossover transition at a temperature of $T=154\pm 9$~MeV~\cite{Bazavov2012,Borsanyi2014,Bazavov2014}. At finite baryon densities only phenomenological models of QCD exist, which are benchmarked by nuclear matter phenomenology around nuclear saturation density $\rho_{\rm sat}\approx 2.7 \times 10^{14}$\,g\,cm$^{-3}$~\cite{Krueger2013} and by perturbative QCD at asymptotic densities~\cite{Kurkela2014}. Those methods, however, are not applicable in the region of the hadron-quark transition. Hence, the nature of the transition to quark matter (crossover or first-order phase transition) remains unclear. Whether the hadron-quark phase transition occurs at conditions which are found in compact stellar objects, e.g., in neutron stars (NS) with central densities of several times $\rho_\mathrm{sat}$, is presently unknown. The very first detection of gravitational waves (GW) from a NS merger~\cite{Abbott2017} highlights the prospect to learn about the presence and the nature of the QCD phase transition in stellar objects, e.g.~\cite{Csaki2018,Paschalidis2018,Most2018a,Han2018,Christian2018,Sieniawska2018,Burgio2018a,Drago2018,Dexheimer2018}.

The merger dynamics and the corresponding GW signal can be divided into an inspiral phase before merging and a postmerger stage~\cite{Faber2012,Baiotti2017,Paschalidis2017,Friedman2018}. The GW signal prior to the merger allows us to measure the tidal deformability of the progenitor stars, which is encoded in the phase evolution of the orbital motion and the corresponding GW signal~\cite{Flanagan2008,Hinderer2008,Read2009,Hinderer2010,Read2013,DelPozzo2013,Wade2014,Agathos2015,Chatziioannou2015,Hotokezaka2016,Chatziioannou2018}. During merging, densities and temperatures increase, and hence the postmerger phase probes a different equation of state (EOS) regime. The associated GW signal contains information about the stellar structure of the remnant. Postmerger oscillation frequencies are correlated with the size of the remnant and with radii of nonrotating cold NSs~\cite{Bauswein2012,Bauswein2012a,Hotokezaka2013a,Takami2014,Bauswein2015}.

In the present work we describe a compelling example of the complementarity of pre- and postmerger GW signals. We demonstrate that the joint detection of GWs from both phases can provide a unique observable signature of a first-order hadron-quark phase transition. Previous works have focused on comparisons between individual models with and without phase transition and on describing differences between these models~\cite{Oechslin2004,Bauswein2010a,Sekiguchi2011,Radice2017,Most2018a}. While these studies have revealed potential indicators of phase transitions, it is not clear whether the differences observed are indeed an unambiguous signature for a phase transition. To identify clear evidence for a phase transition it is indispensable to ensure that a particular signature can {\it only} be caused by the presence of a phase transition. Unless this criterion is met, any observational indication of a phase transition would be degenerate with the uncertainty of the hadronic EOS.

The novelty of our work lies precisely in the fact that we describe a scenario that allows us to uniquely discriminate an EOS with a strong first-order phase transition. To this end we provide evidence that {\it all} possible hadronic EOS models yield a different observational signature. We achieve this by considering a large, representative sample of hadronic EOSs that exhibit a clearly distinguishable behavior. In this sense we provide here for the first time an observable signature of a first-order phase transition in NS mergers.

Two aspects are critical. First, a potential signature of a phase transition should involve quantities which are measurable with sufficient precision in future experiments. This has been shown for the tidal deformability~\cite{Read2009,Read2013,DelPozzo2013,Wade2014,Agathos2015,Chatziioannou2015,Hotokezaka2016,Abbott2017,TheLIGOScientificCollaboration2018a,Abbott2018,Chatziioannou2018,De2018,Carney2018} and postmerger GW frequencies~\cite{Clark2014,Clark2016,Chatziioannou2017,Bose2018,Yang2018,Torres-Rivas2018}. Second, the observable quantities under consideration should be determined from theoretical models or simulations with sufficient precision to allow for an interpretation of the measurements. In contrast to for instance the remnant life time and the precise phase evolution in the postmerger phase, the tidal deformability during inspiral and the oscillation frequencies of the postmerger remnant can be determined with relatively high reliability~\cite{Faber2012,Baiotti2017,Paschalidis2017,Friedman2018,Duez2019}. We remark that identifying the impact of a phase transition on the tidal deformability in binaries where at least one component contains a quark core, would require highly precise measurements of the masses and tidal deformabilities apart from the problem that massive stars with quark core may be less abundant.

{\it Equations of state:}~In this work we present NS merger simulations with the novel temperature-dependent, microscopic hadron-quark hybrid EOS DD2F-SF of Ref.~\cite{Fischer2018}. Among other purely hadronic EOS models, we consider a nucleonic reference EOS (DD2F)~\cite{Typel2005,Typel2010,Alvarez-Castillo2016} and corresponding hybrid EOSs with a phase transition to deconfined quark matter (DD2F-SF) of~\cite{Fischer2018}. The latter employ the classical two-phase construction, which features a strong first-order phase transition within the standard Maxwell approach. The stiffening of the quark phase admits gravitationally stable stellar configurations with extended quark matter cores, so-called hybrid stars. We consider different choices of parameters for the description of the quark phase resulting in seven specific hybrid EOSs, which cover a variaty of different models, i.e. with different onset densities and different density jumps. We dub these EOSs DD2F-SF-n with $n \in \{1,2,3,4,5,6,7\}$. Below we use the acronym DD2F-SF to refer to all seven hybrid models. Details of the microphysical model for DD2F and DD2F-SF are provided in the Supplemental Material along with information about 15 other EOSs, which serve as representative sample of purely hadronic models. (The Supplemental Material includes additional references~\cite{Hempel2012,Hempel2010,Kaltenborn2017a,Nambu1961,Klevansky1992,Ruester2005,Blaschke2005,Bastian2018,Alvarez-Castillo2016,Benic2015,Klaehn2015,Typel2016,Akmal1998,Banik2014,Goriely2010,Typel2010,Wiringa1988,Shen2011,Lattimer1991,Lalazissis1997a,Steiner2013,Douchin2001,Sugahara1994a,Toki1995,Bauswein2012a,Bauswein2013a,Bauswein2014a,Fortin2018,Marques2017,Alford2005,Read2009a,Danielewicz2002,Tsang2018,Lattimer2013,Krueger2013,Antoniadis2013,Arzoumanian2018a,Abbott2017,Bauswein2017,De2018,Abbott2018} with information on the models and some astrophysical and nuclear physics constraints, which are met by DD2F.) Three of these purely hadronic EOSs include a 2nd order phase transition to hyperonic matter. Additionally, we employ the EOSs ALF2 and ALF4 from~\cite{Read2009a}, which resemble models with a more continuous transition to quark matter (with vanishing latent heat)~\cite{Alford2005}.

{\it Simulations: }We perform NS merger simulations with a relativistic smooth particle hydrodynamics code, which imposes the conformal flatness condition~\cite{Wilson1996,Isenberg1980} to solve the Einstein equations (see~\cite{Oechslin2002,Oechslin2007,Bauswein2010a} for details and e.g.~\cite{Bauswein2012a,Hotokezaka2013a,Takami2014} for a comparison of GW frequencies with grid-based codes solving the full field equations). The calculations start from circular quasi-equilibrium orbits with non-spinning stars a few revolutions before merging. The stars are initially in beta-equilibrium at zero temperature. During the evolution temperature effects are taken into account selfconsistently if provided by the EOS. For some EOSs where the temperature dependence is not available, we employ an approximate treatment of thermal effects, which requires to choose a coefficient $\Gamma_\mathrm{th}$ (see e.g.~\cite{Bauswein2010}). It regulates the strength of thermal pressure support. We adopt $\Gamma_\mathrm{th}=1.75$, which reproduces results with fully temperature dependent EOSs relatively well~\cite{Bauswein2010}.

We focus on merger simulations for equal-mass systems with a total mass of $M_{\rm tot}=2.7~$M$_\odot$, which is comparable to the total mass of GW170817~\cite{Abbott2017,Abbott2018,TheLIGOScientificCollaboration2018a}. This represents a likely binary configuration according to pulsar observations and population synthesis studies~\cite{Dominik2012,Lattimer2012}. We emphasize that in future the binary component masses will be measured with good precision for events which are sufficiently close to allow an extraction of EOS effects from the GW signal~\cite{Rodriguez2014,Farr2016}. This justifies to focus on fixed binary masses in our investigation.
\begin{figure}
\includegraphics[width=\columnwidth]{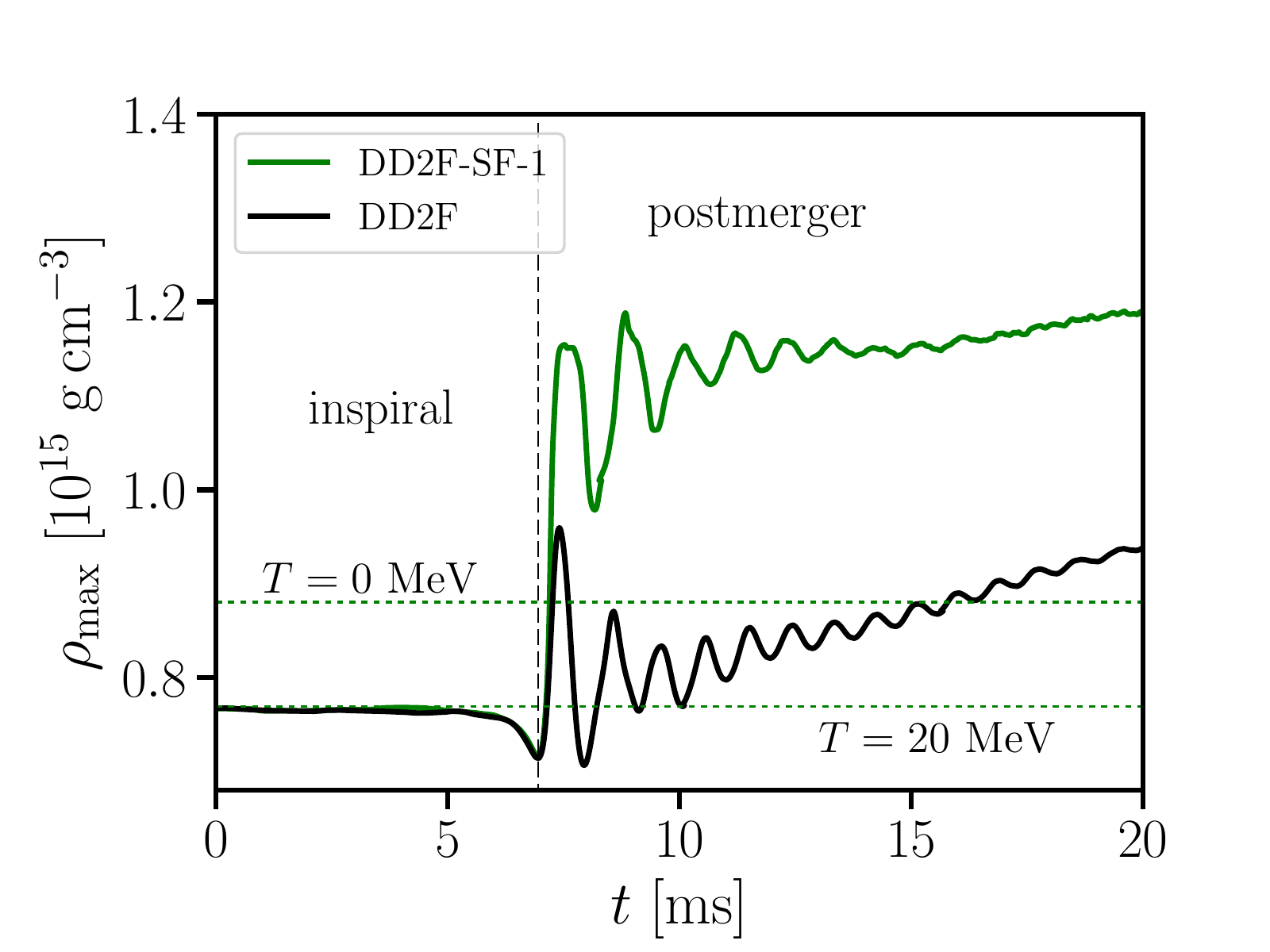}
\caption{Evolution of the maximum rest-mass density comparing DD2F-SF-1 (green) and DD2F (black) for 1.35-1.35~$M_\odot$ mergers (solid curves). Horizontal dotted green lines mark the onset density $\rho_{\rm onset}$ of the phase transition for DD2F-SF-1 at $T=0$ and at 20~MeV.}
\label{fig:rhomax}
\end{figure}

We start with a exemplary discussion of DD2F-SF-1 noting that the other models of the DD2F-SF class behave similarly. Figure~\ref{fig:rhomax} displays the evolution of the maximum rest-mass density as function of time for 1.35-1.35~$M_\odot$ simulations with the DD2F-SF-1 (green) and the purely hadronic counterpart DD2F (black). The dotted horizontal green lines indicate the onset density $\rho_\mathrm{onset}$ of the phase transition at $T=0$ and 20~MeV for beta-equilibrium. During the inspiral phase the central density of the stars is below the transition density and the two systems evolve identically. The two stars merge at about 7~ms and form a single central object associated with a steep increase of the maximum rest-mass density. For the quark matter EOS the density rises above the threshold for the hadron-quark phase transition, reaching the pure quark matter phase. A quark core forms in the center of the merger remnant. The mass enclosed inside the quark matter core comprises about 20--30\% of the total mass. The maximum density in the calculation with the purely hadronic EOS always remains below that of DD2F-SF-1. The stronger density increase in the model with quark matter is a direct consequence of the density jump across the phase transition and the stiffening only at higher densities.

\begin{figure}
\includegraphics[width=\columnwidth]{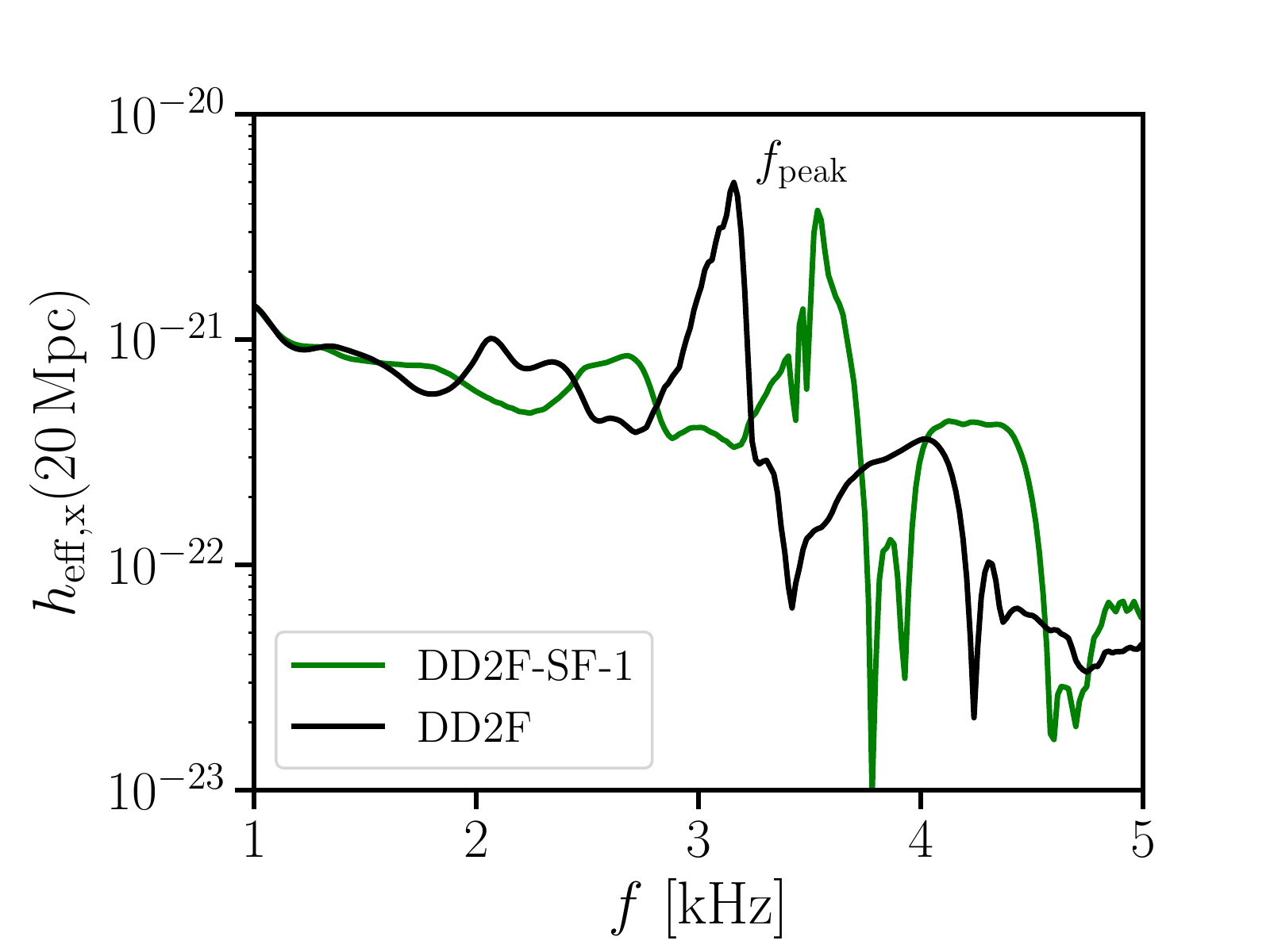}
\caption{GW spectrum of the cross polarization at a distance of 20~Mpc along the polar axis comparing the DD2F-SF-1 EOS (green curve) and the DD2F EOS (black curve).}
\label{fig:spectrum}
\end{figure}

{\it GW spectrum:} The different evolution of the mergers with and without phase transition to quark matter is reflected in the GW signal. Figure~\ref{fig:spectrum} shows the GW spectra of the cross polarization at a distance of 20~Mpc along the polar axis comparing the DD2F-SF-1 EOS (green) and the DD2F EOS (black). During the pre-merger phase the GW signals reach a maximum frequency of about 1.7~kHz, and the GW spectra are similar below this frequency. The high-frequency content of the spectra is shaped by the postmerger stage and significant differences between the two simulations are apparent. In particular, the frequency $f_\mathrm{peak}$ of the dominant oscillation of the postmerger phase is clearly different. This peak is a robust and generic feature that occurs in all simulations which do not directly form a black hole after merging~\cite{Shibata2005a,Shibata2006,Oechslin2007a,Hotokezaka2011,Bauswein2012a,Bauswein2013}. 

The frequency of the main peak depends sensitively on the EOS~\cite{Zhuge1996,Shibata2005a,Shibata2006,Oechslin2007a}. It has been found~\cite{Bauswein2012,Bauswein2012a} that $f_\mathrm{peak}$ scales tightly with radii $R$ of nonrotating cold NSs for different fixed binary masses (cf. Figs.~9--12 and 22--24 in~\cite{Bauswein2012a}). In turn, these relations $f_\mathrm{peak}(R)$ offer the possibility to determine NS radii from a measurement of the dominant postmerger GW frequency~\cite{Clark2014,Clark2016,Chatziioannou2017,Bose2018,Yang2018}.

\begin{figure}
\includegraphics[width=\columnwidth]{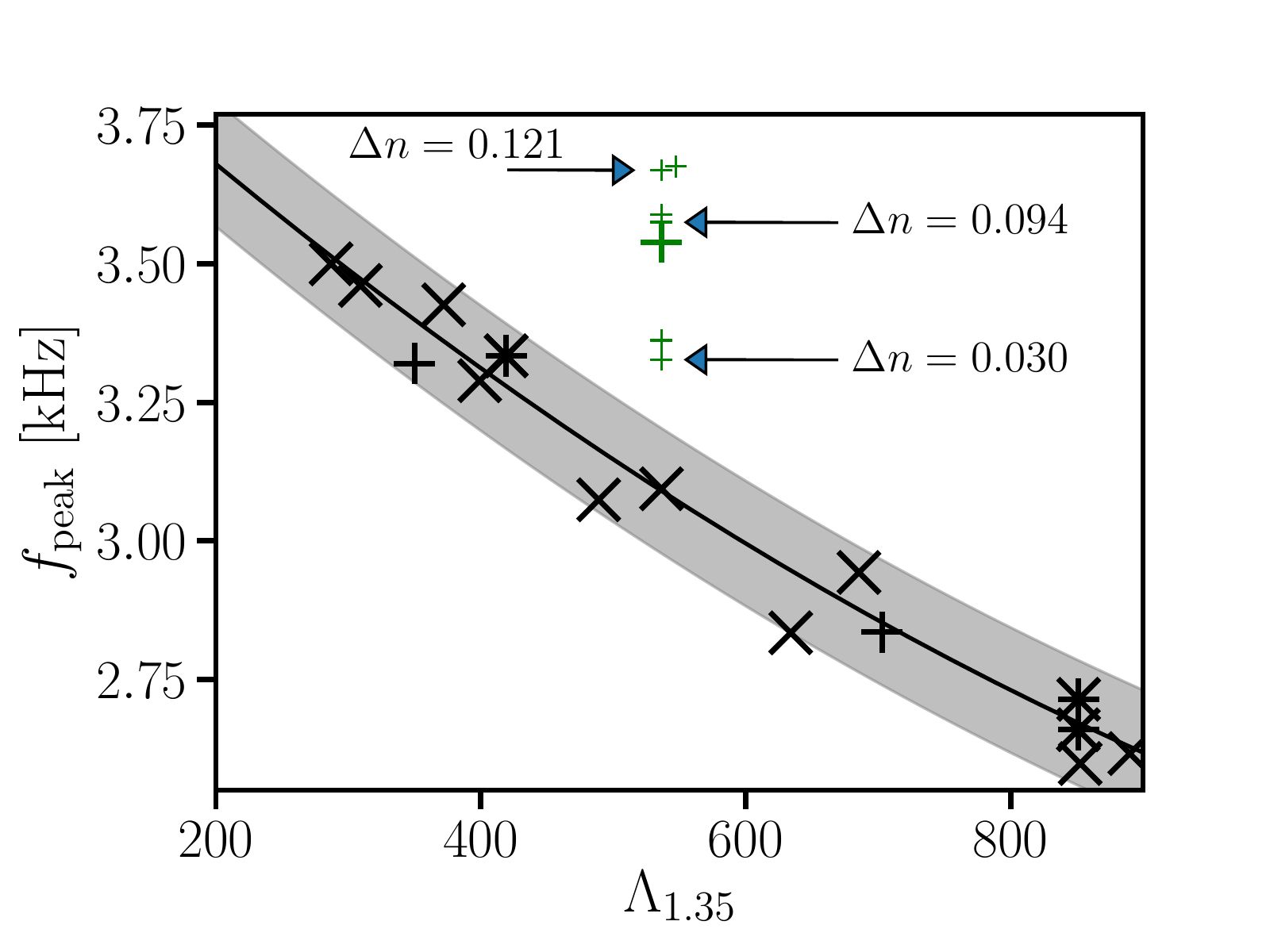}
\caption{Dominant postmerger GW frequency $f_\mathrm{peak}$ as function of tidal deformability $\Lambda$ for 1.35-1.35~$M_\odot$ mergers. The DD2F-SF models with a phase transition to deconfined quark matter (green symbols) appear as clear outliers (big symbol for DD2F-SF-1). Solid curve displays the least square fit Eq.~\eqref{eq:fit} for all purely hadronic EOSs (including three models with hyperons marked by asterisks). ALF2 and ALF4 are marked by black plus signs. EOSs incompatible with GW170817 are not shown. Arrows mark DD2F-SF models 3, 6 and 7, which feature differently strong density jumps $\Delta n$ (in $\mathrm{fm^{-3}}$) with roughly the same onset density and stiffness of quark matter.}
\label{fig:fpeaklam}
\end{figure}

Moreover, during the inspiral phase of NS mergers finite-size effects are measurable and encoded in the tidal deformability $\Lambda=\frac{2}{3}k_2\left(\frac{R}{M}\right)^5$ with the tidal Love number $k_2$~\cite{Hinderer2008,Hinderer2010}. Considering the strong dependence of $\Lambda$ on NS radii, it is clear that $f_\mathrm{peak}$ also correlates with the tidal deformability of NSs (see Fig.~\ref{fig:fpeaklam} and~\cite{Bernuzzi2015,Rezzolla2016} for plots with the tidal coupling constant including different total binary masses). It is conceivable that $\Lambda$ will be measured with significantly better precision in future observations compared to GW170817, which resulted in a measurement uncertainty on $\Lambda$ of a 1.4~$M_\odot$ NS of about 510 at the 90\% level~\cite{Abbott2017,TheLIGOScientificCollaboration2018a,Abbott2018}. For instance, an event similar to GW170817 would reduce this error by a factor of about 3 once the detectors reach their design sensitivity~\cite{Read2009,Read2013,DelPozzo2013,Wade2014,Agathos2015,Chatziioannou2015,Hotokezaka2016,Chatziioannou2018}. Similarly, it is expected that the dominant postmerger frequency will be measured to within a few 10~Hz in future nearby events with the projected improvements for the current generation of detectors~\cite{Clark2014,Clark2016,Chatziioannou2017,Bose2018,Yang2018,Torres-Rivas2018}.

{\it Observational signature of phase transitions: }
In Fig.~\ref{fig:fpeaklam} we show the dominant postmerger frequency $f_\mathrm{peak}$ as function of the tidal deformability $\Lambda_{1.35}=\Lambda(1.35~M_\odot)$ for the 1.35-1.35~$M_\odot$ mergers for all EOSs of this study. As anticipated, $f_\mathrm{peak}$ scales tightly with the tidal deformability for all EOS models (black symbols). There is only one exception: the DD2F-SF EOSs lead to significantly higher peak frequencies of 3.3~kHz to 3.7~kHz (green symbols). The purely hadronic counterpart of these EOS models without phase transition yields a peak frequency of only 3.098~kHz, while the tidal deformability parameters are identical for both types of EOSs. 

Excluding the hybrid models DD2F-SF, ALF2 and ALF4 we obtain a least square fit
\begin{equation}
f_\mathrm{peak}=(6.486\times 10^{-7}\, \Lambda^2 \,- \,2.231\times10^{-3}\,\Lambda\,+\,4.1 )~\rm kHz~,
\label{eq:fit}
\end{equation}
for all purely hadronic EOSs (solid curve in Fig.~\ref{fig:fpeaklam}). The maximum deviation between data (black symbols) and the fit~Eq.~\eqref{eq:fit} is 113~Hz (grey band in Fig.~\ref{fig:fpeaklam}), with an average scatter of 44~Hz~\footnote{The fit parameters as well as the deviations from the fit depend slightly on the chosen sample of EOS models; employing a large set of EOSs we expect to diminish any bias.}. In comparison, for the DD2F-SF-1 model the peak frequency is 448~Hz above the value which is expected from the $f_\mathrm{peak}(\Lambda)$ fit for the given tidal deformability of this EOS. 

A deviation of nearly 0.5~kHz is significant also if we assume a measurement accuracy of the tidal deformability of 100--200 and of several tens of~Hz for the peak frequency. These error bars can be achieved within the next years for events with distances similar to that of GW170817~\cite{Read2009,DelPozzo2013,Read2013,Wade2014,Agathos2015,Chatziioannou2015,Clark2014,Clark2016,Hotokezaka2016,Chatziioannou2017,Bose2018,Yang2018,Chatziioannou2018,Torres-Rivas2018,TheLIGOScientificCollaboration2018a,Dudi2018}. Note that actually the mass ratio $q=M_1/M_2$ and the combined tidal deformability $\tilde{\Lambda}$ are measured during the inspiral (for equal mass systems $\tilde\Lambda=\Lambda$). We also remark that neither $\tilde\Lambda$ nor the postmerger frequencies are too strongly affected by small variations of $q$, which we confirm by additional simulations for $q=0.8$. These simulations yield $f_\mathrm{peak}=3.79$~kHz and $f_\mathrm{peak}=3.01$~kHz for DD2F-SF-1 and DD2F, respectively, and thus the deviations are even larger. The combined tidal deformability of an asymmetric merger with $q=0.8$ is to within 5\% identical to the one of the equal-mass binary of the same total mass. Hence, small uncertainties in the determination of the mass ratio do not affect our ability to discern models with and without high-density phase transitions.

Three EOSs of our sample include a 2nd order phase transition to hyperonic matter (BHBLP~\cite{Banik2014}, SFHOY~\cite{Fortin2018} and DD2Y~\cite{Marques2017}). These EOSs follow closely the $f_\mathrm{peak}-\Lambda$ relation similarly to purely nucleonic EOSs. This is in line with the simulations for BHBLP in Ref.~\cite{Radice2017} showing no significant frequency shift compared to the nucleonic reference model. Similarly, the postmerger frequencies of the calculations with the ALF2 and ALF4 EOSs (involving continuous transitions without density jump) are consistent with the $f_\mathrm{peak}-\Lambda$ relation (black plus signs in Fig.~\ref{fig:fpeaklam}).

This indicates that only a sufficiently strong first-order phase transition (to deconfined quark matter) with a significant impact on the stellar structure (see Figs.~2 and~3 in the Supplemental Material) can alter the postmerger GW signal in such a way that a measurable deviation from the $f_\mathrm{peak}-\Lambda$ relation occurs. In these cases the formation of a quark matter core in the early postmerger phase leads to a stronger compactification of the remnant and thus to higher oscillation frequencies. The effect is less pronounced for phase transitions which are weaker in the sense that the resulting mass-radius relations deviate less from that of the purely hadronic reference model such as DD2F-SF-4 and DD2F-SF-7 (see Supplemental Material). This is quantitatively supported by considering the increase of $f_\mathrm{peak}$ as function of the density jump $\Delta n$ across the phase transition while approximately fixing other EOS parameters which regulate the onset density and the stiffness of quark matter (symbols marked by arrows in Fig.~\ref{fig:fpeaklam}). The deviation from the $\Lambda-f_\mathrm{peak}$ relation is stronger for larger density jumps, and weakens for less drastic transitions. We thus explicitly stress that we do not expect that every 1st order phase transition would lead to such clearly observable features, but that there is a class of viable hybrid star models that do exhibit the described signature. This would thus be indicative for a transition because the signature cannot result from a purely hadronic EOS as our representative sample of hadronic models shows. Note that at least in principle, any transition which is formally not first order but which is able to resemble a strong softening of the EOS in a transition region as our DD2F-SF, could lead to a similar impact on the stellar structure and thus an increase of $f_\mathrm{peak}$.

A measured peak frequency being consistent with the $f_\mathrm{peak}(\Lambda)$ fit, rules out a strong first-order phase transition as in DD2F-SF and points to either purely hadronic matter or a weak imprint of the phase transition in the probed density regime (cf. Fig.~\ref{fig:rhomax}). Clearly, an agreement with Eq.~\eqref{eq:fit} cannot inform about phase transitions at higher densities 
and about phase transitions which are that strong that they rapidly induce the collapse of the merger remnant (see below).

To understand which density regimes are probed during the postmerger evolution, we extract the largest value $\rho_\mathrm{max}^\mathrm{max}$ of the maximum rest-mass density $\rho_\mathrm{max}(t)$ during the first few milliseconds after merging. In Fig.~1 $\rho_\mathrm{max}^\mathrm{max}$ is reached at 8.8~ms for DD2F-SF-1 and at 7.4~ms for DD2F. $\rho_\mathrm{max}(t)$ can exceed $\rho_\mathrm{max}^\mathrm{max}$ at later times, but here we are interested in the initial phase when the postmerger GW emission is strongest.

\begin{figure}
\includegraphics[width=8.cm]{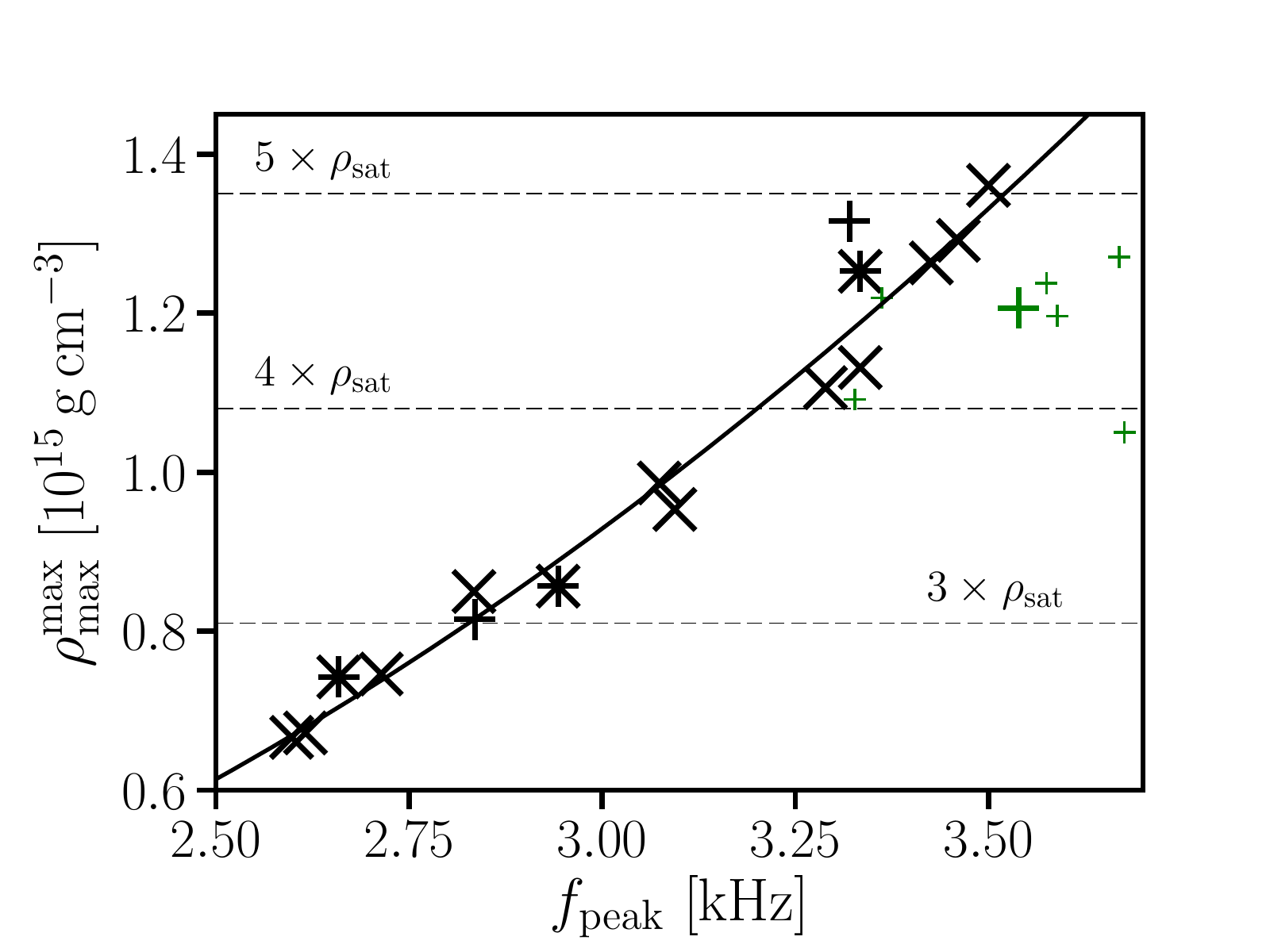}
\caption{Maximum rest-mass density $\rho_\mathrm{max}^\mathrm{max}$ during the first milliseconds of the postmerger phase as function of the dominant postmerger GW frequency $f_\mathrm{peak}$ for 1.35-1.35~$M_\odot$ mergers. Green symbols display results for DD2F-SF (big symbol for DD2F-SF-1). Asterisks indicate models with hyperons. Black plus signs display ALF2/4. Solid curve is a second order polynomial least square fit to the data excluding hybrid EOSs.}
\label{fig:rhomaxfpeak}
\end{figure}

Figure~\ref{fig:rhomaxfpeak} displays $\rho_\mathrm{max}^\mathrm{max}$ as function of $f_\mathrm{peak}$ for all 1.35-1.35~$M_\odot$ simulations. The figure reveals a correlation between  $\rho_\mathrm{max}^\mathrm{max}$ and $f_\mathrm{peak}$, which can be approximated by the least square fit
\begin{equation}\label{eq:rhomax}
\rho_\mathrm{max}^\mathrm{max}=\left( a\,f_\mathrm{peak}^2 + b\,f_\mathrm{peak} + c \right)~\mathrm{g\,cm^{-3}}
\end{equation}
with $f_\mathrm{peak}$ in kHz and $a=1.89\times 10^{14}$, $b=- 4.13\times 10^{14}$ and $c=4.66\times 10^{14}$ (excluding hybrid models). This result shows that a measurement of $f_\mathrm{peak}$ can serve as a proxy for the highest rest-mass density which is reached during the initial phase of the postmerger evolution.

If the dominant postmerger GW frequency is in agreement with the $f_\mathrm{peak}(\Lambda)$ fit, Eq.~\eqref{eq:rhomax} approximately determines up to which rest-mass density no strong first-order phase transition of similar type as the ones in DD2F-SF occurred for $M_\mathrm{tot}=2.7~M_\odot$ (see extended discussion in Supplemental Material). 

{\it Relation to other works: } It is instructive to compare our finding with the merger simulations of absolutely stable strange stars~\cite{Bodmer1971,Witten1984,Alcock1986,Haensel1986}, which do not feature a phase transition at supernuclear densities but a large density jump at the surface. The calculation for the model EOS MIT40 in~\cite{Bauswein2012} yields $f_{\rm peak}=2.62$~kHz, while the tidal deformability $\Lambda(1.35~M_\odot)=1161.7$ for this EOS. This model shows a somewhat weaker but similar trend as DD2F-SF in Fig.~\ref{fig:fpeaklam}. In principle, a deviation from the fit~Eq.~\eqref{eq:fit} may thus also be characteristic for absolutely stable strange quark matter~\cite{Bodmer1971,Witten1984}. However, this particular model of absolutely stable strange quark matter is incompatible with existing constraints on $\Lambda$, and it is likely that a merger of absolutely stable strange stars~\cite{Bauswein2009,Bauswein2010a} would lead to an electromagnetic counterpart different from that of GW170817~\cite{Abbott2017b}. We thus suspect that such a scenario would be distinguishable from the collision of two hybrid stars as described in this study.

Recently, Ref.~\cite{Most2018a} used the model EOS CMF of~\cite{Dexheimer2010} for hadronic and quark matter in merger simulations. Compared to our DD2F-SF, the phase transition of CMF has a very different impact on the stellar structure and consequently on merger simulations (according to Fig. 5 in~\cite{Dexheimer2010} this EOS does not yield gravitationally stable hybrid stars with extended quark matter cores). We find a massive gravitationally stable quark matter core with a strong imprint on the postmerger GW frequency for DD2F-SF. In comparison, the CMF EOS leads to a small quark matter fraction during most of the postmerger evolution. Only at late times the quark matter fraction increases and immediately induces the gravitational collapse of the remnant. Hence, the influence on the GW frequency is significantly weaker compared to our model. In comparison to its purely hadronic reference model, the CMF EOS results in an earlier collapse of the remnant and a dephasing of about 3 radian within 30 cycles. The postmerger frequency is thus shifted only slightly. Such signatures cannot be easily interpreted as being an unambiguous feature for the occurrence of quark matter. A similar phase and frequency shift and a shorter remnant life time can as well be expected from a purely hadronic EOS, being somewhat softer at higher densities compared to the CMF hadronic reference model.

In the Supplemental Material we discuss our findings in the context of empirical relations between $f_{\rm peak}$ and radii of nonrotating NSs~\cite{Bauswein2012,Bauswein2012a,Hotokezaka2013a}.

{\it Summary and conclusions:} Within this work we describe a way to detect a strong first-order phase transition in NSs, complementary to efforts at the future experimental facilities FAIR at GSI and NICA in Dubna dedicated to the study of compressed matter in heavy-ion collisions~\cite{Friman2011,Blaschke2016}. Our scenario involves quantities which have been shown to be measurable in future GW detections. We provide evidence that the described signature can only be related to a strong first-order phase transition by showing that a representative set of hadronic EOS models behaves differently. These results highlight the complementarity of the information which can be obtained from the inspiral and the postmerger phase of NS mergers. It stresses the importance of kHz GW astronomy both with current second-generation~\cite{LIGOScientificCollaboration2015,LIGOScientificCollaboration2015a,Acernese2015,Collaboration2017b} and proposed third-generation detectors like~\cite{Punturo2010,Hild2011,Miller2015}. Future work should consider a larger class of EOS models with a hadron-quark phase transition to determine under which conditions a clearly distinguishable imprint on the GW signal can be identified. We will also investigate other observables like electromagnetic counterparts and secondary features of the GW spectrum.

\acknowledgements{Acknowledgements: AB acknowledges support by the European Research Council (ERC) under the European Union's Horizon 2020 research and innovation programme under grant agreement No. 759253 and the Klaus-Tschira Foundation. NUB and TF acknowledge support from the Polish National Science Center (NCN) under grant no. UMO-2016/23/B/ST2/00720. DB acknowledges support through the Russian Science Foundation under project No. 17-12-01427 and the MEPhI Academic Excellence Project under contract No.~02.a03.21.0005. We acknowledge stimulating discussions during the EMMI Rapid Reaction Task Force: The physics of neutron star mergers at GSI/FAIR and the support of networking activities by the COST Actions CA15213 ``THOR'', CA16117 ``ChETEC'' and CA16214 ``PHAROS''.}


\end{bibunit}


\clearpage
\newpage
\begin{bibunit}
\setcounter{figure}{0}
\setcounter{equation}{0}
\setcounter{page}{1}

\onecolumngrid
\begin{center}
{\large \bf Supplemental Material}
\vspace*{1cm}
\twocolumngrid
\end{center}

\section{Equations of state}

We provide here information about the underlying model for the DD2F and the DD2F-SF equations of state (EOSs) as well as the set of candidate EOSs, which serve as a representative sample of purely hadronic EOSs.

The DD2F EOS is based on the relativistic mean-field approach with density dependent couplings \cite{Typel2005,Typel2010,Alvarez-Castillo2016}, which is consistent with the EOS constraint derived from an analysis of transverse and elliptic flow data of heavy-ion collision experiments~\cite{Danielewicz2002,Tsang2018}. At low densities and temperatures, the presence of nuclear clusters is taken into account consistently within the modified nuclear statistical equilibrium model of Ref.~\cite{Hempel2010,Hempel2012}. DD2F is consistent with all presently known constraints, e.g., neutron matter from chiral effective field theory~\cite{Krueger2013}, the nuclear symmetry energy and its slope~\cite{Lattimer2013,Oertel2017}, the maximum mass of nonrotating neutron stars (NSs)~\cite{Antoniadis2013,Arzoumanian2018a}, and stellar parameters in agreement with the analysis of GW170817~\cite{Abbott2017,Bauswein2017,De2018,Abbott2018}.

The quark-matter EOS in the high-density regime of DD2F-SF is based on the phenomenological two-flavor string-flip model (SF), derived within the density-functional formalism depending on scalar and vector quark densities (for details see Ref.~\cite{Kaltenborn2017} and references therein). Deconfinement is considered via an effective string potential, which distinguishes SF from common chiral quark-matter approaches, e.g., models of the Nambu-Jona-Lasinio type~\cite{Nambu1961,Klevansky1992,Ruester2005,Blaschke2005} where (de)confinement is absent. A medium-dependent reduction of the string tension is modeled via a Gaussian functional~\cite{Kaltenborn2017}. Divergent quark masses suppress quark degrees of freedom at low densities. The SF model includes an additional dependence on the isovector-vector density, i.e. the equivalent to $\rho$-meson interactions in hadronic matter~\cite{Fischer2018}.

\begin{figure}
\includegraphics[width=8.cm]{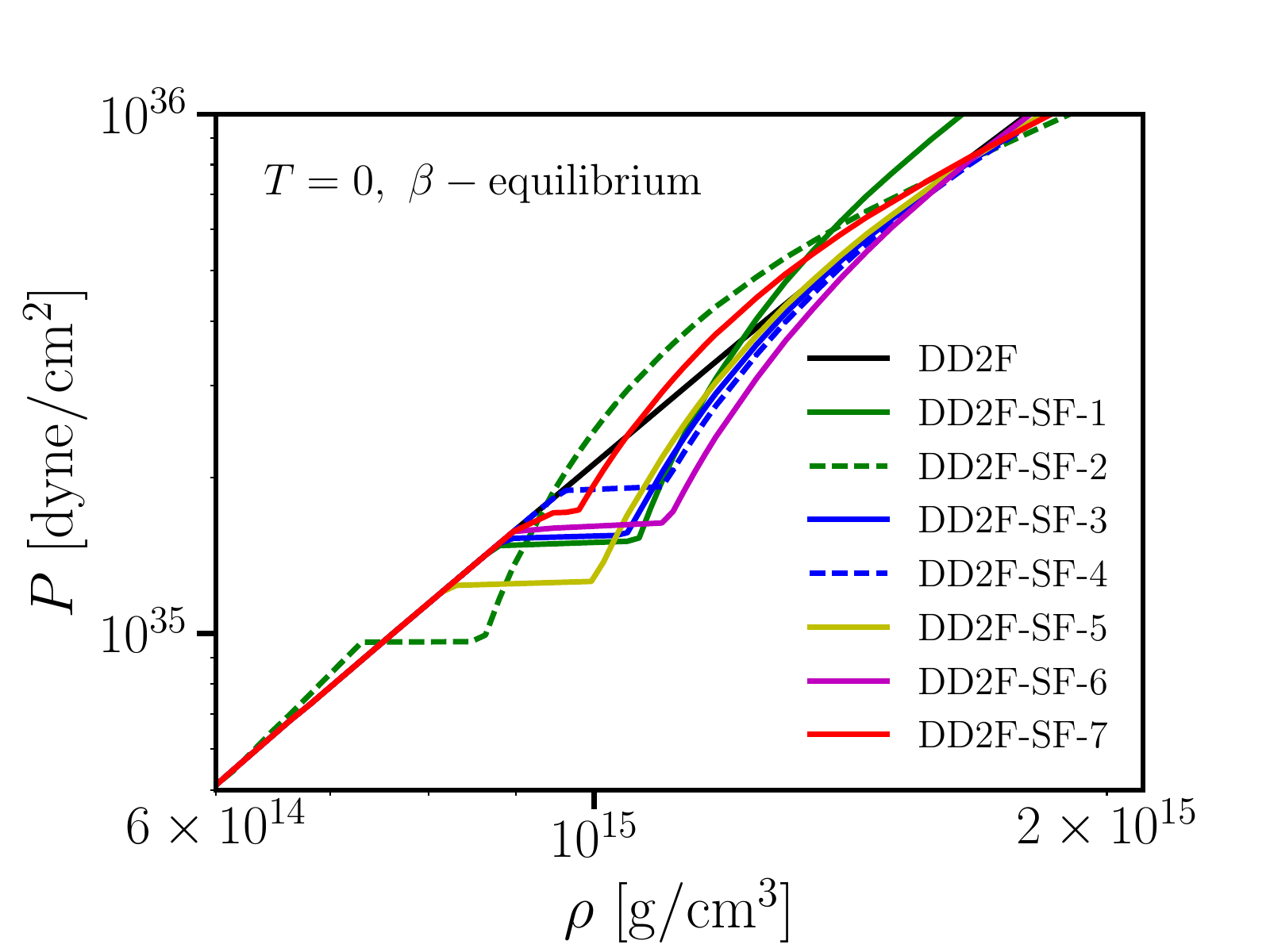}
\caption{Pressure as function of the rest-mass density for different hybrid EOSs of the DD2F-SF class. Black curve displays the purely hadronic reference model DD2F.}
\label{fig:eos}
\end{figure}

\begin{figure}
\includegraphics[width=8.cm]{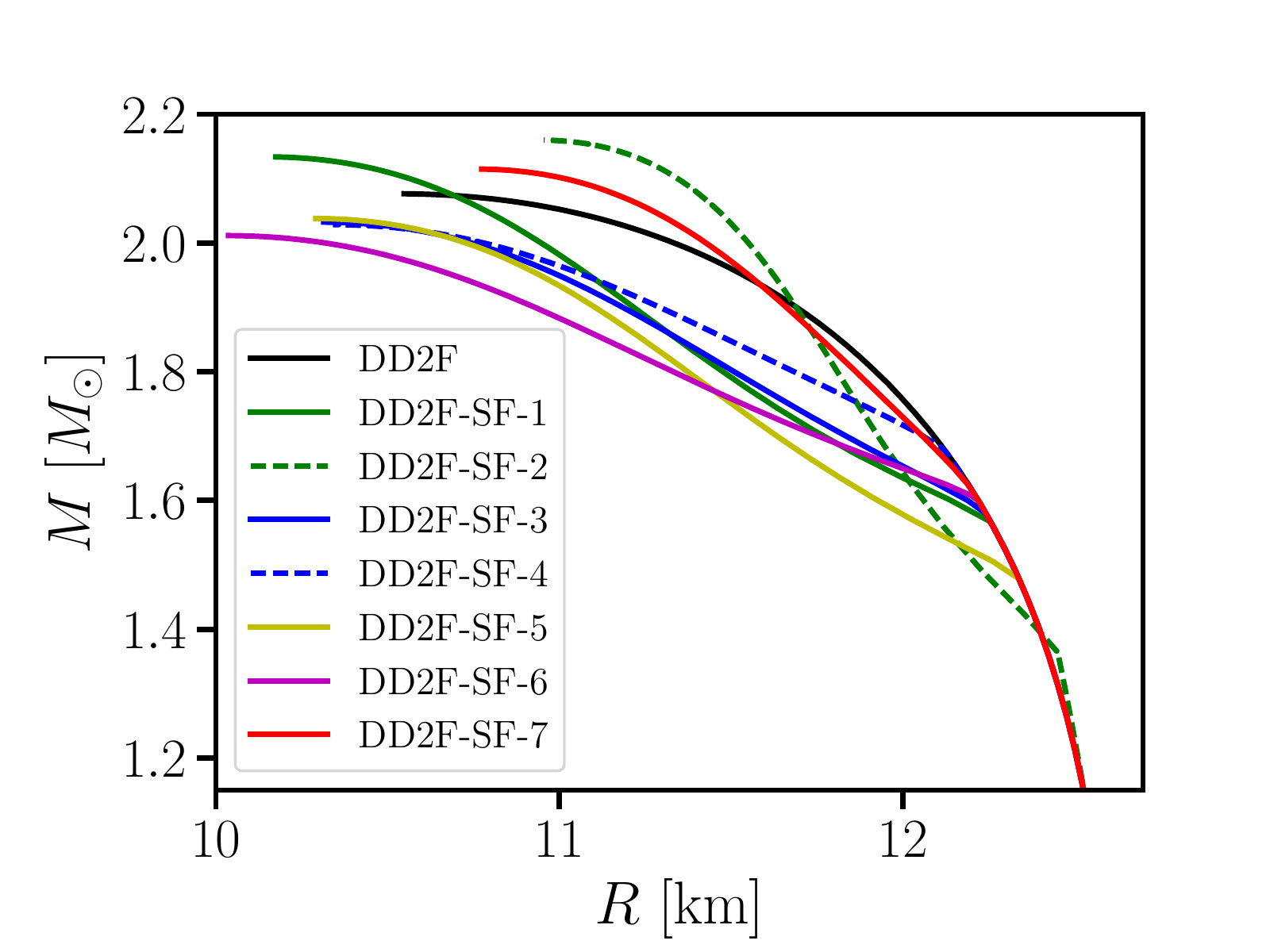}
\caption{Mass-radius relations for the DD2F-SF EOSs employed in this study. $M$ is the gravitational mass, $R$ the circumferential eigen radius for nonrotating cold NSs. Solid green (black) curve displays the $M$-$R$ relation for DD2F-SF-1 (DD2F).}
\label{fig:tovquark}
\end{figure}

\begin{figure}
\includegraphics[width=8.cm]{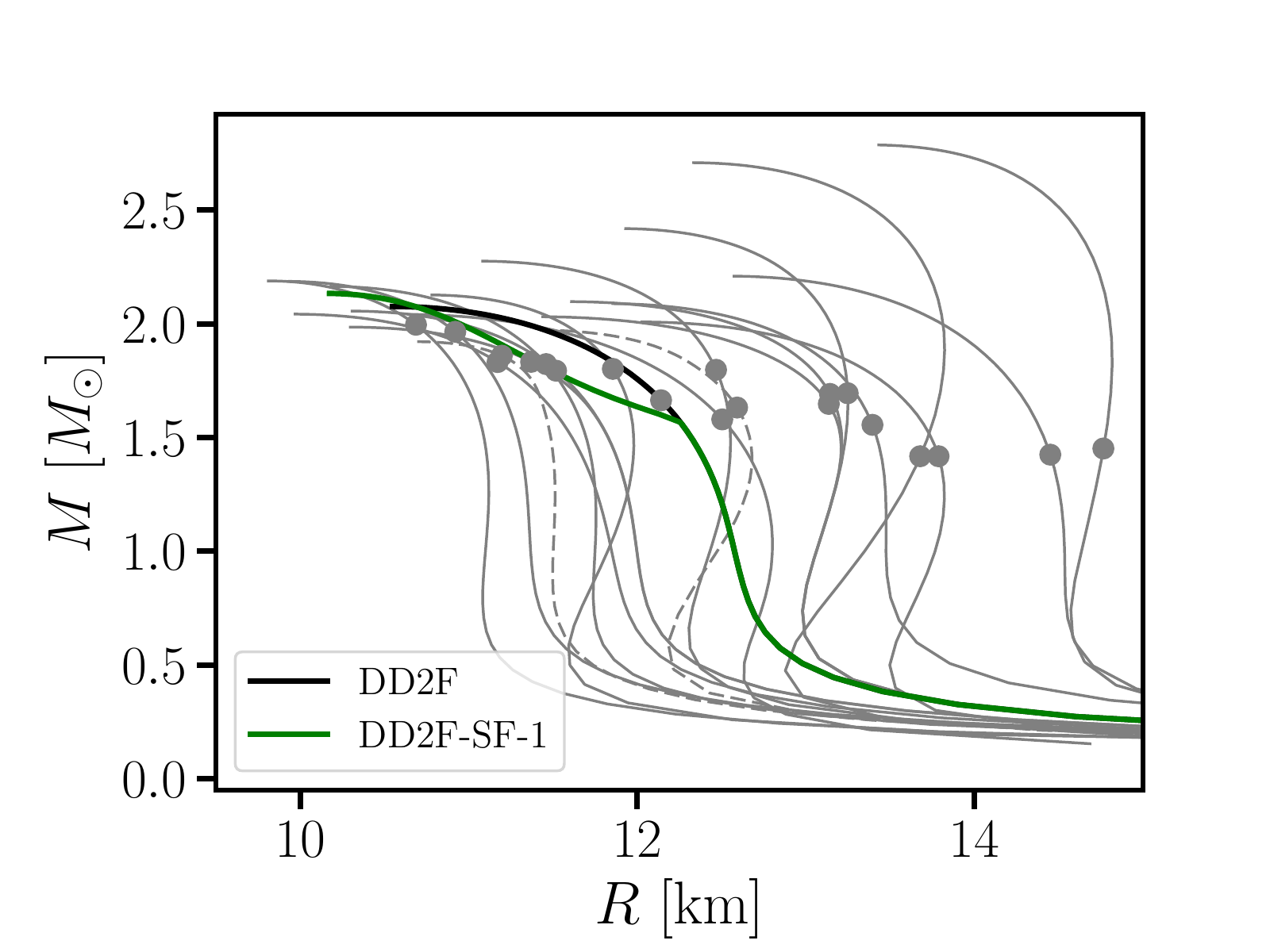}
\caption{Mass-radius relations for the model EOSs employed in this study. $M$ is the gravitational mass, $R$ the circumferential eigen radius for nonrotating cold NSs. Green (black) curve displays the $M$-$R$ relation for DD2F-SF-1 (DD2F). Gray dashed curves correspond to ALF2 and ALF4. Gray dots indicate $M_\mathrm{fid}$, which corresponds to the stellar configuration whose central rest-mass density equals the maximum density of the early postmerger evolution in a 1.35-1.35~$M_\odot$ simulation with the same EOS.}
\label{fig:tov}
\end{figure}

In this work we employ seven different sets of SF parameters~\cite{Kaltenborn2017,Bastian2018,Fischer2018} listed in Tab.~\ref{tab1} and we call the resulting EOS models DD2F-SF-n with $n \in \{1,2,3,4,5,6,7\}$. We use the acronym DD2F-SF if we refer to the whole class of all seven hybrid models. In the main article we focus on the exemplary hybrid model DD2F-SF-1, which was also considered in~\cite{Fischer2018}. The SF parameters of our seven quark matter EOSs correspond to different onset and final densities of the first-order phase transition, which are provided in Tab.~\ref{tab1} (see also Fig.~\ref{fig:eos}). These phase boundaries of DD2F-SF have a mild temperature dependence for the relevant range, e.g., for DD2F-SF-1 at $T=20$~MeV we have $\rho_{\rm onset}=2.90\times\rho_{\rm sat}$ and $\rho_{\rm final}=3.81\times\rho_{\rm sat}$ with $\rho_\mathrm{sat}=2.7\times 10^{14}~\mathrm{g\,cm^{-3}}$ being the nuclear saturation density (to be compared with the values for $T=0$ in Tab.~\ref{tab1}). The first-order phase transition leads to a significant softening of the EOS in the phase transition region, which represents a phase where hadrons and quarks coexist. Vector repulsion, including higher-order terms, in the pure quark matter phase is essential for stable stellar configurations~\cite{Benic2015,Klaehn2015}.  The chosen SF parameters lead also to a variation of the stiffness of the quark matter EOS (see Fig.~\ref{fig:eos}).

Our parameter choices yield maximum masses for nonrotating stars between 2.01~$M_\odot$ and 2.16~M$_\odot$ for the different DD2F-SF models (see Tab.~\ref{tab1}). The different properties of the quark phase (onset densities, density jumps and quark phase stiffness) are also apparent in the resulting mass-radius relations of nonrotating cold stars for DD2F-SF, which are shown in Fig.~\ref{fig:tovquark} together with the purely hadronic reference model DD2F (black curve). DD2F-SF-1 as reference is indicated by a solid green line. Note that for DD2F-SF-2 we employ a slightly modified variant of the hadronic DD2F which includes an excluded volume modeling~\cite{Typel2016}. This leads to minor modifications of the hadronic phase just below the onset density (see Figs.~\ref{fig:eos} and Fig.~\ref{fig:tovquark}) and is responsible for the slightly larger tidal deformability of DD2F-SF-2 in Fig.~3 of the main paper.

The stellar properties of our reference models DD2F and DD2F-SF-1 are also displayed in Fig.~\ref{fig:tov}. The figure includes mass-radius relations of other EOS models (grey lines), which serve as representative sample of hadronic EOSs in this study. This set includes APR~\cite{Akmal1998}, BHBLP~\cite{Banik2014}, BSK20~\cite{Goriely2010}, BSK21~\cite{Goriely2010}, DD2~\cite{Hempel2010,Typel2010}, {eosUU}~\cite{Wiringa1988}, GS2~\cite{Shen2011}, LS220~\cite{Lattimer1991}, LS375~\cite{Lattimer1991}, NL3~\cite{Hempel2010,Lalazissis1997a}, SFHO~\cite{Steiner2013}, SFHX~\cite{Steiner2013}, Sly4~\cite{Douchin2001}, TM1~\cite{Sugahara1994a,Hempel2012} and TMA~\cite{Toki1995,Hempel2012} (see~\cite{Bauswein2012a,Bauswein2013a,Bauswein2014a} for the meaning of the acronyms and more details about the different EOSs; GS2, LS375, NL3, TM1 and TMA are incompatible with the 90\% credible level of the tidal deformability constraint deduced from GW170817~\cite{Abbott2017,Abbott2018,De2018}). Additionally, we consider modified versions of SFHO and DD2 with a 2nd order phase transition to hyperonic matter~\cite{Fortin2018,Marques2017}, which we refer to as SFHOY and DD2Y, respectively. Hyperonic interactions for these models have been chosen to be compatible with hypernuclear data and a cold NS maximum mass of 2 $M_\odot$, such that these EOSs fulfill all presently available constraints. We also investigate the two models ALF2 and ALF4~\cite{Alford2005,Read2009a} (implemented as piecewise polytropes), which resemble hybrid EOSs with a more continuous transition to quark matter. As discussed in~\cite{Alford2005} these models (gray dashed curves in Fig.~\ref{fig:tov}) do not show qualitative differences in the mass-radius relations compared to purely hadronic EOSs.

\begin{table*}
\caption{ \label{tab1} Different hybrid EOS models of the DD2F-SF class employed in this study. $D_0$, $\alpha$, $a$, $b$, $c$, $\rho_1$ are SF parameters as defined in~\cite{Kaltenborn2017,Bastian2018,Fischer2018}. $n_\mathrm{onset}$ and $\Delta n$ are the onset baryon density and baryon density jump of the phase transition (in neutrinoless beta equilibrium and for zero temperature). $M_\mathrm{onset}$ is the lowest NS mass with a quark matter core. $M_\mathrm{max}$ is the maximum mass of cold nonrotating NSs. $f_\mathrm{peak}$ denotes the dominant postmerger GW frequency.}
 \begin{ruledtabular}
 \begin{tabular}{|l|l|l|l|l| l|l|l|l|l| l|l|}
\hline    
EOS & $\sqrt{D_0}$ & $\alpha$ & $a$ & $b$ & $c$ & $\rho_1$ & $n_\mathrm{onset}$ & $\Delta n$ & $M_\mathrm{onset}$ & $M_\mathrm{max}$ & $f_\mathrm{peak}$ \\ 

& $(\mathrm{MeV})$ & $(\mathrm{fm^6})$ & $(\mathrm{MeV\,fm^3})$ & $(\mathrm{MeV\,fm^9})$ & $(\mathrm{fm^6})$ & $(\mathrm{MeV\,fm^3})$ & $(\mathrm{fm^{-3}})$ & $(\mathrm{fm^{-3}})$ & $(M_\odot)$ & $(M_\odot)$ & $(\mathrm{kHz})$ \\ \hline

DD2-SF-1 & 265 & 0.39 & -4.0 & 1.6 & 0.025 & 80.0 & 0.533 & 0.106& 1.57 & 2.13 & 3.54 \\ 
DD2-SF-2 & 250 & 0.60 & 10.0 & 0.0 & 0.000 & 80.0 & 0.466 & 0.057& 1.37 & 2.16 & 3.68 \\ 
DD2-SF-3 & 240 & 0.36 & 1.0  & 0.5 & 0.015 & 80.0 & 0.538 & 0.094& 1.58 & 2.03 & 3.58 \\ 
DD2-SF-4 & 240 & 0.34 & 1.0  & 0.5 & 0.015 & 80.0 & 0.580 & 0.082& 1.68 & 2.03 & 3.36 \\ 
DD2-SF-5 & 240 & 0.38 & 1.0  & 0.5 & 0.015 & 80.0 & 0.499 & 0.108& 1.48 & 2.04 & 3.59 \\ 
DD2-SF-6 & 240 & 0.30 & -3.0 & 0.8 & 0.015 & 80.0 & 0.545 & 0.121& 1.60 & 2.01 & 3.67 \\ 
DD2-SF-7 & 240 & 0.47 & 7.0  & 0.2 & 0.015 & 80.0 & 0.562 & 0.030& 1.62 & 2.11 & 3.33 \\ 

\end{tabular}
 \end{ruledtabular} 
\end{table*}

\section{Postmerger densities}
In the main article we show that a measurement of the dominant postmerger gravitational-wave (GW) frequency can be used to estimate the highest rest-mass density $\rho_\mathrm{max}^\mathrm{max}$ which occurs during the early postmerger evolution (see Fig.~4 in the main article). For softer EOSs higher densities are reached in the postmerger phase. This information can be mapped to nonrotating stellar configurations and roughly determines up to which NS mass the presence of a strong phase transition is probed by the postmerger GW emission of 1.35-1.35~$M_\odot$ binaries as described in the main part.

\begin{figure}
\includegraphics[width=8.cm]{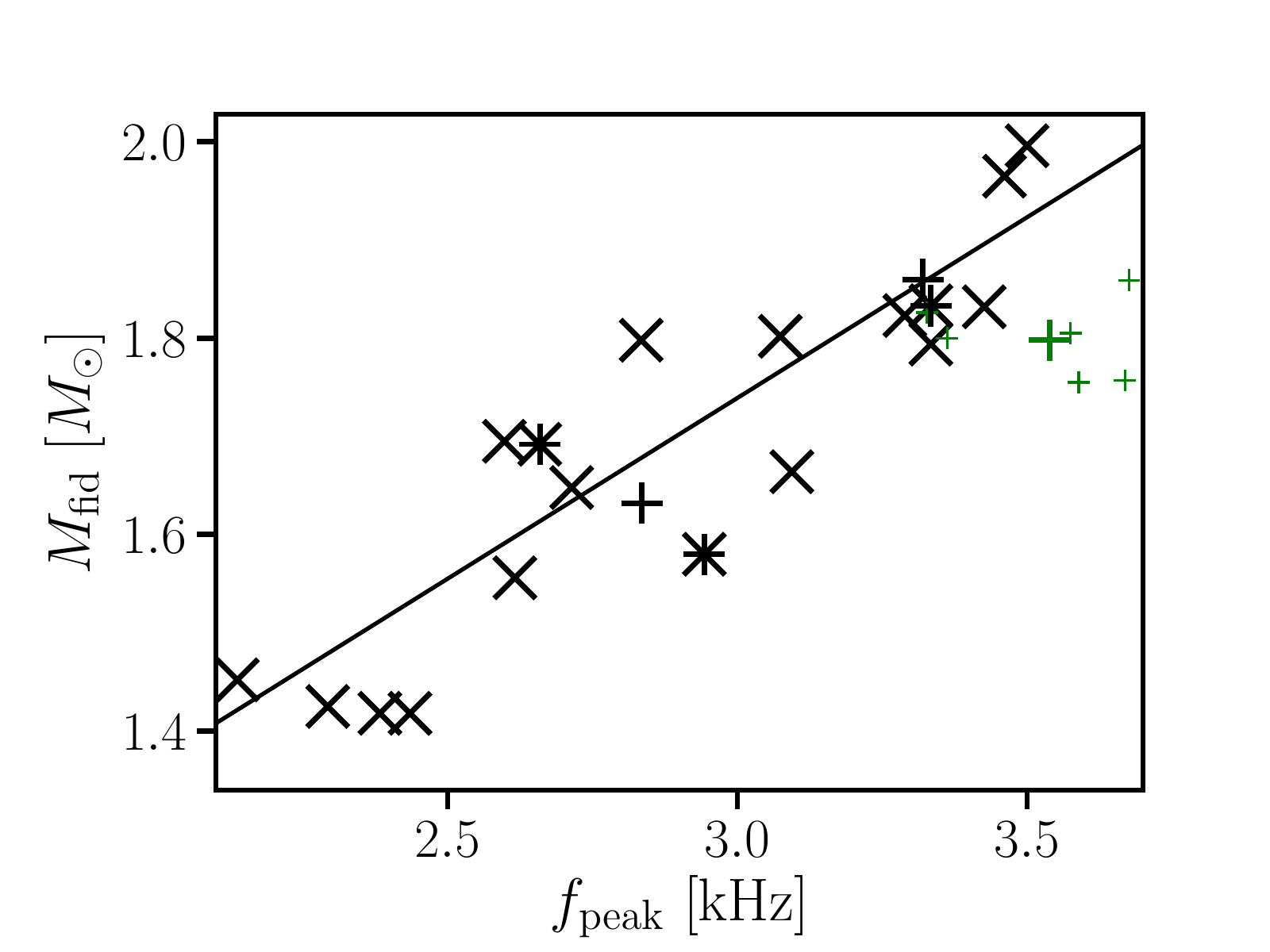}
\caption{Gravitational mass $M_\mathrm{fid}$ of nonrotating NSs whose central rest-mass density equals the maximum rest-mass density $\rho_\mathrm{max}^\mathrm{max}$ during the first few milliseconds of the postmerger evolution, for 1.35-1.35~$M_\odot$ mergers producing postmerger GW emission with frequency $f_\mathrm{peak}$. Green symbols display results for DD2F-SF (big green plus sign for DD2F-SF-1). Solid curve is a second order polynomial least square fit to the data excluding hybrid EOSs. Asterisks mark models with hyperonic matter. Black plus signs indicate ALF2 and ALF4. Models incompatible with GW170817 are not shown.}
\label{fig:fidmfpeak}
\end{figure}

To this end we identify the nonrotating stellar configuration with a gravitational mass $M_\mathrm{fid}=M(\rho_\mathrm{max}^\mathrm{max})$ whose central rest-mass density equals $\rho_\mathrm{max}^\mathrm{max}$. Figure~\ref{fig:fidmfpeak} shows $M_\mathrm{fid}$ as function of $f_\mathrm{peak}$ for all 1.35-1.35~$M_\odot$ simulations. We also plot $M_\mathrm{fid}$ as gray dots on the corresponding mass-radius relations of the different EOSs investigated in this study. The dots indicate $M_\mathrm{fid}$ for 1.35-1.35~$M_\odot$ mergers and illustrate which NS mass regime is probed by the postmerger remnant. Events with higher total binary masses $M_\mathrm{tot}$ would lead to higher densities in the postmerger phase. Consequently, $M_\mathrm{fid}$ increases with $M_\mathrm{tot}$.

\section{Neutron star radius measurements from $f_\mathrm{peak}$}
\begin{figure}
\includegraphics[width=8.cm]{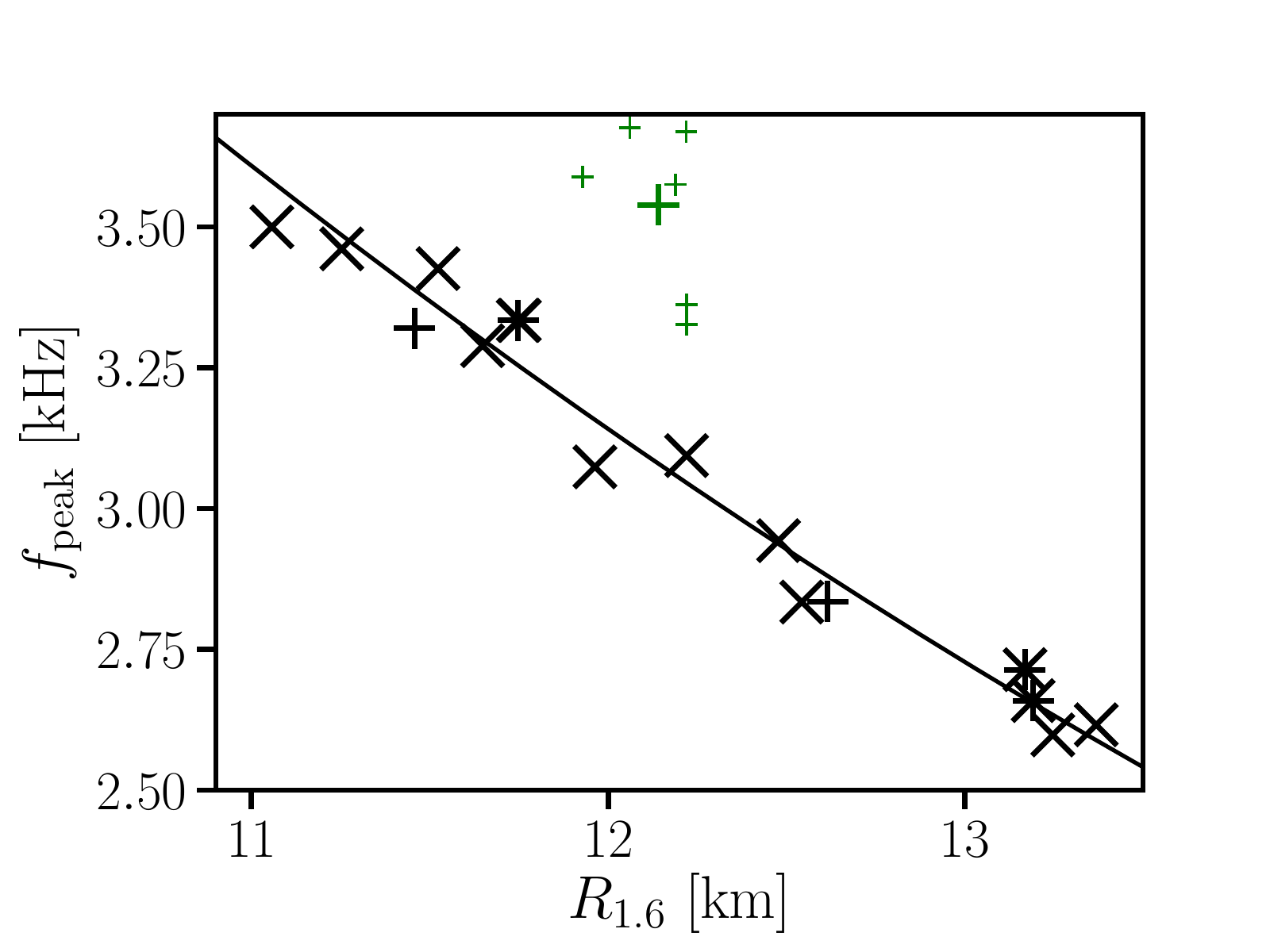}
\caption{Dominant postmerger GW frequency $f_\mathrm{peak}$ as function of the radius $R_{1.6}$ of a nonrotating NS with 1.6~$M_\odot$ for 1.35-1.35~$M_\odot$ binaries. The DD2F-SF models are shown by green symbols (big green plus sign for DD2F-SF-1). Asterisks mark hyperonic EOSs.  Black plus signs indicate ALF2 and ALF4. The solid curve provides a second order polynomial least square fit to the data (black symbols, excluding hybrid EOSs). Models incompatible with GW170817 are not shown.}
\label{fig:fpeakr16}
\end{figure}

We briefly comment on the empirically found relations between $f_{\rm peak}$ and radii $R$ of a nonrotating NS~\cite{Bauswein2012,Bauswein2012a,Hotokezaka2013a}, which can be employed for accurate and robust NS radius measurements under the assumption of purely hadronic EOSs~\cite{Clark2014,Clark2016,Chatziioannou2017}.
Our results in the main article show that such relations do not generically hold for EOSs with a strong first-order phase transition to quark matter since such models give rise to generally higher frequencies relative to the $f_\mathrm{peak}(R)$ relation formed by purely hadronic EOSs. This is visible in Fig.~\ref{fig:fpeakr16} for the relation between $f_\mathrm{peak}$ and the radius of a nonrotating NS with 1.6~$M_\odot$. If there is evidence for the presence of a strong phase transition, a measurement of $f_\mathrm{peak}$ thus only establishes an accurate lower bound on NS radii. The actual radius may then be up to about 1~km larger than the one inferred from $f_\mathrm{peak}(R)$~relations of purely hadronic EOSs if the merger remnant contains a large quark matter core as for our 1.35-1.35~$M_\odot$ mergers with DD2F-SF. (The deviation of the DD2F-SF models in $f_\mathrm{peak}(R)$ relations is larger for $R=R(1.35~M_\odot)$ and gets smaller for $R=R(1.8~M_\odot)$ since the latter radius reflects the occurrence of quark matter.)

It is likely that beside the signature uncovered in this work, additional information about the presence of a strong first-order phase transition will become available either by other astronomical measurements (e.g. neutrino signals and other observables of near-by core-collapse supernovae~\cite{Fischer2018}) or by the merger observation itself. For instance, we find that the slope, $\frac{d f_\mathrm{peak}}{d M_\mathrm{tot}}$, for mergers involving quark matter like the DD2F-SF is significantly steeper compared to the slope of purely hadronic models with comparable $f_\mathrm{peak}$ (cf. Fig.~1 in Ref.~\cite{Bauswein2014a}). Here we compare DD2F-SF-1 and the hadronic models APR~\cite{Akmal1998}, eosUU~\cite{Wiringa1988} and SLy4~\cite{Douchin2001}, which lead to peak frequencies in the range between 3.54 and 3.43~kHz for $M_\mathrm{tot}=2.7~M_\odot$. 
For DD2F-SF-1, the slope~\footnote{We determine $\frac{d f_\mathrm{peak}}{d M_\mathrm{tot}}$ by $(f_\mathrm{peak}(2.7\,M_\odot)-f_\mathrm{peak}(2.6\,M_\odot))/0.1M_\odot$.} equals 3.6~kHz/$M_\odot$ compared to 0.55~kHz/$M_\odot$, 0.28~kHz/$M_\odot$ and 1.56~kHz/$M_\odot$ for APR, eosUU and SLy4.
Observationally, the determination of $\frac{d f_\mathrm{peak}}{d M_\mathrm{tot}}$ requires two measurements of $f_\mathrm{peak}$ for different binary masses~\cite{Bauswein2014a}.


\begin{thebibliography}{122}
\expandafter\ifx\csname natexlab\endcsname\relax\def\natexlab#1{#1}\fi
\expandafter\ifx\csname bibnamefont\endcsname\relax
  \def\bibnamefont#1{#1}\fi
\expandafter\ifx\csname bibfnamefont\endcsname\relax
  \def\bibfnamefont#1{#1}\fi
\expandafter\ifx\csname citenamefont\endcsname\relax
  \def\citenamefont#1{#1}\fi
\expandafter\ifx\csname url\endcsname\relax
  \def\url#1{\texttt{#1}}\fi
\expandafter\ifx\csname urlprefix\endcsname\relax\def\urlprefix{URL }\fi
\providecommand{\bibinfo}[2]{#2}
\providecommand{\eprint}[2][]{\url{#2}}

\bibitem[{\citenamefont{{Bazavov} et~al.}(2012)\citenamefont{{Bazavov}, {Ding},
  {Hegde}, {Kaczmarek}, {Karsch}, {Laermann}, {Mukherjee}, {Petreczky},
  {Schmidt}, {Smith} et~al.}}]{Bazavov2012}
\bibinfo{author}{\bibfnamefont{A.}~\bibnamefont{{Bazavov}}},
  \bibinfo{author}{\bibfnamefont{H.-T.} \bibnamefont{{Ding}}},
  \bibinfo{author}{\bibfnamefont{P.}~\bibnamefont{{Hegde}}},
  \bibinfo{author}{\bibfnamefont{O.}~\bibnamefont{{Kaczmarek}}},
  \bibinfo{author}{\bibfnamefont{F.}~\bibnamefont{{Karsch}}},
  \bibinfo{author}{\bibfnamefont{E.}~\bibnamefont{{Laermann}}},
  \bibinfo{author}{\bibfnamefont{S.}~\bibnamefont{{Mukherjee}}},
  \bibinfo{author}{\bibfnamefont{P.}~\bibnamefont{{Petreczky}}},
  \bibinfo{author}{\bibfnamefont{C.}~\bibnamefont{{Schmidt}}},
  \bibinfo{author}{\bibfnamefont{D.}~\bibnamefont{{Smith}}},
  \bibnamefont{et~al.}, \bibinfo{journal}{\prl} \textbf{\bibinfo{volume}{109}},
  \bibinfo{eid}{192302} (\bibinfo{year}{2012}).

\bibitem[{\citenamefont{{Bors{\'a}nyi}
  et~al.}(2014)\citenamefont{{Bors{\'a}nyi}, {Fodor}, {Hoelbling}, {Katz},
  {Krieg}, and {Szab{\'o}}}}]{Borsanyi2014}
\bibinfo{author}{\bibfnamefont{S.}~\bibnamefont{{Bors{\'a}nyi}}},
  \bibinfo{author}{\bibfnamefont{Z.}~\bibnamefont{{Fodor}}},
  \bibinfo{author}{\bibfnamefont{C.}~\bibnamefont{{Hoelbling}}},
  \bibinfo{author}{\bibfnamefont{S.~D.} \bibnamefont{{Katz}}},
  \bibinfo{author}{\bibfnamefont{S.}~\bibnamefont{{Krieg}}}, \bibnamefont{and}
  \bibinfo{author}{\bibfnamefont{K.~K.} \bibnamefont{{Szab{\'o}}}},
  \bibinfo{journal}{Physics Letters B} \textbf{\bibinfo{volume}{730}},
  \bibinfo{pages}{99} (\bibinfo{year}{2014}).

\bibitem[{\citenamefont{{Bazavov} et~al.}(2014)\citenamefont{{Bazavov},
  {Bhattacharya}, {DeTar}, {Ding}, {Gottlieb}, {Gupta}, {Hegde}, {Heller},
  {Karsch}, {Laermann} et~al.}}]{Bazavov2014}
\bibinfo{author}{\bibfnamefont{A.}~\bibnamefont{{Bazavov}}},
  \bibinfo{author}{\bibfnamefont{T.}~\bibnamefont{{Bhattacharya}}},
  \bibinfo{author}{\bibfnamefont{C.}~\bibnamefont{{DeTar}}},
  \bibinfo{author}{\bibfnamefont{H.-T.} \bibnamefont{{Ding}}},
  \bibinfo{author}{\bibfnamefont{S.}~\bibnamefont{{Gottlieb}}},
  \bibinfo{author}{\bibfnamefont{R.}~\bibnamefont{{Gupta}}},
  \bibinfo{author}{\bibfnamefont{P.}~\bibnamefont{{Hegde}}},
  \bibinfo{author}{\bibfnamefont{U.~M.} \bibnamefont{{Heller}}},
  \bibinfo{author}{\bibfnamefont{F.}~\bibnamefont{{Karsch}}},
  \bibinfo{author}{\bibfnamefont{E.}~\bibnamefont{{Laermann}}},
  \bibnamefont{et~al.}, \bibinfo{journal}{\prd} \textbf{\bibinfo{volume}{90}},
  \bibinfo{eid}{094503} (\bibinfo{year}{2014}).

\bibitem[{\citenamefont{{Kr{\"u}ger} et~al.}(2013)\citenamefont{{Kr{\"u}ger},
  {Tews}, {Hebeler}, and {Schwenk}}}]{Krueger2013}
\bibinfo{author}{\bibfnamefont{T.}~\bibnamefont{{Kr{\"u}ger}}},
  \bibinfo{author}{\bibfnamefont{I.}~\bibnamefont{{Tews}}},
  \bibinfo{author}{\bibfnamefont{K.}~\bibnamefont{{Hebeler}}},
  \bibnamefont{and}
  \bibinfo{author}{\bibfnamefont{A.}~\bibnamefont{{Schwenk}}},
  \bibinfo{journal}{\prc} \textbf{\bibinfo{volume}{88}}, \bibinfo{eid}{025802}
  (\bibinfo{year}{2013}).

\bibitem[{\citenamefont{Kurkela et~al.}(2014)\citenamefont{Kurkela, Fraga,
  Schaffner-Bielich, and Vuorinen}}]{Kurkela2014}
\bibinfo{author}{\bibfnamefont{A.}~\bibnamefont{Kurkela}},
  \bibinfo{author}{\bibfnamefont{E.~S.} \bibnamefont{Fraga}},
  \bibinfo{author}{\bibfnamefont{J.}~\bibnamefont{Schaffner-Bielich}},
  \bibnamefont{and} \bibinfo{author}{\bibfnamefont{A.}~\bibnamefont{Vuorinen}},
  \bibinfo{journal}{\apj} \textbf{\bibinfo{volume}{789}}, \bibinfo{pages}{127}
  (\bibinfo{year}{2014}).

\bibitem[{\citenamefont{Abbott et~al.}(2017{\natexlab{a}})\citenamefont{Abbott,
  Abbott, Abbott, Acernese, Ackley, Adams, Adams, Addesso, Adhikari, Adya
  et~al.}}]{Abbott2017}
\bibinfo{author}{\bibfnamefont{B.~P.} \bibnamefont{Abbott}},
  \bibinfo{author}{\bibfnamefont{R.}~\bibnamefont{Abbott}},
  \bibinfo{author}{\bibfnamefont{T.~D.} \bibnamefont{Abbott}},
  \bibinfo{author}{\bibfnamefont{F.}~\bibnamefont{Acernese}},
  \bibinfo{author}{\bibfnamefont{K.}~\bibnamefont{Ackley}},
  \bibinfo{author}{\bibfnamefont{C.}~\bibnamefont{Adams}},
  \bibinfo{author}{\bibfnamefont{T.}~\bibnamefont{Adams}},
  \bibinfo{author}{\bibfnamefont{P.}~\bibnamefont{Addesso}},
  \bibinfo{author}{\bibfnamefont{R.~X.} \bibnamefont{Adhikari}},
  \bibinfo{author}{\bibfnamefont{V.~B.} \bibnamefont{Adya}},
  \bibnamefont{et~al.} (\bibinfo{collaboration}{LIGO Scientific Collaboration
  and Virgo Collaboration}), \bibinfo{journal}{\prl}
  \textbf{\bibinfo{volume}{119}}, \bibinfo{pages}{161101}
  (\bibinfo{year}{2017}{\natexlab{a}}).

\bibitem[{\citenamefont{Cs{\'a}ki et~al.}(2018)\citenamefont{Cs{\'a}ki,
  Er{\"o}ncel, Hubisz, Rigo, and Terning}}]{Csaki2018}
\bibinfo{author}{\bibfnamefont{C.}~\bibnamefont{Cs{\'a}ki}},
  \bibinfo{author}{\bibfnamefont{C.}~\bibnamefont{Er{\"o}ncel}},
  \bibinfo{author}{\bibfnamefont{J.}~\bibnamefont{Hubisz}},
  \bibinfo{author}{\bibfnamefont{G.}~\bibnamefont{Rigo}}, \bibnamefont{and}
  \bibinfo{author}{\bibfnamefont{J.}~\bibnamefont{Terning}},
  \bibinfo{journal}{Journal of High Energy Physics}
  \textbf{\bibinfo{volume}{9}}, \bibinfo{eid}{87} (\bibinfo{year}{2018}).

\bibitem[{\citenamefont{Paschalidis et~al.}(2018)\citenamefont{Paschalidis,
  Yagi, Alvarez-Castillo, Blaschke, and Sedrakian}}]{Paschalidis2018}
\bibinfo{author}{\bibfnamefont{V.}~\bibnamefont{Paschalidis}},
  \bibinfo{author}{\bibfnamefont{K.}~\bibnamefont{Yagi}},
  \bibinfo{author}{\bibfnamefont{D.}~\bibnamefont{Alvarez-Castillo}},
  \bibinfo{author}{\bibfnamefont{D.~B.} \bibnamefont{Blaschke}},
  \bibnamefont{and}
  \bibinfo{author}{\bibfnamefont{A.}~\bibnamefont{Sedrakian}},
  \bibinfo{journal}{\prd} \textbf{\bibinfo{volume}{97}}, \bibinfo{eid}{084038}
  (\bibinfo{year}{2018}).

\bibitem[{\citenamefont{Most et~al.}(2018)\citenamefont{Most, Papenfort,
  Dexheimer, Hanauske, Schramm, St{\"o}cker, and Rezzolla}}]{Most2018a}
\bibinfo{author}{\bibfnamefont{E.~R.} \bibnamefont{Most}},
  \bibinfo{author}{\bibfnamefont{L.~J.} \bibnamefont{Papenfort}},
  \bibinfo{author}{\bibfnamefont{V.}~\bibnamefont{Dexheimer}},
  \bibinfo{author}{\bibfnamefont{M.}~\bibnamefont{Hanauske}},
  \bibinfo{author}{\bibfnamefont{S.}~\bibnamefont{Schramm}},
  \bibinfo{author}{\bibfnamefont{H.}~\bibnamefont{St{\"o}cker}},
  \bibnamefont{and} \bibinfo{author}{\bibfnamefont{L.}~\bibnamefont{Rezzolla}},
  \bibinfo{journal}{ArXiv e-prints}  (\bibinfo{year}{2018}),
  \eprint{1807.03684}.

\bibitem[{\citenamefont{Han and Steiner}(2018)}]{Han2018}
\bibinfo{author}{\bibfnamefont{S.}~\bibnamefont{Han}} \bibnamefont{and}
  \bibinfo{author}{\bibfnamefont{A.~W.} \bibnamefont{Steiner}},
  \bibinfo{journal}{arXiv e-prints}  (\bibinfo{year}{2018}),
  \eprint{1810.10967}.

\bibitem[{\citenamefont{Christian et~al.}(2018)\citenamefont{Christian, Zacchi,
  and Schaffner-Bielich}}]{Christian2018}
\bibinfo{author}{\bibfnamefont{J.-E.} \bibnamefont{Christian}},
  \bibinfo{author}{\bibfnamefont{A.}~\bibnamefont{Zacchi}}, \bibnamefont{and}
  \bibinfo{author}{\bibfnamefont{J.}~\bibnamefont{Schaffner-Bielich}},
  \bibinfo{journal}{arXiv e-prints}  (\bibinfo{year}{2018}),
  \eprint{1809.03333}.

\bibitem[{\citenamefont{Sieniawska et~al.}(2018)\citenamefont{Sieniawska,
  Turcza{\'n}ski, Bejger, and Leszek~Zdunik}}]{Sieniawska2018}
\bibinfo{author}{\bibfnamefont{M.}~\bibnamefont{Sieniawska}},
  \bibinfo{author}{\bibfnamefont{W.}~\bibnamefont{Turcza{\'n}ski}},
  \bibinfo{author}{\bibfnamefont{M.}~\bibnamefont{Bejger}}, \bibnamefont{and}
  \bibinfo{author}{\bibfnamefont{J.}~\bibnamefont{Leszek~Zdunik}},
  \bibinfo{journal}{arXiv e-prints}  (\bibinfo{year}{2018}),
  \eprint{1807.11581}.

\bibitem[{\citenamefont{Burgio et~al.}(2018)\citenamefont{Burgio, Drago,
  Pagliara, Schulze, and Wei}}]{Burgio2018a}
\bibinfo{author}{\bibfnamefont{G.~F.} \bibnamefont{Burgio}},
  \bibinfo{author}{\bibfnamefont{A.}~\bibnamefont{Drago}},
  \bibinfo{author}{\bibfnamefont{G.}~\bibnamefont{Pagliara}},
  \bibinfo{author}{\bibfnamefont{H.-J.} \bibnamefont{Schulze}},
  \bibnamefont{and} \bibinfo{author}{\bibfnamefont{J.-B.} \bibnamefont{Wei}},
  \bibinfo{journal}{\apj} \textbf{\bibinfo{volume}{860}}, \bibinfo{eid}{139}
  (\bibinfo{year}{2018}).

\bibitem[{\citenamefont{Drago and Pagliara}(2018)}]{Drago2018}
\bibinfo{author}{\bibfnamefont{A.}~\bibnamefont{Drago}} \bibnamefont{and}
  \bibinfo{author}{\bibfnamefont{G.}~\bibnamefont{Pagliara}},
  \bibinfo{journal}{\apjl} \textbf{\bibinfo{volume}{852}}, \bibinfo{eid}{L32}
  (\bibinfo{year}{2018}).

\bibitem[{\citenamefont{Dexheimer et~al.}(2018)\citenamefont{Dexheimer,
  de~Oliveira~Gomes, Schramm, and Pais}}]{Dexheimer2018}
\bibinfo{author}{\bibfnamefont{V.}~\bibnamefont{Dexheimer}},
  \bibinfo{author}{\bibfnamefont{R.}~\bibnamefont{de~Oliveira~Gomes}},
  \bibinfo{author}{\bibfnamefont{S.}~\bibnamefont{Schramm}}, \bibnamefont{and}
  \bibinfo{author}{\bibfnamefont{H.}~\bibnamefont{Pais}},
  \bibinfo{journal}{arXiv e-prints}  (\bibinfo{year}{2018}),
  \eprint{1810.06109}.

\bibitem[{\citenamefont{Faber and Rasio}(2012)}]{Faber2012}
\bibinfo{author}{\bibfnamefont{J.~A.} \bibnamefont{Faber}} \bibnamefont{and}
  \bibinfo{author}{\bibfnamefont{F.~A.} \bibnamefont{Rasio}},
  \bibinfo{journal}{\lrr} \textbf{\bibinfo{volume}{15}}, \bibinfo{eid}{8}
  (\bibinfo{year}{2012}).

\bibitem[{\citenamefont{Baiotti and Rezzolla}(2017)}]{Baiotti2017}
\bibinfo{author}{\bibfnamefont{L.}~\bibnamefont{Baiotti}} \bibnamefont{and}
  \bibinfo{author}{\bibfnamefont{L.}~\bibnamefont{Rezzolla}},
  \bibinfo{journal}{Reports on Progress in Physics}
  \textbf{\bibinfo{volume}{80}}, \bibinfo{eid}{096901} (\bibinfo{year}{2017}).

\bibitem[{\citenamefont{Paschalidis and Stergioulas}(2017)}]{Paschalidis2017}
\bibinfo{author}{\bibfnamefont{V.}~\bibnamefont{Paschalidis}} \bibnamefont{and}
  \bibinfo{author}{\bibfnamefont{N.}~\bibnamefont{Stergioulas}},
  \bibinfo{journal}{\lrr} \textbf{\bibinfo{volume}{20}}, \bibinfo{eid}{7}
  (\bibinfo{year}{2017}).

\bibitem[{\citenamefont{Friedman}(2018)}]{Friedman2018}
\bibinfo{author}{\bibfnamefont{J.~L.} \bibnamefont{Friedman}},
  \bibinfo{journal}{International Journal of Modern Physics D}
  \textbf{\bibinfo{volume}{27}}, \bibinfo{eid}{1843018} (\bibinfo{year}{2018}).

\bibitem[{\citenamefont{{Flanagan} and {Hinderer}}(2008)}]{Flanagan2008}
\bibinfo{author}{\bibfnamefont{{\'E}.~{\'E}.} \bibnamefont{{Flanagan}}}
  \bibnamefont{and}
  \bibinfo{author}{\bibfnamefont{T.}~\bibnamefont{{Hinderer}}},
  \bibinfo{journal}{\prd} \textbf{\bibinfo{volume}{77}}, \bibinfo{eid}{021502}
  (\bibinfo{year}{2008}).

\bibitem[{\citenamefont{Hinderer}(2008)}]{Hinderer2008}
\bibinfo{author}{\bibfnamefont{T.}~\bibnamefont{Hinderer}},
  \bibinfo{journal}{\apj} \textbf{\bibinfo{volume}{677}}, \bibinfo{pages}{1216}
  (\bibinfo{year}{2008}).

\bibitem[{\citenamefont{Read et~al.}(2009{\natexlab{a}})\citenamefont{Read,
  Markakis, Shibata, Ury{\= u}, Creighton, and Friedman}}]{Read2009}
\bibinfo{author}{\bibfnamefont{J.~S.} \bibnamefont{Read}},
  \bibinfo{author}{\bibfnamefont{C.}~\bibnamefont{Markakis}},
  \bibinfo{author}{\bibfnamefont{M.}~\bibnamefont{Shibata}},
  \bibinfo{author}{\bibfnamefont{K.}~\bibnamefont{Ury{\= u}}},
  \bibinfo{author}{\bibfnamefont{J.~D.~E.} \bibnamefont{Creighton}},
  \bibnamefont{and} \bibinfo{author}{\bibfnamefont{J.~L.}
  \bibnamefont{Friedman}}, \bibinfo{journal}{\prd}
  \textbf{\bibinfo{volume}{79}}, \bibinfo{eid}{124033}
  (\bibinfo{year}{2009}{\natexlab{a}}).

\bibitem[{\citenamefont{{Hinderer} et~al.}(2010)\citenamefont{{Hinderer},
  {Lackey}, {Lang}, and {Read}}}]{Hinderer2010}
\bibinfo{author}{\bibfnamefont{T.}~\bibnamefont{{Hinderer}}},
  \bibinfo{author}{\bibfnamefont{B.~D.} \bibnamefont{{Lackey}}},
  \bibinfo{author}{\bibfnamefont{R.~N.} \bibnamefont{{Lang}}},
  \bibnamefont{and} \bibinfo{author}{\bibfnamefont{J.~S.}
  \bibnamefont{{Read}}}, \bibinfo{journal}{\prd} \textbf{\bibinfo{volume}{81}},
  \bibinfo{eid}{123016} (\bibinfo{year}{2010}).

\bibitem[{\citenamefont{{Read} et~al.}(2013)\citenamefont{{Read}, {Baiotti},
  {Creighton}, {Friedman}, {Giacomazzo}, {Kyutoku}, {Markakis}, {Rezzolla},
  {Shibata}, and {Taniguchi}}}]{Read2013}
\bibinfo{author}{\bibfnamefont{J.~S.} \bibnamefont{{Read}}},
  \bibinfo{author}{\bibfnamefont{L.}~\bibnamefont{{Baiotti}}},
  \bibinfo{author}{\bibfnamefont{J.~D.~E.} \bibnamefont{{Creighton}}},
  \bibinfo{author}{\bibfnamefont{J.~L.} \bibnamefont{{Friedman}}},
  \bibinfo{author}{\bibfnamefont{B.}~\bibnamefont{{Giacomazzo}}},
  \bibinfo{author}{\bibfnamefont{K.}~\bibnamefont{{Kyutoku}}},
  \bibinfo{author}{\bibfnamefont{C.}~\bibnamefont{{Markakis}}},
  \bibinfo{author}{\bibfnamefont{L.}~\bibnamefont{{Rezzolla}}},
  \bibinfo{author}{\bibfnamefont{M.}~\bibnamefont{{Shibata}}},
  \bibnamefont{and}
  \bibinfo{author}{\bibfnamefont{K.}~\bibnamefont{{Taniguchi}}},
  \bibinfo{journal}{\prd} \textbf{\bibinfo{volume}{88}}, \bibinfo{eid}{044042}
  (\bibinfo{year}{2013}).

\bibitem[{\citenamefont{{Del Pozzo} et~al.}(2013)\citenamefont{{Del Pozzo},
  {Li}, {Agathos}, {Van Den Broeck}, and {Vitale}}}]{DelPozzo2013}
\bibinfo{author}{\bibfnamefont{W.}~\bibnamefont{{Del Pozzo}}},
  \bibinfo{author}{\bibfnamefont{T.~G.~F.} \bibnamefont{{Li}}},
  \bibinfo{author}{\bibfnamefont{M.}~\bibnamefont{{Agathos}}},
  \bibinfo{author}{\bibfnamefont{C.}~\bibnamefont{{Van Den Broeck}}},
  \bibnamefont{and} \bibinfo{author}{\bibfnamefont{S.}~\bibnamefont{{Vitale}}},
  \bibinfo{journal}{\prl} \textbf{\bibinfo{volume}{111}}, \bibinfo{eid}{071101}
  (\bibinfo{year}{2013}).

\bibitem[{\citenamefont{{Wade} et~al.}(2014)\citenamefont{{Wade}, {Creighton},
  {Ochsner}, {Lackey}, {Farr}, {Littenberg}, and {Raymond}}}]{Wade2014}
\bibinfo{author}{\bibfnamefont{L.}~\bibnamefont{{Wade}}},
  \bibinfo{author}{\bibfnamefont{J.~D.~E.} \bibnamefont{{Creighton}}},
  \bibinfo{author}{\bibfnamefont{E.}~\bibnamefont{{Ochsner}}},
  \bibinfo{author}{\bibfnamefont{B.~D.} \bibnamefont{{Lackey}}},
  \bibinfo{author}{\bibfnamefont{B.~F.} \bibnamefont{{Farr}}},
  \bibinfo{author}{\bibfnamefont{T.~B.} \bibnamefont{{Littenberg}}},
  \bibnamefont{and}
  \bibinfo{author}{\bibfnamefont{V.}~\bibnamefont{{Raymond}}},
  \bibinfo{journal}{\prd} \textbf{\bibinfo{volume}{89}}, \bibinfo{eid}{103012}
  (\bibinfo{year}{2014}).

\bibitem[{\citenamefont{{Agathos} et~al.}(2015)\citenamefont{{Agathos},
  {Meidam}, {Del Pozzo}, {Li}, {Tompitak}, {Veitch}, {Vitale}, and {Van Den
  Broeck}}}]{Agathos2015}
\bibinfo{author}{\bibfnamefont{M.}~\bibnamefont{{Agathos}}},
  \bibinfo{author}{\bibfnamefont{J.}~\bibnamefont{{Meidam}}},
  \bibinfo{author}{\bibfnamefont{W.}~\bibnamefont{{Del Pozzo}}},
  \bibinfo{author}{\bibfnamefont{T.~G.~F.} \bibnamefont{{Li}}},
  \bibinfo{author}{\bibfnamefont{M.}~\bibnamefont{{Tompitak}}},
  \bibinfo{author}{\bibfnamefont{J.}~\bibnamefont{{Veitch}}},
  \bibinfo{author}{\bibfnamefont{S.}~\bibnamefont{{Vitale}}}, \bibnamefont{and}
  \bibinfo{author}{\bibfnamefont{C.}~\bibnamefont{{Van Den Broeck}}},
  \bibinfo{journal}{\prd} \textbf{\bibinfo{volume}{92}}, \bibinfo{eid}{023012}
  (\bibinfo{year}{2015}).

\bibitem[{\citenamefont{Chatziioannou et~al.}(2015)\citenamefont{Chatziioannou,
  Yagi, Klein, Cornish, and Yunes}}]{Chatziioannou2015}
\bibinfo{author}{\bibfnamefont{K.}~\bibnamefont{Chatziioannou}},
  \bibinfo{author}{\bibfnamefont{K.}~\bibnamefont{Yagi}},
  \bibinfo{author}{\bibfnamefont{A.}~\bibnamefont{Klein}},
  \bibinfo{author}{\bibfnamefont{N.}~\bibnamefont{Cornish}}, \bibnamefont{and}
  \bibinfo{author}{\bibfnamefont{N.}~\bibnamefont{Yunes}},
  \bibinfo{journal}{\prd} \textbf{\bibinfo{volume}{92}}, \bibinfo{eid}{104008}
  (\bibinfo{year}{2015}).

\bibitem[{\citenamefont{Hotokezaka et~al.}(2016)\citenamefont{Hotokezaka,
  Kyutoku, Sekiguchi, and Shibata}}]{Hotokezaka2016}
\bibinfo{author}{\bibfnamefont{K.}~\bibnamefont{Hotokezaka}},
  \bibinfo{author}{\bibfnamefont{K.}~\bibnamefont{Kyutoku}},
  \bibinfo{author}{\bibfnamefont{Y.-i.} \bibnamefont{Sekiguchi}},
  \bibnamefont{and} \bibinfo{author}{\bibfnamefont{M.}~\bibnamefont{Shibata}},
  \bibinfo{journal}{\prd} \textbf{\bibinfo{volume}{93}}, \bibinfo{eid}{064082}
  (\bibinfo{year}{2016}).

\bibitem[{\citenamefont{{Chatziioannou}
  et~al.}(2018)\citenamefont{{Chatziioannou}, {Haster}, and
  {Zimmerman}}}]{Chatziioannou2018}
\bibinfo{author}{\bibfnamefont{K.}~\bibnamefont{{Chatziioannou}}},
  \bibinfo{author}{\bibfnamefont{C.-J.} \bibnamefont{{Haster}}},
  \bibnamefont{and}
  \bibinfo{author}{\bibfnamefont{A.}~\bibnamefont{{Zimmerman}}},
  \bibinfo{journal}{\prd} \textbf{\bibinfo{volume}{97}}, \bibinfo{eid}{104036}
  (\bibinfo{year}{2018}).

\bibitem[{\citenamefont{{Bauswein} and {Janka}}(2012)}]{Bauswein2012}
\bibinfo{author}{\bibfnamefont{A.}~\bibnamefont{{Bauswein}}} \bibnamefont{and}
  \bibinfo{author}{\bibfnamefont{H.-T.} \bibnamefont{{Janka}}},
  \bibinfo{journal}{\prl} \textbf{\bibinfo{volume}{108}}, \bibinfo{eid}{011101}
  (\bibinfo{year}{2012}).

\bibitem[{\citenamefont{{Bauswein} et~al.}(2012)\citenamefont{{Bauswein},
  {Janka}, {Hebeler}, and {Schwenk}}}]{Bauswein2012a}
\bibinfo{author}{\bibfnamefont{A.}~\bibnamefont{{Bauswein}}},
  \bibinfo{author}{\bibfnamefont{H.-T.} \bibnamefont{{Janka}}},
  \bibinfo{author}{\bibfnamefont{K.}~\bibnamefont{{Hebeler}}},
  \bibnamefont{and}
  \bibinfo{author}{\bibfnamefont{A.}~\bibnamefont{{Schwenk}}},
  \bibinfo{journal}{\prd} \textbf{\bibinfo{volume}{86}}, \bibinfo{eid}{063001}
  (\bibinfo{year}{2012}).

\bibitem[{\citenamefont{{Hotokezaka} et~al.}(2013)\citenamefont{{Hotokezaka},
  {Kiuchi}, {Kyutoku}, {Muranushi}, {Sekiguchi}, {Shibata}, and
  {Taniguchi}}}]{Hotokezaka2013a}
\bibinfo{author}{\bibfnamefont{K.}~\bibnamefont{{Hotokezaka}}},
  \bibinfo{author}{\bibfnamefont{K.}~\bibnamefont{{Kiuchi}}},
  \bibinfo{author}{\bibfnamefont{K.}~\bibnamefont{{Kyutoku}}},
  \bibinfo{author}{\bibfnamefont{T.}~\bibnamefont{{Muranushi}}},
  \bibinfo{author}{\bibfnamefont{Y.}~\bibnamefont{{Sekiguchi}}},
  \bibinfo{author}{\bibfnamefont{M.}~\bibnamefont{{Shibata}}},
  \bibnamefont{and}
  \bibinfo{author}{\bibfnamefont{K.}~\bibnamefont{{Taniguchi}}},
  \bibinfo{journal}{\prd} \textbf{\bibinfo{volume}{88}}, \bibinfo{eid}{044026}
  (\bibinfo{year}{2013}).

\bibitem[{\citenamefont{Takami et~al.}(2014)\citenamefont{Takami, Rezzolla, and
  Baiotti}}]{Takami2014}
\bibinfo{author}{\bibfnamefont{K.}~\bibnamefont{Takami}},
  \bibinfo{author}{\bibfnamefont{L.}~\bibnamefont{Rezzolla}}, \bibnamefont{and}
  \bibinfo{author}{\bibfnamefont{L.}~\bibnamefont{Baiotti}},
  \bibinfo{journal}{\prl} \textbf{\bibinfo{volume}{113}}, \bibinfo{eid}{091104}
  (\bibinfo{year}{2014}).

\bibitem[{\citenamefont{{Bauswein} and {Stergioulas}}(2015)}]{Bauswein2015}
\bibinfo{author}{\bibfnamefont{A.}~\bibnamefont{{Bauswein}}} \bibnamefont{and}
  \bibinfo{author}{\bibfnamefont{N.}~\bibnamefont{{Stergioulas}}},
  \bibinfo{journal}{\prd} \textbf{\bibinfo{volume}{91}}, \bibinfo{eid}{124056}
  (\bibinfo{year}{2015}).

\bibitem[{\citenamefont{Oechslin et~al.}(2004)\citenamefont{Oechslin, Ury{\=
  u}, Poghosyan, and Thielemann}}]{Oechslin2004}
\bibinfo{author}{\bibfnamefont{R.}~\bibnamefont{Oechslin}},
  \bibinfo{author}{\bibfnamefont{K.}~\bibnamefont{Ury{\= u}}},
  \bibinfo{author}{\bibfnamefont{G.}~\bibnamefont{Poghosyan}},
  \bibnamefont{and} \bibinfo{author}{\bibfnamefont{F.~K.}
  \bibnamefont{Thielemann}}, \bibinfo{journal}{\mnras}
  \textbf{\bibinfo{volume}{349}}, \bibinfo{pages}{1469} (\bibinfo{year}{2004}).

\bibitem[{\citenamefont{Bauswein
  et~al.}(2010{\natexlab{a}})\citenamefont{Bauswein, Oechslin, and
  Janka}}]{Bauswein2010a}
\bibinfo{author}{\bibfnamefont{A.}~\bibnamefont{Bauswein}},
  \bibinfo{author}{\bibfnamefont{R.}~\bibnamefont{Oechslin}}, \bibnamefont{and}
  \bibinfo{author}{\bibfnamefont{H.-T.} \bibnamefont{Janka}},
  \bibinfo{journal}{\prd} \textbf{\bibinfo{volume}{81}}, \bibinfo{eid}{024012}
  (\bibinfo{year}{2010}{\natexlab{a}}).

\bibitem[{\citenamefont{Sekiguchi et~al.}(2011)\citenamefont{Sekiguchi, Kiuchi,
  Kyutoku, and Shibata}}]{Sekiguchi2011}
\bibinfo{author}{\bibfnamefont{Y.}~\bibnamefont{Sekiguchi}},
  \bibinfo{author}{\bibfnamefont{K.}~\bibnamefont{Kiuchi}},
  \bibinfo{author}{\bibfnamefont{K.}~\bibnamefont{Kyutoku}}, \bibnamefont{and}
  \bibinfo{author}{\bibfnamefont{M.}~\bibnamefont{Shibata}},
  \bibinfo{journal}{Physical Review Letters} \textbf{\bibinfo{volume}{107}},
  \bibinfo{eid}{211101} (\bibinfo{year}{2011}).

\bibitem[{\citenamefont{Radice et~al.}(2017)\citenamefont{Radice, Bernuzzi,
  Del~Pozzo, Roberts, and Ott}}]{Radice2017}
\bibinfo{author}{\bibfnamefont{D.}~\bibnamefont{Radice}},
  \bibinfo{author}{\bibfnamefont{S.}~\bibnamefont{Bernuzzi}},
  \bibinfo{author}{\bibfnamefont{W.}~\bibnamefont{Del~Pozzo}},
  \bibinfo{author}{\bibfnamefont{L.~F.} \bibnamefont{Roberts}},
  \bibnamefont{and} \bibinfo{author}{\bibfnamefont{C.~D.} \bibnamefont{Ott}},
  \bibinfo{journal}{\apjl} \textbf{\bibinfo{volume}{842}}, \bibinfo{eid}{L10}
  (\bibinfo{year}{2017}).

\bibitem[{\citenamefont{{The LIGO Scientific Collaboration}
  et~al.}(2018)\citenamefont{{The LIGO Scientific Collaboration}, {the Virgo
  Collaboration}, {Abbott}, {Abbott}, {Abbott}, {Acernese}, {Ackley}, {Adams},
  {Adams}, {Addesso} et~al.}}]{TheLIGOScientificCollaboration2018a}
\bibinfo{author}{\bibnamefont{{The LIGO Scientific Collaboration}}},
  \bibinfo{author}{\bibnamefont{{the Virgo Collaboration}}},
  \bibinfo{author}{\bibfnamefont{B.~P.} \bibnamefont{{Abbott}}},
  \bibinfo{author}{\bibfnamefont{R.}~\bibnamefont{{Abbott}}},
  \bibinfo{author}{\bibfnamefont{T.~D.} \bibnamefont{{Abbott}}},
  \bibinfo{author}{\bibfnamefont{F.}~\bibnamefont{{Acernese}}},
  \bibinfo{author}{\bibfnamefont{K.}~\bibnamefont{{Ackley}}},
  \bibinfo{author}{\bibfnamefont{C.}~\bibnamefont{{Adams}}},
  \bibinfo{author}{\bibfnamefont{T.}~\bibnamefont{{Adams}}},
  \bibinfo{author}{\bibfnamefont{P.}~\bibnamefont{{Addesso}}},
  \bibnamefont{et~al.}, \bibinfo{journal}{ArXiv e-prints}
  (\bibinfo{year}{2018}), \eprint{1805.11579}.

\bibitem[{\citenamefont{Abbott et~al.}(2018)\citenamefont{Abbott, Abbott,
  Abbott, Acernese, Ackley, Adams, Adams, Addesso, Adhikari, Adya
  et~al.}}]{Abbott2018}
\bibinfo{author}{\bibfnamefont{B.~P.} \bibnamefont{Abbott}},
  \bibinfo{author}{\bibfnamefont{R.}~\bibnamefont{Abbott}},
  \bibinfo{author}{\bibfnamefont{T.~D.} \bibnamefont{Abbott}},
  \bibinfo{author}{\bibfnamefont{F.}~\bibnamefont{Acernese}},
  \bibinfo{author}{\bibfnamefont{K.}~\bibnamefont{Ackley}},
  \bibinfo{author}{\bibfnamefont{C.}~\bibnamefont{Adams}},
  \bibinfo{author}{\bibfnamefont{T.}~\bibnamefont{Adams}},
  \bibinfo{author}{\bibfnamefont{P.}~\bibnamefont{Addesso}},
  \bibinfo{author}{\bibfnamefont{R.~X.} \bibnamefont{Adhikari}},
  \bibinfo{author}{\bibfnamefont{V.~B.} \bibnamefont{Adya}},
  \bibnamefont{et~al.}, \bibinfo{journal}{\prl} \textbf{\bibinfo{volume}{121}},
  \bibinfo{eid}{161101} (\bibinfo{year}{2018}).

\bibitem[{\citenamefont{De et~al.}(2018)\citenamefont{De, Finstad, Lattimer,
  Brown, Berger, and Biwer}}]{De2018}
\bibinfo{author}{\bibfnamefont{S.}~\bibnamefont{De}},
  \bibinfo{author}{\bibfnamefont{D.}~\bibnamefont{Finstad}},
  \bibinfo{author}{\bibfnamefont{J.~M.} \bibnamefont{Lattimer}},
  \bibinfo{author}{\bibfnamefont{D.~A.} \bibnamefont{Brown}},
  \bibinfo{author}{\bibfnamefont{E.}~\bibnamefont{Berger}}, \bibnamefont{and}
  \bibinfo{author}{\bibfnamefont{C.~M.} \bibnamefont{Biwer}},
  \bibinfo{journal}{\prl} \textbf{\bibinfo{volume}{121}}, \bibinfo{eid}{091102}
  (\bibinfo{year}{2018}).

\bibitem[{\citenamefont{Carney et~al.}(2018)\citenamefont{Carney, Wade, and
  Irwin}}]{Carney2018}
\bibinfo{author}{\bibfnamefont{M.~F.} \bibnamefont{Carney}},
  \bibinfo{author}{\bibfnamefont{L.~E.} \bibnamefont{Wade}}, \bibnamefont{and}
  \bibinfo{author}{\bibfnamefont{B.~S.} \bibnamefont{Irwin}},
  \bibinfo{journal}{\prd} \textbf{\bibinfo{volume}{98}}, \bibinfo{eid}{063004}
  (\bibinfo{year}{2018}).

\bibitem[{\citenamefont{{Clark} et~al.}(2014)\citenamefont{{Clark}, {Bauswein},
  {Cadonati}, {Janka}, {Pankow}, and {Stergioulas}}}]{Clark2014}
\bibinfo{author}{\bibfnamefont{J.}~\bibnamefont{{Clark}}},
  \bibinfo{author}{\bibfnamefont{A.}~\bibnamefont{{Bauswein}}},
  \bibinfo{author}{\bibfnamefont{L.}~\bibnamefont{{Cadonati}}},
  \bibinfo{author}{\bibfnamefont{H.-T.} \bibnamefont{{Janka}}},
  \bibinfo{author}{\bibfnamefont{C.}~\bibnamefont{{Pankow}}}, \bibnamefont{and}
  \bibinfo{author}{\bibfnamefont{N.}~\bibnamefont{{Stergioulas}}},
  \bibinfo{journal}{\prd} \textbf{\bibinfo{volume}{90}}, \bibinfo{eid}{062004}
  (\bibinfo{year}{2014}).

\bibitem[{\citenamefont{{Clark} et~al.}(2016)\citenamefont{{Clark}, {Bauswein},
  {Stergioulas}, and {Shoemaker}}}]{Clark2016}
\bibinfo{author}{\bibfnamefont{J.~A.} \bibnamefont{{Clark}}},
  \bibinfo{author}{\bibfnamefont{A.}~\bibnamefont{{Bauswein}}},
  \bibinfo{author}{\bibfnamefont{N.}~\bibnamefont{{Stergioulas}}},
  \bibnamefont{and}
  \bibinfo{author}{\bibfnamefont{D.}~\bibnamefont{{Shoemaker}}},
  \bibinfo{journal}{Classical and Quantum Gravity}
  \textbf{\bibinfo{volume}{33}}, \bibinfo{eid}{085003} (\bibinfo{year}{2016}).
  
\bibitem[{\citenamefont{Chatziioannou et~al.}(2017)\citenamefont{Chatziioannou,
  Clark, Bauswein, Millhouse, Littenberg, and Cornish}}]{Chatziioannou2017}
\bibinfo{author}{\bibfnamefont{K.}~\bibnamefont{Chatziioannou}},
  \bibinfo{author}{\bibfnamefont{J.~A.} \bibnamefont{Clark}},
  \bibinfo{author}{\bibfnamefont{A.}~\bibnamefont{Bauswein}},
  \bibinfo{author}{\bibfnamefont{M.}~\bibnamefont{Millhouse}},
  \bibinfo{author}{\bibfnamefont{T.~B.} \bibnamefont{Littenberg}},
  \bibnamefont{and} \bibinfo{author}{\bibfnamefont{N.}~\bibnamefont{Cornish}},
  \bibinfo{journal}{\prd} \textbf{\bibinfo{volume}{96}}, \bibinfo{eid}{124035}
  (\bibinfo{year}{2017}).

\bibitem[{\citenamefont{Bose et~al.}(2018)\citenamefont{Bose, Chakravarti,
  Rezzolla, Sathyaprakash, and Takami}}]{Bose2018}
\bibinfo{author}{\bibfnamefont{S.}~\bibnamefont{Bose}},
  \bibinfo{author}{\bibfnamefont{K.}~\bibnamefont{Chakravarti}},
  \bibinfo{author}{\bibfnamefont{L.}~\bibnamefont{Rezzolla}},
  \bibinfo{author}{\bibfnamefont{B.~S.} \bibnamefont{Sathyaprakash}},
  \bibnamefont{and} \bibinfo{author}{\bibfnamefont{K.}~\bibnamefont{Takami}},
  \bibinfo{journal}{\prl} \textbf{\bibinfo{volume}{120}}, \bibinfo{eid}{031102}
  (\bibinfo{year}{2018}).

\bibitem[{\citenamefont{Yang et~al.}(2018)\citenamefont{Yang, Paschalidis,
  Yagi, Lehner, Pretorius, and Yunes}}]{Yang2018}
\bibinfo{author}{\bibfnamefont{H.}~\bibnamefont{Yang}},
  \bibinfo{author}{\bibfnamefont{V.}~\bibnamefont{Paschalidis}},
  \bibinfo{author}{\bibfnamefont{K.}~\bibnamefont{Yagi}},
  \bibinfo{author}{\bibfnamefont{L.}~\bibnamefont{Lehner}},
  \bibinfo{author}{\bibfnamefont{F.}~\bibnamefont{Pretorius}},
  \bibnamefont{and} \bibinfo{author}{\bibfnamefont{N.}~\bibnamefont{Yunes}},
  \bibinfo{journal}{\prd} \textbf{\bibinfo{volume}{97}}, \bibinfo{eid}{024049}
  (\bibinfo{year}{2018}).

\bibitem[{\citenamefont{Torres-Rivas et~al.}(2018)\citenamefont{Torres-Rivas,
  Chatziioannou, Bauswein, and Clark}}]{Torres-Rivas2018}
\bibinfo{author}{\bibfnamefont{A.}~\bibnamefont{Torres-Rivas}},
  \bibinfo{author}{\bibfnamefont{K.}~\bibnamefont{Chatziioannou}},
  \bibinfo{author}{\bibfnamefont{A.}~\bibnamefont{Bauswein}}, \bibnamefont{and}
  \bibinfo{author}{\bibfnamefont{J.~A.} \bibnamefont{Clark}},
  \bibinfo{journal}{arXiv e-prints}  (\bibinfo{year}{2018}),
  \eprint{1811.08931}.

\bibitem[{\citenamefont{Duez and Zlochower}(2019)}]{Duez2019}
\bibinfo{author}{\bibfnamefont{M.~D.} \bibnamefont{Duez}} \bibnamefont{and}
  \bibinfo{author}{\bibfnamefont{Y.}~\bibnamefont{Zlochower}},
  \bibinfo{journal}{Reports on Progress in Physics}
  \textbf{\bibinfo{volume}{82}}, \bibinfo{eid}{016902} (\bibinfo{year}{2019}).

\bibitem[{\citenamefont{Fischer et~al.}(2018)\citenamefont{Fischer, Bastian,
  Wu, Baklanov, Sorokina, Blinnikov, Typel, Kl{\"a}hn, and
  Blaschke}}]{Fischer2018}
\bibinfo{author}{\bibfnamefont{T.}~\bibnamefont{Fischer}},
  \bibinfo{author}{\bibfnamefont{N.-U.~F.} \bibnamefont{Bastian}},
  \bibinfo{author}{\bibfnamefont{M.-R.} \bibnamefont{Wu}},
  \bibinfo{author}{\bibfnamefont{P.}~\bibnamefont{Baklanov}},
  \bibinfo{author}{\bibfnamefont{E.}~\bibnamefont{Sorokina}},
  \bibinfo{author}{\bibfnamefont{S.}~\bibnamefont{Blinnikov}},
  \bibinfo{author}{\bibfnamefont{S.}~\bibnamefont{Typel}},
  \bibinfo{author}{\bibfnamefont{T.}~\bibnamefont{Kl{\"a}hn}},
  \bibnamefont{and} \bibinfo{author}{\bibfnamefont{D.~B.}
  \bibnamefont{Blaschke}}, \bibinfo{journal}{Nature Astronomy}
  \textbf{\bibinfo{volume}{2}}, \bibinfo{pages}{980} (\bibinfo{year}{2018}).

\bibitem[{\citenamefont{{Typel}}(2005)}]{Typel2005}
\bibinfo{author}{\bibfnamefont{S.}~\bibnamefont{{Typel}}},
  \bibinfo{journal}{\prc} \textbf{\bibinfo{volume}{71}}, \bibinfo{eid}{064301}
  (\bibinfo{year}{2005}).

\bibitem[{\citenamefont{{Typel} et~al.}(2010)\citenamefont{{Typel},
  {R{\"o}pke}, {Kl{\"a}hn}, {Blaschke}, and {Wolter}}}]{Typel2010}
\bibinfo{author}{\bibfnamefont{S.}~\bibnamefont{{Typel}}},
  \bibinfo{author}{\bibfnamefont{G.}~\bibnamefont{{R{\"o}pke}}},
  \bibinfo{author}{\bibfnamefont{T.}~\bibnamefont{{Kl{\"a}hn}}},
  \bibinfo{author}{\bibfnamefont{D.}~\bibnamefont{{Blaschke}}},
  \bibnamefont{and} \bibinfo{author}{\bibfnamefont{H.~H.}
  \bibnamefont{{Wolter}}}, \bibinfo{journal}{\prc}
  \textbf{\bibinfo{volume}{81}}, \bibinfo{eid}{015803} (\bibinfo{year}{2010}).

\bibitem[{\citenamefont{Alvarez-Castillo
  et~al.}(2016)\citenamefont{Alvarez-Castillo, Ayriyan, Benic, Blaschke,
  Grigorian, and Typel}}]{Alvarez-Castillo2016}
\bibinfo{author}{\bibfnamefont{D.}~\bibnamefont{Alvarez-Castillo}},
  \bibinfo{author}{\bibfnamefont{A.}~\bibnamefont{Ayriyan}},
  \bibinfo{author}{\bibfnamefont{S.}~\bibnamefont{Benic}},
  \bibinfo{author}{\bibfnamefont{D.}~\bibnamefont{Blaschke}},
  \bibinfo{author}{\bibfnamefont{H.}~\bibnamefont{Grigorian}},
  \bibnamefont{and} \bibinfo{author}{\bibfnamefont{S.}~\bibnamefont{Typel}},
  \bibinfo{journal}{European Physical Journal A} \textbf{\bibinfo{volume}{52}},
  \bibinfo{eid}{69} (\bibinfo{year}{2016}).

\bibitem[{\citenamefont{{Hempel} et~al.}(2012)\citenamefont{{Hempel},
  {Fischer}, {Schaffner-Bielich}, and {Liebend{\"o}rfer}}}]{Hempel2012}
\bibinfo{author}{\bibfnamefont{M.}~\bibnamefont{{Hempel}}},
  \bibinfo{author}{\bibfnamefont{T.}~\bibnamefont{{Fischer}}},
  \bibinfo{author}{\bibfnamefont{J.}~\bibnamefont{{Schaffner-Bielich}}},
  \bibnamefont{and}
  \bibinfo{author}{\bibfnamefont{M.}~\bibnamefont{{Liebend{\"o}rfer}}},
  \bibinfo{journal}{\apj} \textbf{\bibinfo{volume}{748}}, \bibinfo{eid}{70}
  (\bibinfo{year}{2012}).

\bibitem[{\citenamefont{{Hempel} and {Schaffner-Bielich}}(2010)}]{Hempel2010}
\bibinfo{author}{\bibfnamefont{M.}~\bibnamefont{{Hempel}}} \bibnamefont{and}
  \bibinfo{author}{\bibfnamefont{J.}~\bibnamefont{{Schaffner-Bielich}}},
  \bibinfo{journal}{Nucl. Phys. A} \textbf{\bibinfo{volume}{837}},
  \bibinfo{pages}{210} (\bibinfo{year}{2010}).

\bibitem[{\citenamefont{{Kaltenborn} et~al.}(2017)\citenamefont{{Kaltenborn},
  {Bastian}, and {Blaschke}}}]{Kaltenborn2017a}
\bibinfo{author}{\bibfnamefont{M.~A.~R.} \bibnamefont{{Kaltenborn}}},
  \bibinfo{author}{\bibfnamefont{N.-U.~F.} \bibnamefont{{Bastian}}},
  \bibnamefont{and} \bibinfo{author}{\bibfnamefont{D.~B.}
  \bibnamefont{{Blaschke}}}, \bibinfo{journal}{\prd}
  \textbf{\bibinfo{volume}{96}}, \bibinfo{eid}{056024} (\bibinfo{year}{2017}).

\bibitem[{\citenamefont{Nambu and Jona-Lasinio}(1961)}]{Nambu1961}
\bibinfo{author}{\bibfnamefont{Y.}~\bibnamefont{Nambu}} \bibnamefont{and}
  \bibinfo{author}{\bibfnamefont{G.}~\bibnamefont{Jona-Lasinio}},
  \bibinfo{journal}{Phys. Rev.} \textbf{\bibinfo{volume}{122}},
  \bibinfo{pages}{345} (\bibinfo{year}{1961}).

\bibitem[{\citenamefont{Klevansky}(1992)}]{Klevansky1992}
\bibinfo{author}{\bibfnamefont{S.}~\bibnamefont{Klevansky}},
  \bibinfo{journal}{Rev. Mod. Phys.} \textbf{\bibinfo{volume}{64}},
  \bibinfo{pages}{649} (\bibinfo{year}{1992}).

\bibitem[{\citenamefont{{R{\"u}ster} et~al.}(2005)\citenamefont{{R{\"u}ster},
  {Werth}, {Buballa}, {Shovkovy}, and {Rischke}}}]{Ruester2005}
\bibinfo{author}{\bibfnamefont{S.~B.} \bibnamefont{{R{\"u}ster}}},
  \bibinfo{author}{\bibfnamefont{V.}~\bibnamefont{{Werth}}},
  \bibinfo{author}{\bibfnamefont{M.}~\bibnamefont{{Buballa}}},
  \bibinfo{author}{\bibfnamefont{I.~A.} \bibnamefont{{Shovkovy}}},
  \bibnamefont{and} \bibinfo{author}{\bibfnamefont{D.~H.}
  \bibnamefont{{Rischke}}}, \bibinfo{journal}{\prd}
  \textbf{\bibinfo{volume}{72}}, \bibinfo{eid}{034004} (\bibinfo{year}{2005}).

\bibitem[{\citenamefont{Blaschke et~al.}(2005)\citenamefont{Blaschke,
  Fredriksson, Grigorian, {\"O}zta{\c s}, and Sandin}}]{Blaschke2005}
\bibinfo{author}{\bibfnamefont{D.}~\bibnamefont{Blaschke}},
  \bibinfo{author}{\bibfnamefont{S.}~\bibnamefont{Fredriksson}},
  \bibinfo{author}{\bibfnamefont{H.}~\bibnamefont{Grigorian}},
  \bibinfo{author}{\bibfnamefont{A.~M.} \bibnamefont{{\"O}zta{\c s}}},
  \bibnamefont{and} \bibinfo{author}{\bibfnamefont{F.}~\bibnamefont{Sandin}},
  \bibinfo{journal}{\prd} \textbf{\bibinfo{volume}{72}}, \bibinfo{eid}{065020}
  (\bibinfo{year}{2005}).

\bibitem[{\citenamefont{{Bastian} et~al.}(2018)\citenamefont{{Bastian},
  {Blaschke}, {Fischer}, and {R{\"o}pke}}}]{Bastian2018}
\bibinfo{author}{\bibfnamefont{N.-U.} \bibnamefont{{Bastian}}},
  \bibinfo{author}{\bibfnamefont{D.}~\bibnamefont{{Blaschke}}},
  \bibinfo{author}{\bibfnamefont{T.}~\bibnamefont{{Fischer}}},
  \bibnamefont{and}
  \bibinfo{author}{\bibfnamefont{G.}~\bibnamefont{{R{\"o}pke}}},
  \bibinfo{journal}{Universe} \textbf{\bibinfo{volume}{4}}, \bibinfo{pages}{67}
  (\bibinfo{year}{2018}).

\bibitem[{\citenamefont{Benic et~al.}(2015)\citenamefont{Benic, Blaschke,
  Alvarez-Castillo, Fischer, and Typel}}]{Benic2015}
\bibinfo{author}{\bibfnamefont{S.}~\bibnamefont{Benic}},
  \bibinfo{author}{\bibfnamefont{D.}~\bibnamefont{Blaschke}},
  \bibinfo{author}{\bibfnamefont{D.~E.} \bibnamefont{Alvarez-Castillo}},
  \bibinfo{author}{\bibfnamefont{T.}~\bibnamefont{Fischer}}, \bibnamefont{and}
  \bibinfo{author}{\bibfnamefont{S.}~\bibnamefont{Typel}},
  \bibinfo{journal}{Astron. Astrophys.} \textbf{\bibinfo{volume}{577}},
  \bibinfo{pages}{A40} (\bibinfo{year}{2015}).

\bibitem[{\citenamefont{{Kl{\"a}hn} and {Fischer}}(2015)}]{Klaehn2015}
\bibinfo{author}{\bibfnamefont{T.}~\bibnamefont{{Kl{\"a}hn}}} \bibnamefont{and}
  \bibinfo{author}{\bibfnamefont{T.}~\bibnamefont{{Fischer}}},
  \bibinfo{journal}{\apj} \textbf{\bibinfo{volume}{810}}, \bibinfo{eid}{134}
  (\bibinfo{year}{2015}).

\bibitem[{\citenamefont{Typel}(2016)}]{Typel2016}
\bibinfo{author}{\bibfnamefont{S.}~\bibnamefont{Typel}},
  \bibinfo{journal}{European Physical Journal A} \textbf{\bibinfo{volume}{52}},
  \bibinfo{eid}{16} (\bibinfo{year}{2016}).

\bibitem[{\citenamefont{Akmal et~al.}(1998)\citenamefont{Akmal, Pandharipande,
  and Ravenhall}}]{Akmal1998}
\bibinfo{author}{\bibfnamefont{A.}~\bibnamefont{Akmal}},
  \bibinfo{author}{\bibfnamefont{V.~R.} \bibnamefont{Pandharipande}},
  \bibnamefont{and} \bibinfo{author}{\bibfnamefont{D.~G.}
  \bibnamefont{Ravenhall}}, \bibinfo{journal}{\prc}
  \textbf{\bibinfo{volume}{58}}, \bibinfo{pages}{1804} (\bibinfo{year}{1998}).

\bibitem[{\citenamefont{Banik et~al.}(2014)\citenamefont{Banik, Hempel, and
  Bandyopadhyay}}]{Banik2014}
\bibinfo{author}{\bibfnamefont{S.}~\bibnamefont{Banik}},
  \bibinfo{author}{\bibfnamefont{M.}~\bibnamefont{Hempel}}, \bibnamefont{and}
  \bibinfo{author}{\bibfnamefont{D.}~\bibnamefont{Bandyopadhyay}},
  \bibinfo{journal}{\apjs} \textbf{\bibinfo{volume}{214}}, \bibinfo{eid}{22}
  (\bibinfo{year}{2014}).

\bibitem[{\citenamefont{{Goriely} et~al.}(2010)\citenamefont{{Goriely},
  {Chamel}, and {Pearson}}}]{Goriely2010}
\bibinfo{author}{\bibfnamefont{S.}~\bibnamefont{{Goriely}}},
  \bibinfo{author}{\bibfnamefont{N.}~\bibnamefont{{Chamel}}}, \bibnamefont{and}
  \bibinfo{author}{\bibfnamefont{J.~M.} \bibnamefont{{Pearson}}},
  \bibinfo{journal}{\prc} \textbf{\bibinfo{volume}{82}}, \bibinfo{eid}{035804}
  (\bibinfo{year}{2010}).

\bibitem[{\citenamefont{Wiringa et~al.}(1988)\citenamefont{Wiringa, Fiks, and
  Fabrocini}}]{Wiringa1988}
\bibinfo{author}{\bibfnamefont{R.~B.} \bibnamefont{Wiringa}},
  \bibinfo{author}{\bibfnamefont{V.}~\bibnamefont{Fiks}}, \bibnamefont{and}
  \bibinfo{author}{\bibfnamefont{A.}~\bibnamefont{Fabrocini}},
  \bibinfo{journal}{\prc} \textbf{\bibinfo{volume}{38}}, \bibinfo{pages}{1010}
  (\bibinfo{year}{1988}).

\bibitem[{\citenamefont{Shen et~al.}(2011)\citenamefont{Shen, Horowitz, and
  Teige}}]{Shen2011}
\bibinfo{author}{\bibfnamefont{G.}~\bibnamefont{Shen}},
  \bibinfo{author}{\bibfnamefont{C.~J.} \bibnamefont{Horowitz}},
  \bibnamefont{and} \bibinfo{author}{\bibfnamefont{S.}~\bibnamefont{Teige}},
  \bibinfo{journal}{\prc} \textbf{\bibinfo{volume}{83}}, \bibinfo{eid}{035802}
  (\bibinfo{year}{2011}).

\bibitem[{\citenamefont{Lattimer and Douglas~Swesty}(1991)}]{Lattimer1991}
\bibinfo{author}{\bibfnamefont{J.~M.} \bibnamefont{Lattimer}} \bibnamefont{and}
  \bibinfo{author}{\bibfnamefont{F.}~\bibnamefont{Douglas~Swesty}},
  \bibinfo{journal}{Nuclear Physics A} \textbf{\bibinfo{volume}{535}},
  \bibinfo{pages}{331} (\bibinfo{year}{1991}).

\bibitem[{\citenamefont{Lalazissis et~al.}(1997)\citenamefont{Lalazissis,
  K{\"o}nig, and Ring}}]{Lalazissis1997a}
\bibinfo{author}{\bibfnamefont{G.~A.} \bibnamefont{Lalazissis}},
  \bibinfo{author}{\bibfnamefont{J.}~\bibnamefont{K{\"o}nig}},
  \bibnamefont{and} \bibinfo{author}{\bibfnamefont{P.}~\bibnamefont{Ring}},
  \bibinfo{journal}{\prc} \textbf{\bibinfo{volume}{55}}, \bibinfo{pages}{540}
  (\bibinfo{year}{1997}).

\bibitem[{\citenamefont{Steiner et~al.}(2013)\citenamefont{Steiner, Hempel, and
  Fischer}}]{Steiner2013}
\bibinfo{author}{\bibfnamefont{A.~W.} \bibnamefont{Steiner}},
  \bibinfo{author}{\bibfnamefont{M.}~\bibnamefont{Hempel}}, \bibnamefont{and}
  \bibinfo{author}{\bibfnamefont{T.}~\bibnamefont{Fischer}},
  \bibinfo{journal}{\apj} \textbf{\bibinfo{volume}{774}}, \bibinfo{eid}{17}
  (\bibinfo{year}{2013}).

\bibitem[{\citenamefont{{Douchin} and {Haensel}}(2001)}]{Douchin2001}
\bibinfo{author}{\bibfnamefont{F.}~\bibnamefont{{Douchin}}} \bibnamefont{and}
  \bibinfo{author}{\bibfnamefont{P.}~\bibnamefont{{Haensel}}},
  \bibinfo{journal}{\aap} \textbf{\bibinfo{volume}{380}}, \bibinfo{pages}{151}
  (\bibinfo{year}{2001}).

\bibitem[{\citenamefont{Sugahara and Toki}(1994)}]{Sugahara1994a}
\bibinfo{author}{\bibfnamefont{Y.}~\bibnamefont{Sugahara}} \bibnamefont{and}
  \bibinfo{author}{\bibfnamefont{H.}~\bibnamefont{Toki}},
  \bibinfo{journal}{Nuclear Physics A} \textbf{\bibinfo{volume}{579}},
  \bibinfo{pages}{557} (\bibinfo{year}{1994}).

\bibitem[{\citenamefont{Toki et~al.}(1995)\citenamefont{Toki, Hirata, Sugahara,
  Sumiyoshi, and Tanihata}}]{Toki1995}
\bibinfo{author}{\bibfnamefont{H.}~\bibnamefont{Toki}},
  \bibinfo{author}{\bibfnamefont{D.}~\bibnamefont{Hirata}},
  \bibinfo{author}{\bibfnamefont{Y.}~\bibnamefont{Sugahara}},
  \bibinfo{author}{\bibfnamefont{K.}~\bibnamefont{Sumiyoshi}},
  \bibnamefont{and} \bibinfo{author}{\bibfnamefont{I.}~\bibnamefont{Tanihata}},
  \bibinfo{journal}{Nuclear Physics A} \textbf{\bibinfo{volume}{588}},
  \bibinfo{pages}{357} (\bibinfo{year}{1995}).

\bibitem[{\citenamefont{{Bauswein}
  et~al.}(2013{\natexlab{a}})\citenamefont{{Bauswein}, {Goriely}, and
  {Janka}}}]{Bauswein2013a}
\bibinfo{author}{\bibfnamefont{A.}~\bibnamefont{{Bauswein}}},
  \bibinfo{author}{\bibfnamefont{S.}~\bibnamefont{{Goriely}}},
  \bibnamefont{and} \bibinfo{author}{\bibfnamefont{H.-T.}
  \bibnamefont{{Janka}}}, \bibinfo{journal}{\apj}
  \textbf{\bibinfo{volume}{773}}, \bibinfo{eid}{78}
  (\bibinfo{year}{2013}{\natexlab{a}}).

\bibitem[{\citenamefont{{Bauswein} et~al.}(2014)\citenamefont{{Bauswein},
  {Stergioulas}, and {Janka}}}]{Bauswein2014a}
\bibinfo{author}{\bibfnamefont{A.}~\bibnamefont{{Bauswein}}},
  \bibinfo{author}{\bibfnamefont{N.}~\bibnamefont{{Stergioulas}}},
  \bibnamefont{and} \bibinfo{author}{\bibfnamefont{H.-T.}
  \bibnamefont{{Janka}}}, \bibinfo{journal}{\prd}
  \textbf{\bibinfo{volume}{90}}, \bibinfo{eid}{023002} (\bibinfo{year}{2014}).

\bibitem[{\citenamefont{Fortin et~al.}(2018)\citenamefont{Fortin, Oertel, and
  Provid{\^e}ncia}}]{Fortin2018}
\bibinfo{author}{\bibfnamefont{M.}~\bibnamefont{Fortin}},
  \bibinfo{author}{\bibfnamefont{M.}~\bibnamefont{Oertel}}, \bibnamefont{and}
  \bibinfo{author}{\bibfnamefont{C.}~\bibnamefont{Provid{\^e}ncia}},
  \bibinfo{journal}{\pasa} \textbf{\bibinfo{volume}{35}}
  (\bibinfo{year}{2018}).

\bibitem[{\citenamefont{Marques et~al.}(2017)\citenamefont{Marques, Oertel,
  Hempel, and Novak}}]{Marques2017}
\bibinfo{author}{\bibfnamefont{M.}~\bibnamefont{Marques}},
  \bibinfo{author}{\bibfnamefont{M.}~\bibnamefont{Oertel}},
  \bibinfo{author}{\bibfnamefont{M.}~\bibnamefont{Hempel}}, \bibnamefont{and}
  \bibinfo{author}{\bibfnamefont{J.}~\bibnamefont{Novak}},
  \bibinfo{journal}{Phys. Rev.} \textbf{\bibinfo{volume}{C96}},
  \bibinfo{pages}{045806} (\bibinfo{year}{2017}).

\bibitem[{\citenamefont{Alford et~al.}(2005)\citenamefont{Alford, Braby, Paris,
  and Reddy}}]{Alford2005}
\bibinfo{author}{\bibfnamefont{M.}~\bibnamefont{Alford}},
  \bibinfo{author}{\bibfnamefont{M.}~\bibnamefont{Braby}},
  \bibinfo{author}{\bibfnamefont{M.}~\bibnamefont{Paris}}, \bibnamefont{and}
  \bibinfo{author}{\bibfnamefont{S.}~\bibnamefont{Reddy}},
  \bibinfo{journal}{\apj} \textbf{\bibinfo{volume}{629}}, \bibinfo{pages}{969}
  (\bibinfo{year}{2005}).

\bibitem[{\citenamefont{Read et~al.}(2009{\natexlab{b}})\citenamefont{Read,
  Lackey, Owen, and Friedman}}]{Read2009a}
\bibinfo{author}{\bibfnamefont{J.~S.} \bibnamefont{Read}},
  \bibinfo{author}{\bibfnamefont{B.~D.} \bibnamefont{Lackey}},
  \bibinfo{author}{\bibfnamefont{B.~J.} \bibnamefont{Owen}}, \bibnamefont{and}
  \bibinfo{author}{\bibfnamefont{J.~L.} \bibnamefont{Friedman}},
  \bibinfo{journal}{\prd} \textbf{\bibinfo{volume}{79}}, \bibinfo{eid}{124032}
  (\bibinfo{year}{2009}{\natexlab{b}}).

\bibitem[{\citenamefont{{Danielewicz} et~al.}(2002)\citenamefont{{Danielewicz},
  {Lacey}, and {Lynch}}}]{Danielewicz2002}
\bibinfo{author}{\bibfnamefont{P.}~\bibnamefont{{Danielewicz}}},
  \bibinfo{author}{\bibfnamefont{R.}~\bibnamefont{{Lacey}}}, \bibnamefont{and}
  \bibinfo{author}{\bibfnamefont{W.~G.} \bibnamefont{{Lynch}}},
  \bibinfo{journal}{Science} \textbf{\bibinfo{volume}{298}},
  \bibinfo{pages}{1592} (\bibinfo{year}{2002}).

\bibitem[{\citenamefont{{Tsang} et~al.}(2018)\citenamefont{{Tsang}, {Tsang},
  {Danielewicz}, {Lynch}, and {Fattoyev}}}]{Tsang2018}
\bibinfo{author}{\bibfnamefont{C.~Y.} \bibnamefont{{Tsang}}},
  \bibinfo{author}{\bibfnamefont{M.~B.} \bibnamefont{{Tsang}}},
  \bibinfo{author}{\bibfnamefont{P.}~\bibnamefont{{Danielewicz}}},
  \bibinfo{author}{\bibfnamefont{W.~G.} \bibnamefont{{Lynch}}},
  \bibnamefont{and} \bibinfo{author}{\bibfnamefont{F.~J.}
  \bibnamefont{{Fattoyev}}}, \bibinfo{journal}{ArXiv e-prints}
  (\bibinfo{year}{2018}), \eprint{1807.06571}.

\bibitem[{\citenamefont{{Lattimer} and {Lim}}(2013)}]{Lattimer2013}
\bibinfo{author}{\bibfnamefont{J.~M.} \bibnamefont{{Lattimer}}}
  \bibnamefont{and} \bibinfo{author}{\bibfnamefont{Y.}~\bibnamefont{{Lim}}},
  \bibinfo{journal}{\apj} \textbf{\bibinfo{volume}{771}}, \bibinfo{eid}{51}
  (\bibinfo{year}{2013}).

\bibitem[{\citenamefont{{Antoniadis} et~al.}(2013)\citenamefont{{Antoniadis},
  {Freire}, {Wex}, {Tauris}, {Lynch}, {van Kerkwijk}, {Kramer}, {Bassa},
  {Dhillon}, {Driebe} et~al.}}]{Antoniadis2013}
\bibinfo{author}{\bibfnamefont{J.}~\bibnamefont{{Antoniadis}}},
  \bibinfo{author}{\bibfnamefont{P.~C.~C.} \bibnamefont{{Freire}}},
  \bibinfo{author}{\bibfnamefont{N.}~\bibnamefont{{Wex}}},
  \bibinfo{author}{\bibfnamefont{T.~M.} \bibnamefont{{Tauris}}},
  \bibinfo{author}{\bibfnamefont{R.~S.} \bibnamefont{{Lynch}}},
  \bibinfo{author}{\bibfnamefont{M.~H.} \bibnamefont{{van Kerkwijk}}},
  \bibinfo{author}{\bibfnamefont{M.}~\bibnamefont{{Kramer}}},
  \bibinfo{author}{\bibfnamefont{C.}~\bibnamefont{{Bassa}}},
  \bibinfo{author}{\bibfnamefont{V.~S.} \bibnamefont{{Dhillon}}},
  \bibinfo{author}{\bibfnamefont{T.}~\bibnamefont{{Driebe}}},
  \bibnamefont{et~al.}, \bibinfo{journal}{Science}
  \textbf{\bibinfo{volume}{340}}, \bibinfo{pages}{448} (\bibinfo{year}{2013}).

\bibitem[{\citenamefont{Arzoumanian et~al.}(2018)\citenamefont{Arzoumanian,
  Brazier, Burke-Spolaor, Chamberlin, Chatterjee, Christy, Cordes, Cornish,
  Crawford, Thankful~Cromartie et~al.}}]{Arzoumanian2018a}
\bibinfo{author}{\bibfnamefont{Z.}~\bibnamefont{Arzoumanian}},
  \bibinfo{author}{\bibfnamefont{A.}~\bibnamefont{Brazier}},
  \bibinfo{author}{\bibfnamefont{S.}~\bibnamefont{Burke-Spolaor}},
  \bibinfo{author}{\bibfnamefont{S.}~\bibnamefont{Chamberlin}},
  \bibinfo{author}{\bibfnamefont{S.}~\bibnamefont{Chatterjee}},
  \bibinfo{author}{\bibfnamefont{B.}~\bibnamefont{Christy}},
  \bibinfo{author}{\bibfnamefont{J.~M.} \bibnamefont{Cordes}},
  \bibinfo{author}{\bibfnamefont{N.~J.} \bibnamefont{Cornish}},
  \bibinfo{author}{\bibfnamefont{F.}~\bibnamefont{Crawford}},
  \bibinfo{author}{\bibfnamefont{H.}~\bibnamefont{Thankful~Cromartie}},
  \bibnamefont{et~al.}, \bibinfo{journal}{\apjs}
  \textbf{\bibinfo{volume}{235}}, \bibinfo{eid}{37} (\bibinfo{year}{2018}).

\bibitem[{\citenamefont{{Bauswein} et~al.}(2017)\citenamefont{{Bauswein},
  {Just}, {Janka}, and {Stergioulas}}}]{Bauswein2017}
\bibinfo{author}{\bibfnamefont{A.}~\bibnamefont{{Bauswein}}},
  \bibinfo{author}{\bibfnamefont{O.}~\bibnamefont{{Just}}},
  \bibinfo{author}{\bibfnamefont{H.-T.} \bibnamefont{{Janka}}},
  \bibnamefont{and}
  \bibinfo{author}{\bibfnamefont{N.}~\bibnamefont{{Stergioulas}}},
  \bibinfo{journal}{\apjl} \textbf{\bibinfo{volume}{850}}, \bibinfo{eid}{L34}
  (\bibinfo{year}{2017}).

\bibitem[{\citenamefont{{Wilson} et~al.}(1996)\citenamefont{{Wilson},
  {Mathews}, and {Marronetti}}}]{Wilson1996}
\bibinfo{author}{\bibfnamefont{J.~R.} \bibnamefont{{Wilson}}},
  \bibinfo{author}{\bibfnamefont{G.~J.} \bibnamefont{{Mathews}}},
  \bibnamefont{and}
  \bibinfo{author}{\bibfnamefont{P.}~\bibnamefont{{Marronetti}}},
  \bibinfo{journal}{\prd} \textbf{\bibinfo{volume}{54}}, \bibinfo{pages}{1317}
  (\bibinfo{year}{1996}).

\bibitem[{\citenamefont{{Isenberg} and {Nester}}(1980)}]{Isenberg1980}
\bibinfo{author}{\bibfnamefont{J.}~\bibnamefont{{Isenberg}}} \bibnamefont{and}
  \bibinfo{author}{\bibfnamefont{J.}~\bibnamefont{{Nester}}}, in
  \emph{\bibinfo{booktitle}{General Relativity and Gravitation. Vol. 1. One
  hundred years after the birth of Albert Einstein. Edited by A. Held. New
  York, NY: Plenum Press, p. 23, 1980}}, edited by
  \bibinfo{editor}{\bibfnamefont{A.}~\bibnamefont{{Held}}}
  (\bibinfo{year}{1980}), p.~\bibinfo{pages}{23}.

\bibitem[{\citenamefont{Oechslin et~al.}(2002)\citenamefont{Oechslin, Rosswog,
  and Thielemann}}]{Oechslin2002}
\bibinfo{author}{\bibfnamefont{R.}~\bibnamefont{Oechslin}},
  \bibinfo{author}{\bibfnamefont{S.}~\bibnamefont{Rosswog}}, \bibnamefont{and}
  \bibinfo{author}{\bibfnamefont{F.-K.} \bibnamefont{Thielemann}},
  \bibinfo{journal}{\prd} \textbf{\bibinfo{volume}{65}}, \bibinfo{eid}{103005}
  (\bibinfo{year}{2002}).

\bibitem[{\citenamefont{Oechslin et~al.}(2007)\citenamefont{Oechslin, Janka,
  and Marek}}]{Oechslin2007}
\bibinfo{author}{\bibfnamefont{R.}~\bibnamefont{Oechslin}},
  \bibinfo{author}{\bibfnamefont{H.-T.} \bibnamefont{Janka}}, \bibnamefont{and}
  \bibinfo{author}{\bibfnamefont{A.}~\bibnamefont{Marek}},
  \bibinfo{journal}{\aap} \textbf{\bibinfo{volume}{467}}, \bibinfo{pages}{395}
  (\bibinfo{year}{2007}).

\bibitem[{\citenamefont{Bauswein
  et~al.}(2010{\natexlab{b}})\citenamefont{Bauswein, Janka, and
  Oechslin}}]{Bauswein2010}
\bibinfo{author}{\bibfnamefont{A.}~\bibnamefont{Bauswein}},
  \bibinfo{author}{\bibfnamefont{H.-T.} \bibnamefont{Janka}}, \bibnamefont{and}
  \bibinfo{author}{\bibfnamefont{R.}~\bibnamefont{Oechslin}},
  \bibinfo{journal}{\prd} \textbf{\bibinfo{volume}{82}}, \bibinfo{eid}{084043}
  (\bibinfo{year}{2010}{\natexlab{b}}).

\bibitem[{\citenamefont{{Dominik} et~al.}(2012)\citenamefont{{Dominik},
  {Belczynski}, {Fryer}, {Holz}, {Berti}, {Bulik}, {Mandel}, and
  {O'Shaughnessy}}}]{Dominik2012}
\bibinfo{author}{\bibfnamefont{M.}~\bibnamefont{{Dominik}}},
  \bibinfo{author}{\bibfnamefont{K.}~\bibnamefont{{Belczynski}}},
  \bibinfo{author}{\bibfnamefont{C.}~\bibnamefont{{Fryer}}},
  \bibinfo{author}{\bibfnamefont{D.~E.} \bibnamefont{{Holz}}},
  \bibinfo{author}{\bibfnamefont{E.}~\bibnamefont{{Berti}}},
  \bibinfo{author}{\bibfnamefont{T.}~\bibnamefont{{Bulik}}},
  \bibinfo{author}{\bibfnamefont{I.}~\bibnamefont{{Mandel}}}, \bibnamefont{and}
  \bibinfo{author}{\bibfnamefont{R.}~\bibnamefont{{O'Shaughnessy}}},
  \bibinfo{journal}{\apj} \textbf{\bibinfo{volume}{759}}, \bibinfo{eid}{52}
  (\bibinfo{year}{2012}).

\bibitem[{\citenamefont{{Lattimer}}(2012)}]{Lattimer2012}
\bibinfo{author}{\bibfnamefont{J.~M.} \bibnamefont{{Lattimer}}},
  \bibinfo{journal}{Annu. Rev. Nucl. Part. Sci.} \textbf{\bibinfo{volume}{62}},
  \bibinfo{pages}{485} (\bibinfo{year}{2012}).

\bibitem[{\citenamefont{{Rodriguez} et~al.}(2014)\citenamefont{{Rodriguez},
  {Farr}, {Raymond}, {Farr}, {Littenberg}, {Fazi}, and
  {Kalogera}}}]{Rodriguez2014}
\bibinfo{author}{\bibfnamefont{C.~L.} \bibnamefont{{Rodriguez}}},
  \bibinfo{author}{\bibfnamefont{B.}~\bibnamefont{{Farr}}},
  \bibinfo{author}{\bibfnamefont{V.}~\bibnamefont{{Raymond}}},
  \bibinfo{author}{\bibfnamefont{W.~M.} \bibnamefont{{Farr}}},
  \bibinfo{author}{\bibfnamefont{T.~B.} \bibnamefont{{Littenberg}}},
  \bibinfo{author}{\bibfnamefont{D.}~\bibnamefont{{Fazi}}}, \bibnamefont{and}
  \bibinfo{author}{\bibfnamefont{V.}~\bibnamefont{{Kalogera}}},
  \bibinfo{journal}{\apj} \textbf{\bibinfo{volume}{784}}, \bibinfo{eid}{119}
  (\bibinfo{year}{2014}).

\bibitem[{\citenamefont{{Farr} et~al.}(2016)\citenamefont{{Farr}, {Berry},
  {Farr}, {Haster}, {Middleton}, {Cannon}, {Graff}, {Hanna}, {Mandel}, {Pankow}
  et~al.}}]{Farr2016}
\bibinfo{author}{\bibfnamefont{B.}~\bibnamefont{{Farr}}},
  \bibinfo{author}{\bibfnamefont{C.~P.~L.} \bibnamefont{{Berry}}},
  \bibinfo{author}{\bibfnamefont{W.~M.} \bibnamefont{{Farr}}},
  \bibinfo{author}{\bibfnamefont{C.-J.} \bibnamefont{{Haster}}},
  \bibinfo{author}{\bibfnamefont{H.}~\bibnamefont{{Middleton}}},
  \bibinfo{author}{\bibfnamefont{K.}~\bibnamefont{{Cannon}}},
  \bibinfo{author}{\bibfnamefont{P.~B.} \bibnamefont{{Graff}}},
  \bibinfo{author}{\bibfnamefont{C.}~\bibnamefont{{Hanna}}},
  \bibinfo{author}{\bibfnamefont{I.}~\bibnamefont{{Mandel}}},
  \bibinfo{author}{\bibfnamefont{C.}~\bibnamefont{{Pankow}}},
  \bibnamefont{et~al.}, \bibinfo{journal}{\apj} \textbf{\bibinfo{volume}{825}},
  \bibinfo{eid}{116} (\bibinfo{year}{2016}).

\bibitem[{\citenamefont{{Shibata} et~al.}(2005)\citenamefont{{Shibata},
  {Taniguchi}, and {Ury{\=u}}}}]{Shibata2005a}
\bibinfo{author}{\bibfnamefont{M.}~\bibnamefont{{Shibata}}},
  \bibinfo{author}{\bibfnamefont{K.}~\bibnamefont{{Taniguchi}}},
  \bibnamefont{and}
  \bibinfo{author}{\bibfnamefont{K.}~\bibnamefont{{Ury{\=u}}}},
  \bibinfo{journal}{\prd} \textbf{\bibinfo{volume}{71}}, \bibinfo{eid}{084021}
  (\bibinfo{year}{2005}).

\bibitem[{\citenamefont{{Shibata} and {Taniguchi}}(2006)}]{Shibata2006}
\bibinfo{author}{\bibfnamefont{M.}~\bibnamefont{{Shibata}}} \bibnamefont{and}
  \bibinfo{author}{\bibfnamefont{K.}~\bibnamefont{{Taniguchi}}},
  \bibinfo{journal}{\prd} \textbf{\bibinfo{volume}{73}}, \bibinfo{eid}{064027}
  (\bibinfo{year}{2006}).

\bibitem[{\citenamefont{{Oechslin} and {Janka}}(2007)}]{Oechslin2007a}
\bibinfo{author}{\bibfnamefont{R.}~\bibnamefont{{Oechslin}}} \bibnamefont{and}
  \bibinfo{author}{\bibfnamefont{H.-T.} \bibnamefont{{Janka}}},
  \bibinfo{journal}{\prl} \textbf{\bibinfo{volume}{99}}, \bibinfo{eid}{121102}
  (\bibinfo{year}{2007}).

\bibitem[{\citenamefont{{Hotokezaka} et~al.}(2011)\citenamefont{{Hotokezaka},
  {Kyutoku}, {Okawa}, {Shibata}, and {Kiuchi}}}]{Hotokezaka2011}
\bibinfo{author}{\bibfnamefont{K.}~\bibnamefont{{Hotokezaka}}},
  \bibinfo{author}{\bibfnamefont{K.}~\bibnamefont{{Kyutoku}}},
  \bibinfo{author}{\bibfnamefont{H.}~\bibnamefont{{Okawa}}},
  \bibinfo{author}{\bibfnamefont{M.}~\bibnamefont{{Shibata}}},
  \bibnamefont{and} \bibinfo{author}{\bibfnamefont{K.}~\bibnamefont{{Kiuchi}}},
  \bibinfo{journal}{\prd} \textbf{\bibinfo{volume}{83}}, \bibinfo{eid}{124008}
  (\bibinfo{year}{2011}).

\bibitem[{\citenamefont{{Bauswein}
  et~al.}(2013{\natexlab{b}})\citenamefont{{Bauswein}, {Baumgarte}, and
  {Janka}}}]{Bauswein2013}
\bibinfo{author}{\bibfnamefont{A.}~\bibnamefont{{Bauswein}}},
  \bibinfo{author}{\bibfnamefont{T.~W.} \bibnamefont{{Baumgarte}}},
  \bibnamefont{and} \bibinfo{author}{\bibfnamefont{H.-T.}
  \bibnamefont{{Janka}}}, \bibinfo{journal}{\prl}
  \textbf{\bibinfo{volume}{111}}, \bibinfo{eid}{131101}
  (\bibinfo{year}{2013}{\natexlab{b}}).

\bibitem[{\citenamefont{Zhuge et~al.}(1996)\citenamefont{Zhuge, Centrella, and
  McMillan}}]{Zhuge1996}
\bibinfo{author}{\bibfnamefont{X.}~\bibnamefont{Zhuge}},
  \bibinfo{author}{\bibfnamefont{J.~M.} \bibnamefont{Centrella}},
  \bibnamefont{and} \bibinfo{author}{\bibfnamefont{S.~L.~W.}
  \bibnamefont{McMillan}}, \bibinfo{journal}{\prd}
  \textbf{\bibinfo{volume}{54}}, \bibinfo{pages}{7261} (\bibinfo{year}{1996}).

\bibitem[{\citenamefont{Bernuzzi et~al.}(2015)\citenamefont{Bernuzzi, Dietrich,
  and Nagar}}]{Bernuzzi2015}
\bibinfo{author}{\bibfnamefont{S.}~\bibnamefont{Bernuzzi}},
  \bibinfo{author}{\bibfnamefont{T.}~\bibnamefont{Dietrich}}, \bibnamefont{and}
  \bibinfo{author}{\bibfnamefont{A.}~\bibnamefont{Nagar}},
  \bibinfo{journal}{\prl} \textbf{\bibinfo{volume}{115}}, \bibinfo{eid}{091101}
  (\bibinfo{year}{2015}).

\bibitem[{\citenamefont{Rezzolla and Takami}(2016)}]{Rezzolla2016}
\bibinfo{author}{\bibfnamefont{L.}~\bibnamefont{Rezzolla}} \bibnamefont{and}
  \bibinfo{author}{\bibfnamefont{K.}~\bibnamefont{Takami}},
  \bibinfo{journal}{\prd} \textbf{\bibinfo{volume}{93}}, \bibinfo{eid}{124051}
  (\bibinfo{year}{2016}).

\bibitem[{\citenamefont{Dudi et~al.}(2018)\citenamefont{Dudi, Pannarale,
  Dietrich, Hannam, Bernuzzi, Ohme, and Br{\"u}gmann}}]{Dudi2018}
\bibinfo{author}{\bibfnamefont{R.}~\bibnamefont{Dudi}},
  \bibinfo{author}{\bibfnamefont{F.}~\bibnamefont{Pannarale}},
  \bibinfo{author}{\bibfnamefont{T.}~\bibnamefont{Dietrich}},
  \bibinfo{author}{\bibfnamefont{M.}~\bibnamefont{Hannam}},
  \bibinfo{author}{\bibfnamefont{S.}~\bibnamefont{Bernuzzi}},
  \bibinfo{author}{\bibfnamefont{F.}~\bibnamefont{Ohme}}, \bibnamefont{and}
  \bibinfo{author}{\bibfnamefont{B.}~\bibnamefont{Br{\"u}gmann}},
  \bibinfo{journal}{\prd} \textbf{\bibinfo{volume}{98}}, \bibinfo{eid}{084061}
  (\bibinfo{year}{2018}).

\bibitem[{\citenamefont{{Bodmer}}(1971)}]{Bodmer1971}
\bibinfo{author}{\bibfnamefont{A.~R.} \bibnamefont{{Bodmer}}},
  \bibinfo{journal}{\prd} \textbf{\bibinfo{volume}{4}}, \bibinfo{pages}{1601}
  (\bibinfo{year}{1971}).

\bibitem[{\citenamefont{{Witten}}(1984)}]{Witten1984}
\bibinfo{author}{\bibfnamefont{E.}~\bibnamefont{{Witten}}},
  \bibinfo{journal}{\prd} \textbf{\bibinfo{volume}{30}}, \bibinfo{pages}{272}
  (\bibinfo{year}{1984}).

\bibitem[{\citenamefont{Alcock et~al.}(1986)\citenamefont{Alcock, Farhi, and
  Olinto}}]{Alcock1986}
\bibinfo{author}{\bibfnamefont{C.}~\bibnamefont{Alcock}},
  \bibinfo{author}{\bibfnamefont{E.}~\bibnamefont{Farhi}}, \bibnamefont{and}
  \bibinfo{author}{\bibfnamefont{A.}~\bibnamefont{Olinto}},
  \bibinfo{journal}{\apj} \textbf{\bibinfo{volume}{310}}, \bibinfo{pages}{261}
  (\bibinfo{year}{1986}).

\bibitem[{\citenamefont{{Haensel} et~al.}(1986)\citenamefont{{Haensel},
  {Zdunik}, and {Schaefer}}}]{Haensel1986}
\bibinfo{author}{\bibfnamefont{P.}~\bibnamefont{{Haensel}}},
  \bibinfo{author}{\bibfnamefont{J.~L.} \bibnamefont{{Zdunik}}},
  \bibnamefont{and}
  \bibinfo{author}{\bibfnamefont{R.}~\bibnamefont{{Schaefer}}},
  \bibinfo{journal}{\aap} \textbf{\bibinfo{volume}{160}}, \bibinfo{pages}{121}
  (\bibinfo{year}{1986}).

\bibitem[{\citenamefont{Bauswein et~al.}(2009)\citenamefont{Bauswein, Janka,
  Oechslin, Pagliara, Sagert, Schaffner-Bielich, Hohle, and
  Neuh{\"a}user}}]{Bauswein2009}
\bibinfo{author}{\bibfnamefont{A.}~\bibnamefont{Bauswein}},
  \bibinfo{author}{\bibfnamefont{H.-T.} \bibnamefont{Janka}},
  \bibinfo{author}{\bibfnamefont{R.}~\bibnamefont{Oechslin}},
  \bibinfo{author}{\bibfnamefont{G.}~\bibnamefont{Pagliara}},
  \bibinfo{author}{\bibfnamefont{I.}~\bibnamefont{Sagert}},
  \bibinfo{author}{\bibfnamefont{J.}~\bibnamefont{Schaffner-Bielich}},
  \bibinfo{author}{\bibfnamefont{M.~M.} \bibnamefont{Hohle}}, \bibnamefont{and}
  \bibinfo{author}{\bibfnamefont{R.}~\bibnamefont{Neuh{\"a}user}},
  \bibinfo{journal}{Physical Review Letters} \textbf{\bibinfo{volume}{103}},
  \bibinfo{eid}{011101} (\bibinfo{year}{2009}).

\bibitem[{\citenamefont{Abbott et~al.}(2017{\natexlab{b}})\citenamefont{Abbott,
  Abbott, Abbott, Acernese, Ackley, Adams, Adams, Addesso, Adhikari, Adya
  et~al.}}]{Abbott2017b}
\bibinfo{author}{\bibfnamefont{B.~P.} \bibnamefont{Abbott}},
  \bibinfo{author}{\bibfnamefont{R.}~\bibnamefont{Abbott}},
  \bibinfo{author}{\bibfnamefont{T.~D.} \bibnamefont{Abbott}},
  \bibinfo{author}{\bibfnamefont{F.}~\bibnamefont{Acernese}},
  \bibinfo{author}{\bibfnamefont{K.}~\bibnamefont{Ackley}},
  \bibinfo{author}{\bibfnamefont{C.}~\bibnamefont{Adams}},
  \bibinfo{author}{\bibfnamefont{T.}~\bibnamefont{Adams}},
  \bibinfo{author}{\bibfnamefont{P.}~\bibnamefont{Addesso}},
  \bibinfo{author}{\bibfnamefont{R.~X.} \bibnamefont{Adhikari}},
  \bibinfo{author}{\bibfnamefont{V.~B.} \bibnamefont{Adya}},
  \bibnamefont{et~al.}, \bibinfo{journal}{\apjl}
  \textbf{\bibinfo{volume}{848}}, \bibinfo{eid}{L12}
  (\bibinfo{year}{2017}{\natexlab{b}}).

\bibitem[{\citenamefont{Dexheimer and Schramm}(2010)}]{Dexheimer2010}
\bibinfo{author}{\bibfnamefont{V.~A.} \bibnamefont{Dexheimer}}
  \bibnamefont{and} \bibinfo{author}{\bibfnamefont{S.}~\bibnamefont{Schramm}},
  \bibinfo{journal}{\prc} \textbf{\bibinfo{volume}{81}}, \bibinfo{eid}{045201}
  (\bibinfo{year}{2010}).

\bibitem[{\citenamefont{{Friman} et~al.}(2011)\citenamefont{{Friman},
  {H{\"o}hne}, {Knoll}, {Leupold}, {Randrup}, {Rapp}, and
  {Senger}}}]{Friman2011}
\bibinfo{editor}{\bibfnamefont{B.}~\bibnamefont{{Friman}}},
  \bibinfo{editor}{\bibfnamefont{C.}~\bibnamefont{{H{\"o}hne}}},
  \bibinfo{editor}{\bibfnamefont{J.}~\bibnamefont{{Knoll}}},
  \bibinfo{editor}{\bibfnamefont{S.}~\bibnamefont{{Leupold}}},
  \bibinfo{editor}{\bibfnamefont{J.}~\bibnamefont{{Randrup}}},
  \bibinfo{editor}{\bibfnamefont{R.}~\bibnamefont{{Rapp}}}, \bibnamefont{and}
  \bibinfo{editor}{\bibfnamefont{P.}~\bibnamefont{{Senger}}}, eds.,
  \emph{\bibinfo{title}{{The CBM Physics Book}}}, vol. \bibinfo{volume}{814} of
  \emph{\bibinfo{series}{Lecture Notes in Physics, Berlin Springer Verlag}}
  (\bibinfo{year}{2011}).

\bibitem[{\citenamefont{{Blaschke} et~al.}(2016)\citenamefont{{Blaschke},
  {Aichelin}, {Bratkovskaya}, {Friese}, {Gazdzicki}, {Randrup}, {Rogachevsky},
  {Teryaev}, and {Toneev}}}]{Blaschke2016}
\bibinfo{author}{\bibfnamefont{D.}~\bibnamefont{{Blaschke}}},
  \bibinfo{author}{\bibfnamefont{J.}~\bibnamefont{{Aichelin}}},
  \bibinfo{author}{\bibfnamefont{E.}~\bibnamefont{{Bratkovskaya}}},
  \bibinfo{author}{\bibfnamefont{V.}~\bibnamefont{{Friese}}},
  \bibinfo{author}{\bibfnamefont{M.}~\bibnamefont{{Gazdzicki}}},
  \bibinfo{author}{\bibfnamefont{J.}~\bibnamefont{{Randrup}}},
  \bibinfo{author}{\bibfnamefont{O.}~\bibnamefont{{Rogachevsky}}},
  \bibinfo{author}{\bibfnamefont{O.}~\bibnamefont{{Teryaev}}},
  \bibnamefont{and} \bibinfo{author}{\bibfnamefont{V.}~\bibnamefont{{Toneev}}},
  \bibinfo{journal}{European Physical Journal A} \textbf{\bibinfo{volume}{52}},
  \bibinfo{eid}{267} (\bibinfo{year}{2016}).

\bibitem[{\citenamefont{{LIGO Scientific Collaboration}
  et~al.}(2015{\natexlab{a}})\citenamefont{{LIGO Scientific Collaboration},
  {Aasi}, {Abbott}, {Abbott}, {Abbott}, {Abernathy}, {Ackley}, {Adams},
  {Adams}, {Addesso} et~al.}}]{LIGOScientificCollaboration2015}
\bibinfo{author}{\bibnamefont{{LIGO Scientific Collaboration}}},
  \bibinfo{author}{\bibfnamefont{J.}~\bibnamefont{{Aasi}}},
  \bibinfo{author}{\bibfnamefont{B.~P.} \bibnamefont{{Abbott}}},
  \bibinfo{author}{\bibfnamefont{R.}~\bibnamefont{{Abbott}}},
  \bibinfo{author}{\bibfnamefont{T.}~\bibnamefont{{Abbott}}},
  \bibinfo{author}{\bibfnamefont{M.~R.} \bibnamefont{{Abernathy}}},
  \bibinfo{author}{\bibfnamefont{K.}~\bibnamefont{{Ackley}}},
  \bibinfo{author}{\bibfnamefont{C.}~\bibnamefont{{Adams}}},
  \bibinfo{author}{\bibfnamefont{T.}~\bibnamefont{{Adams}}},
  \bibinfo{author}{\bibfnamefont{P.}~\bibnamefont{{Addesso}}},
  \bibnamefont{et~al.}, \bibinfo{journal}{Classical and Quantum Gravity}
  \textbf{\bibinfo{volume}{32}}, \bibinfo{eid}{074001}
  (\bibinfo{year}{2015}{\natexlab{a}}).

\bibitem[{\citenamefont{{LIGO Scientific Collaboration}
  et~al.}(2015{\natexlab{b}})\citenamefont{{LIGO Scientific Collaboration},
  {Aasi}, {Abbott}, {Abbott}, {Abbott}, {Abernathy}, {Ackley}, {Adams},
  {Adams}, {Addesso} et~al.}}]{LIGOScientificCollaboration2015a}
\bibinfo{author}{\bibnamefont{{LIGO Scientific Collaboration}}},
  \bibinfo{author}{\bibfnamefont{J.}~\bibnamefont{{Aasi}}},
  \bibinfo{author}{\bibfnamefont{B.~P.} \bibnamefont{{Abbott}}},
  \bibinfo{author}{\bibfnamefont{R.}~\bibnamefont{{Abbott}}},
  \bibinfo{author}{\bibfnamefont{T.}~\bibnamefont{{Abbott}}},
  \bibinfo{author}{\bibfnamefont{M.~R.} \bibnamefont{{Abernathy}}},
  \bibinfo{author}{\bibfnamefont{K.}~\bibnamefont{{Ackley}}},
  \bibinfo{author}{\bibfnamefont{C.}~\bibnamefont{{Adams}}},
  \bibinfo{author}{\bibfnamefont{T.}~\bibnamefont{{Adams}}},
  \bibinfo{author}{\bibfnamefont{P.}~\bibnamefont{{Addesso}}},
  \bibnamefont{et~al.}, \bibinfo{journal}{Classical and Quantum Gravity}
  \textbf{\bibinfo{volume}{32}}, \bibinfo{eid}{074001}
  (\bibinfo{year}{2015}{\natexlab{b}}).

\bibitem[{\citenamefont{{Acernese} et~al.}(2015)\citenamefont{{Acernese},
  {Agathos}, {Agatsuma}, {Aisa}, {Allemandou}, {Allocca}, {Amarni}, {Astone},
  {Balestri}, {Ballardin} et~al.}}]{Acernese2015}
\bibinfo{author}{\bibfnamefont{F.}~\bibnamefont{{Acernese}}},
  \bibinfo{author}{\bibfnamefont{M.}~\bibnamefont{{Agathos}}},
  \bibinfo{author}{\bibfnamefont{K.}~\bibnamefont{{Agatsuma}}},
  \bibinfo{author}{\bibfnamefont{D.}~\bibnamefont{{Aisa}}},
  \bibinfo{author}{\bibfnamefont{N.}~\bibnamefont{{Allemandou}}},
  \bibinfo{author}{\bibfnamefont{A.}~\bibnamefont{{Allocca}}},
  \bibinfo{author}{\bibfnamefont{J.}~\bibnamefont{{Amarni}}},
  \bibinfo{author}{\bibfnamefont{P.}~\bibnamefont{{Astone}}},
  \bibinfo{author}{\bibfnamefont{G.}~\bibnamefont{{Balestri}}},
  \bibinfo{author}{\bibfnamefont{G.}~\bibnamefont{{Ballardin}}},
  \bibnamefont{et~al.}, \bibinfo{journal}{Classical and Quantum Gravity}
  \textbf{\bibinfo{volume}{32}}, \bibinfo{eid}{024001} (\bibinfo{year}{2015}).

\bibitem[{\citenamefont{Collaboration}(2017)}]{Collaboration2017b}
\bibinfo{author}{\bibfnamefont{L.~S.} \bibnamefont{Collaboration}},
  \bibinfo{journal}{LIGO Document Control Center}  (\bibinfo{year}{2017}),
  \urlprefix\url{https://dcc.ligo.org/LIGO-T1700231/public}.

\bibitem[{\citenamefont{Punturo et~al.}(2010)\citenamefont{Punturo, Abernathy,
  Acernese, Allen, Andersson, Arun, Barone, Barr, Barsuglia, Beker
  et~al.}}]{Punturo2010}
\bibinfo{author}{\bibfnamefont{M.}~\bibnamefont{Punturo}},
  \bibinfo{author}{\bibfnamefont{M.}~\bibnamefont{Abernathy}},
  \bibinfo{author}{\bibfnamefont{F.}~\bibnamefont{Acernese}},
  \bibinfo{author}{\bibfnamefont{B.}~\bibnamefont{Allen}},
  \bibinfo{author}{\bibfnamefont{N.}~\bibnamefont{Andersson}},
  \bibinfo{author}{\bibfnamefont{K.}~\bibnamefont{Arun}},
  \bibinfo{author}{\bibfnamefont{F.}~\bibnamefont{Barone}},
  \bibinfo{author}{\bibfnamefont{B.}~\bibnamefont{Barr}},
  \bibinfo{author}{\bibfnamefont{M.}~\bibnamefont{Barsuglia}},
  \bibinfo{author}{\bibfnamefont{M.}~\bibnamefont{Beker}},
  \bibnamefont{et~al.}, \bibinfo{journal}{Classical and Quantum Gravity}
  \textbf{\bibinfo{volume}{27}}, \bibinfo{pages}{084007}
  (\bibinfo{year}{2010}),
  \urlprefix\url{http://stacks.iop.org/0264-9381/27/i=8/a=084007}.

\bibitem[{\citenamefont{Hild et~al.}(2011)\citenamefont{Hild, Abernathy,
  Acernese, Amaro-Seoane, Andersson, Arun, Barone, Barr, Barsuglia, Beker
  et~al.}}]{Hild2011}
\bibinfo{author}{\bibfnamefont{S.}~\bibnamefont{Hild}},
  \bibinfo{author}{\bibfnamefont{M.}~\bibnamefont{Abernathy}},
  \bibinfo{author}{\bibfnamefont{F.}~\bibnamefont{Acernese}},
  \bibinfo{author}{\bibfnamefont{P.}~\bibnamefont{Amaro-Seoane}},
  \bibinfo{author}{\bibfnamefont{N.}~\bibnamefont{Andersson}},
  \bibinfo{author}{\bibfnamefont{K.}~\bibnamefont{Arun}},
  \bibinfo{author}{\bibfnamefont{F.}~\bibnamefont{Barone}},
  \bibinfo{author}{\bibfnamefont{B.}~\bibnamefont{Barr}},
  \bibinfo{author}{\bibfnamefont{M.}~\bibnamefont{Barsuglia}},
  \bibinfo{author}{\bibfnamefont{M.}~\bibnamefont{Beker}},
  \bibnamefont{et~al.}, \bibinfo{journal}{Classical and Quantum Gravity}
  \textbf{\bibinfo{volume}{28}}, \bibinfo{pages}{094013}
  (\bibinfo{year}{2011}),
  \urlprefix\url{http://stacks.iop.org/0264-9381/28/i=9/a=094013}.

\bibitem[{\citenamefont{{Miller} et~al.}(2015)\citenamefont{{Miller},
  {Barsotti}, {Vitale}, {Fritschel}, {Evans}, and {Sigg}}}]{Miller2015}
\bibinfo{author}{\bibfnamefont{J.}~\bibnamefont{{Miller}}},
  \bibinfo{author}{\bibfnamefont{L.}~\bibnamefont{{Barsotti}}},
  \bibinfo{author}{\bibfnamefont{S.}~\bibnamefont{{Vitale}}},
  \bibinfo{author}{\bibfnamefont{P.}~\bibnamefont{{Fritschel}}},
  \bibinfo{author}{\bibfnamefont{M.}~\bibnamefont{{Evans}}}, \bibnamefont{and}
  \bibinfo{author}{\bibfnamefont{D.}~\bibnamefont{{Sigg}}},
  \bibinfo{journal}{\prd} \textbf{\bibinfo{volume}{91}}, \bibinfo{eid}{062005}
  (\bibinfo{year}{2015}).

\end{thebibliography}

\begin{thebibliography}{49}
\expandafter\ifx\csname natexlab\endcsname\relax\def\natexlab#1{#1}\fi
\expandafter\ifx\csname bibnamefont\endcsname\relax
  \def\bibnamefont#1{#1}\fi
\expandafter\ifx\csname bibfnamefont\endcsname\relax
  \def\bibfnamefont#1{#1}\fi
\expandafter\ifx\csname citenamefont\endcsname\relax
  \def\citenamefont#1{#1}\fi
\expandafter\ifx\csname url\endcsname\relax
  \def\url#1{\texttt{#1}}\fi
\expandafter\ifx\csname urlprefix\endcsname\relax\def\urlprefix{URL }\fi
\providecommand{\bibinfo}[2]{#2}
\providecommand{\eprint}[2][]{\url{#2}}

\bibitem[{\citenamefont{{Typel}}(2005)}]{Typel2005}
\bibinfo{author}{\bibfnamefont{S.}~\bibnamefont{{Typel}}},
  \bibinfo{journal}{\prc} \textbf{\bibinfo{volume}{71}}, \bibinfo{eid}{064301}
  (\bibinfo{year}{2005}).

\bibitem[{\citenamefont{{Typel} et~al.}(2010)\citenamefont{{Typel},
  {R{\"o}pke}, {Kl{\"a}hn}, {Blaschke}, and {Wolter}}}]{Typel2010}
\bibinfo{author}{\bibfnamefont{S.}~\bibnamefont{{Typel}}},
  \bibinfo{author}{\bibfnamefont{G.}~\bibnamefont{{R{\"o}pke}}},
  \bibinfo{author}{\bibfnamefont{T.}~\bibnamefont{{Kl{\"a}hn}}},
  \bibinfo{author}{\bibfnamefont{D.}~\bibnamefont{{Blaschke}}},
  \bibnamefont{and} \bibinfo{author}{\bibfnamefont{H.~H.}
  \bibnamefont{{Wolter}}}, \bibinfo{journal}{\prc}
  \textbf{\bibinfo{volume}{81}}, \bibinfo{eid}{015803} (\bibinfo{year}{2010}).

\bibitem[{\citenamefont{Alvarez-Castillo
  et~al.}(2016)\citenamefont{Alvarez-Castillo, Ayriyan, Benic, Blaschke,
  Grigorian, and Typel}}]{Alvarez-Castillo2016}
\bibinfo{author}{\bibfnamefont{D.}~\bibnamefont{Alvarez-Castillo}},
  \bibinfo{author}{\bibfnamefont{A.}~\bibnamefont{Ayriyan}},
  \bibinfo{author}{\bibfnamefont{S.}~\bibnamefont{Benic}},
  \bibinfo{author}{\bibfnamefont{D.}~\bibnamefont{Blaschke}},
  \bibinfo{author}{\bibfnamefont{H.}~\bibnamefont{Grigorian}},
  \bibnamefont{and} \bibinfo{author}{\bibfnamefont{S.}~\bibnamefont{Typel}},
  \bibinfo{journal}{European Physical Journal A} \textbf{\bibinfo{volume}{52}},
  \bibinfo{eid}{69} (\bibinfo{year}{2016}).

\bibitem[{\citenamefont{{Danielewicz} et~al.}(2002)\citenamefont{{Danielewicz},
  {Lacey}, and {Lynch}}}]{Danielewicz2002}
\bibinfo{author}{\bibfnamefont{P.}~\bibnamefont{{Danielewicz}}},
  \bibinfo{author}{\bibfnamefont{R.}~\bibnamefont{{Lacey}}}, \bibnamefont{and}
  \bibinfo{author}{\bibfnamefont{W.~G.} \bibnamefont{{Lynch}}},
  \bibinfo{journal}{Science} \textbf{\bibinfo{volume}{298}},
  \bibinfo{pages}{1592} (\bibinfo{year}{2002}).

\bibitem[{\citenamefont{{Tsang} et~al.}(2018)\citenamefont{{Tsang}, {Tsang},
  {Danielewicz}, {Lynch}, and {Fattoyev}}}]{Tsang2018}
\bibinfo{author}{\bibfnamefont{C.~Y.} \bibnamefont{{Tsang}}},
  \bibinfo{author}{\bibfnamefont{M.~B.} \bibnamefont{{Tsang}}},
  \bibinfo{author}{\bibfnamefont{P.}~\bibnamefont{{Danielewicz}}},
  \bibinfo{author}{\bibfnamefont{W.~G.} \bibnamefont{{Lynch}}},
  \bibnamefont{and} \bibinfo{author}{\bibfnamefont{F.~J.}
  \bibnamefont{{Fattoyev}}}, \bibinfo{journal}{ArXiv e-prints}
  (\bibinfo{year}{2018}), \eprint{1807.06571}.

\bibitem[{\citenamefont{{Hempel} and {Schaffner-Bielich}}(2010)}]{Hempel2010}
\bibinfo{author}{\bibfnamefont{M.}~\bibnamefont{{Hempel}}} \bibnamefont{and}
  \bibinfo{author}{\bibfnamefont{J.}~\bibnamefont{{Schaffner-Bielich}}},
  \bibinfo{journal}{Nucl. Phys. A} \textbf{\bibinfo{volume}{837}},
  \bibinfo{pages}{210} (\bibinfo{year}{2010}).

\bibitem[{\citenamefont{{Hempel} et~al.}(2012)\citenamefont{{Hempel},
  {Fischer}, {Schaffner-Bielich}, and {Liebend{\"o}rfer}}}]{Hempel2012}
\bibinfo{author}{\bibfnamefont{M.}~\bibnamefont{{Hempel}}},
  \bibinfo{author}{\bibfnamefont{T.}~\bibnamefont{{Fischer}}},
  \bibinfo{author}{\bibfnamefont{J.}~\bibnamefont{{Schaffner-Bielich}}},
  \bibnamefont{and}
  \bibinfo{author}{\bibfnamefont{M.}~\bibnamefont{{Liebend{\"o}rfer}}},
  \bibinfo{journal}{\apj} \textbf{\bibinfo{volume}{748}}, \bibinfo{eid}{70}
  (\bibinfo{year}{2012}).

\bibitem[{\citenamefont{{Kr{\"u}ger} et~al.}(2013)\citenamefont{{Kr{\"u}ger},
  {Tews}, {Hebeler}, and {Schwenk}}}]{Krueger2013}
\bibinfo{author}{\bibfnamefont{T.}~\bibnamefont{{Kr{\"u}ger}}},
  \bibinfo{author}{\bibfnamefont{I.}~\bibnamefont{{Tews}}},
  \bibinfo{author}{\bibfnamefont{K.}~\bibnamefont{{Hebeler}}},
  \bibnamefont{and}
  \bibinfo{author}{\bibfnamefont{A.}~\bibnamefont{{Schwenk}}},
  \bibinfo{journal}{\prc} \textbf{\bibinfo{volume}{88}}, \bibinfo{eid}{025802}
  (\bibinfo{year}{2013}).

\bibitem[{\citenamefont{{Lattimer} and {Lim}}(2013)}]{Lattimer2013}
\bibinfo{author}{\bibfnamefont{J.~M.} \bibnamefont{{Lattimer}}}
  \bibnamefont{and} \bibinfo{author}{\bibfnamefont{Y.}~\bibnamefont{{Lim}}},
  \bibinfo{journal}{\apj} \textbf{\bibinfo{volume}{771}}, \bibinfo{eid}{51}
  (\bibinfo{year}{2013}).

\bibitem[{\citenamefont{{Oertel} et~al.}(2017)\citenamefont{{Oertel}, {Hempel},
  {Kl{\"a}hn}, and {Typel}}}]{Oertel2017}
\bibinfo{author}{\bibfnamefont{M.}~\bibnamefont{{Oertel}}},
  \bibinfo{author}{\bibfnamefont{M.}~\bibnamefont{{Hempel}}},
  \bibinfo{author}{\bibfnamefont{T.}~\bibnamefont{{Kl{\"a}hn}}},
  \bibnamefont{and} \bibinfo{author}{\bibfnamefont{S.}~\bibnamefont{{Typel}}},
  \bibinfo{journal}{Reviews of Modern Physics} \textbf{\bibinfo{volume}{89}},
  \bibinfo{eid}{015007} (\bibinfo{year}{2017}).

\bibitem[{\citenamefont{{Antoniadis} et~al.}(2013)\citenamefont{{Antoniadis},
  {Freire}, {Wex}, {Tauris}, {Lynch}, {van Kerkwijk}, {Kramer}, {Bassa},
  {Dhillon}, {Driebe} et~al.}}]{Antoniadis2013}
\bibinfo{author}{\bibfnamefont{J.}~\bibnamefont{{Antoniadis}}},
  \bibinfo{author}{\bibfnamefont{P.~C.~C.} \bibnamefont{{Freire}}},
  \bibinfo{author}{\bibfnamefont{N.}~\bibnamefont{{Wex}}},
  \bibinfo{author}{\bibfnamefont{T.~M.} \bibnamefont{{Tauris}}},
  \bibinfo{author}{\bibfnamefont{R.~S.} \bibnamefont{{Lynch}}},
  \bibinfo{author}{\bibfnamefont{M.~H.} \bibnamefont{{van Kerkwijk}}},
  \bibinfo{author}{\bibfnamefont{M.}~\bibnamefont{{Kramer}}},
  \bibinfo{author}{\bibfnamefont{C.}~\bibnamefont{{Bassa}}},
  \bibinfo{author}{\bibfnamefont{V.~S.} \bibnamefont{{Dhillon}}},
  \bibinfo{author}{\bibfnamefont{T.}~\bibnamefont{{Driebe}}},
  \bibnamefont{et~al.}, \bibinfo{journal}{Science}
  \textbf{\bibinfo{volume}{340}}, \bibinfo{pages}{448} (\bibinfo{year}{2013}).

\bibitem[{\citenamefont{Arzoumanian et~al.}(2018)\citenamefont{Arzoumanian,
  Brazier, Burke-Spolaor, Chamberlin, Chatterjee, Christy, Cordes, Cornish,
  Crawford, Thankful~Cromartie et~al.}}]{Arzoumanian2018a}
\bibinfo{author}{\bibfnamefont{Z.}~\bibnamefont{Arzoumanian}},
  \bibinfo{author}{\bibfnamefont{A.}~\bibnamefont{Brazier}},
  \bibinfo{author}{\bibfnamefont{S.}~\bibnamefont{Burke-Spolaor}},
  \bibinfo{author}{\bibfnamefont{S.}~\bibnamefont{Chamberlin}},
  \bibinfo{author}{\bibfnamefont{S.}~\bibnamefont{Chatterjee}},
  \bibinfo{author}{\bibfnamefont{B.}~\bibnamefont{Christy}},
  \bibinfo{author}{\bibfnamefont{J.~M.} \bibnamefont{Cordes}},
  \bibinfo{author}{\bibfnamefont{N.~J.} \bibnamefont{Cornish}},
  \bibinfo{author}{\bibfnamefont{F.}~\bibnamefont{Crawford}},
  \bibinfo{author}{\bibfnamefont{H.}~\bibnamefont{Thankful~Cromartie}},
  \bibnamefont{et~al.}, \bibinfo{journal}{\apjs}
  \textbf{\bibinfo{volume}{235}}, \bibinfo{eid}{37} (\bibinfo{year}{2018}).

\bibitem[{\citenamefont{Abbott et~al.}(2017)\citenamefont{Abbott, Abbott,
  Abbott, Acernese, Ackley, Adams, Adams, Addesso, Adhikari, Adya
  et~al.}}]{Abbott2017}
\bibinfo{author}{\bibfnamefont{B.~P.} \bibnamefont{Abbott}},
  \bibinfo{author}{\bibfnamefont{R.}~\bibnamefont{Abbott}},
  \bibinfo{author}{\bibfnamefont{T.~D.} \bibnamefont{Abbott}},
  \bibinfo{author}{\bibfnamefont{F.}~\bibnamefont{Acernese}},
  \bibinfo{author}{\bibfnamefont{K.}~\bibnamefont{Ackley}},
  \bibinfo{author}{\bibfnamefont{C.}~\bibnamefont{Adams}},
  \bibinfo{author}{\bibfnamefont{T.}~\bibnamefont{Adams}},
  \bibinfo{author}{\bibfnamefont{P.}~\bibnamefont{Addesso}},
  \bibinfo{author}{\bibfnamefont{R.~X.} \bibnamefont{Adhikari}},
  \bibinfo{author}{\bibfnamefont{V.~B.} \bibnamefont{Adya}},
  \bibnamefont{et~al.} (\bibinfo{collaboration}{LIGO Scientific Collaboration
  and Virgo Collaboration}), \bibinfo{journal}{\prl}
  \textbf{\bibinfo{volume}{119}}, \bibinfo{pages}{161101}
  (\bibinfo{year}{2017}).

\bibitem[{\citenamefont{{Bauswein} et~al.}(2017)\citenamefont{{Bauswein},
  {Just}, {Janka}, and {Stergioulas}}}]{Bauswein2017}
\bibinfo{author}{\bibfnamefont{A.}~\bibnamefont{{Bauswein}}},
  \bibinfo{author}{\bibfnamefont{O.}~\bibnamefont{{Just}}},
  \bibinfo{author}{\bibfnamefont{H.-T.} \bibnamefont{{Janka}}},
  \bibnamefont{and}
  \bibinfo{author}{\bibfnamefont{N.}~\bibnamefont{{Stergioulas}}},
  \bibinfo{journal}{\apjl} \textbf{\bibinfo{volume}{850}}, \bibinfo{eid}{L34}
  (\bibinfo{year}{2017}).

\bibitem[{\citenamefont{De et~al.}(2018)\citenamefont{De, Finstad, Lattimer,
  Brown, Berger, and Biwer}}]{De2018}
\bibinfo{author}{\bibfnamefont{S.}~\bibnamefont{De}},
  \bibinfo{author}{\bibfnamefont{D.}~\bibnamefont{Finstad}},
  \bibinfo{author}{\bibfnamefont{J.~M.} \bibnamefont{Lattimer}},
  \bibinfo{author}{\bibfnamefont{D.~A.} \bibnamefont{Brown}},
  \bibinfo{author}{\bibfnamefont{E.}~\bibnamefont{Berger}}, \bibnamefont{and}
  \bibinfo{author}{\bibfnamefont{C.~M.} \bibnamefont{Biwer}},
  \bibinfo{journal}{\prl} \textbf{\bibinfo{volume}{121}}, \bibinfo{eid}{091102}
  (\bibinfo{year}{2018}).

\bibitem[{\citenamefont{Abbott et~al.}(2018)\citenamefont{Abbott, Abbott,
  Abbott, Acernese, Ackley, Adams, Adams, Addesso, Adhikari, Adya
  et~al.}}]{Abbott2018}
\bibinfo{author}{\bibfnamefont{B.~P.} \bibnamefont{Abbott}},
  \bibinfo{author}{\bibfnamefont{R.}~\bibnamefont{Abbott}},
  \bibinfo{author}{\bibfnamefont{T.~D.} \bibnamefont{Abbott}},
  \bibinfo{author}{\bibfnamefont{F.}~\bibnamefont{Acernese}},
  \bibinfo{author}{\bibfnamefont{K.}~\bibnamefont{Ackley}},
  \bibinfo{author}{\bibfnamefont{C.}~\bibnamefont{Adams}},
  \bibinfo{author}{\bibfnamefont{T.}~\bibnamefont{Adams}},
  \bibinfo{author}{\bibfnamefont{P.}~\bibnamefont{Addesso}},
  \bibinfo{author}{\bibfnamefont{R.~X.} \bibnamefont{Adhikari}},
  \bibinfo{author}{\bibfnamefont{V.~B.} \bibnamefont{Adya}},
  \bibnamefont{et~al.}, \bibinfo{journal}{\prl} \textbf{\bibinfo{volume}{121}},
  \bibinfo{eid}{161101} (\bibinfo{year}{2018}).

\bibitem[{\citenamefont{{Kaltenborn} et~al.}(2017)\citenamefont{{Kaltenborn},
  {Bastian}, and {Blaschke}}}]{Kaltenborn2017}
\bibinfo{author}{\bibfnamefont{M.~A.~R.} \bibnamefont{{Kaltenborn}}},
  \bibinfo{author}{\bibfnamefont{N.-U.~F.} \bibnamefont{{Bastian}}},
  \bibnamefont{and} \bibinfo{author}{\bibfnamefont{D.~B.}
  \bibnamefont{{Blaschke}}}, \bibinfo{journal}{\prd}
  \textbf{\bibinfo{volume}{96}}, \bibinfo{eid}{056024} (\bibinfo{year}{2017}).

\bibitem[{\citenamefont{Nambu and Jona-Lasinio}(1961)}]{Nambu1961}
\bibinfo{author}{\bibfnamefont{Y.}~\bibnamefont{Nambu}} \bibnamefont{and}
  \bibinfo{author}{\bibfnamefont{G.}~\bibnamefont{Jona-Lasinio}},
  \bibinfo{journal}{Phys. Rev.} \textbf{\bibinfo{volume}{122}},
  \bibinfo{pages}{345} (\bibinfo{year}{1961}).

\bibitem[{\citenamefont{Klevansky}(1992)}]{Klevansky1992}
\bibinfo{author}{\bibfnamefont{S.}~\bibnamefont{Klevansky}},
  \bibinfo{journal}{Rev. Mod. Phys.} \textbf{\bibinfo{volume}{64}},
  \bibinfo{pages}{649} (\bibinfo{year}{1992}).

\bibitem[{\citenamefont{{R{\"u}ster} et~al.}(2005)\citenamefont{{R{\"u}ster},
  {Werth}, {Buballa}, {Shovkovy}, and {Rischke}}}]{Ruester2005}
\bibinfo{author}{\bibfnamefont{S.~B.} \bibnamefont{{R{\"u}ster}}},
  \bibinfo{author}{\bibfnamefont{V.}~\bibnamefont{{Werth}}},
  \bibinfo{author}{\bibfnamefont{M.}~\bibnamefont{{Buballa}}},
  \bibinfo{author}{\bibfnamefont{I.~A.} \bibnamefont{{Shovkovy}}},
  \bibnamefont{and} \bibinfo{author}{\bibfnamefont{D.~H.}
  \bibnamefont{{Rischke}}}, \bibinfo{journal}{\prd}
  \textbf{\bibinfo{volume}{72}}, \bibinfo{eid}{034004} (\bibinfo{year}{2005}).

\bibitem[{\citenamefont{Blaschke et~al.}(2005)\citenamefont{Blaschke,
  Fredriksson, Grigorian, {\"O}zta{\c s}, and Sandin}}]{Blaschke2005}
\bibinfo{author}{\bibfnamefont{D.}~\bibnamefont{Blaschke}},
  \bibinfo{author}{\bibfnamefont{S.}~\bibnamefont{Fredriksson}},
  \bibinfo{author}{\bibfnamefont{H.}~\bibnamefont{Grigorian}},
  \bibinfo{author}{\bibfnamefont{A.~M.} \bibnamefont{{\"O}zta{\c s}}},
  \bibnamefont{and} \bibinfo{author}{\bibfnamefont{F.}~\bibnamefont{Sandin}},
  \bibinfo{journal}{\prd} \textbf{\bibinfo{volume}{72}}, \bibinfo{eid}{065020}
  (\bibinfo{year}{2005}).

\bibitem[{\citenamefont{Fischer et~al.}(2018)\citenamefont{Fischer, Bastian,
  Wu, Baklanov, Sorokina, Blinnikov, Typel, Kl{\"a}hn, and
  Blaschke}}]{Fischer2018}
\bibinfo{author}{\bibfnamefont{T.}~\bibnamefont{Fischer}},
  \bibinfo{author}{\bibfnamefont{N.-U.~F.} \bibnamefont{Bastian}},
  \bibinfo{author}{\bibfnamefont{M.-R.} \bibnamefont{Wu}},
  \bibinfo{author}{\bibfnamefont{P.}~\bibnamefont{Baklanov}},
  \bibinfo{author}{\bibfnamefont{E.}~\bibnamefont{Sorokina}},
  \bibinfo{author}{\bibfnamefont{S.}~\bibnamefont{Blinnikov}},
  \bibinfo{author}{\bibfnamefont{S.}~\bibnamefont{Typel}},
  \bibinfo{author}{\bibfnamefont{T.}~\bibnamefont{Kl{\"a}hn}},
  \bibnamefont{and} \bibinfo{author}{\bibfnamefont{D.~B.}
  \bibnamefont{Blaschke}}, \bibinfo{journal}{Nature Astronomy}
  \textbf{\bibinfo{volume}{2}}, \bibinfo{pages}{980} (\bibinfo{year}{2018}).

\bibitem[{\citenamefont{{Bastian} et~al.}(2018)\citenamefont{{Bastian},
  {Blaschke}, {Fischer}, and {R{\"o}pke}}}]{Bastian2018}
\bibinfo{author}{\bibfnamefont{N.-U.} \bibnamefont{{Bastian}}},
  \bibinfo{author}{\bibfnamefont{D.}~\bibnamefont{{Blaschke}}},
  \bibinfo{author}{\bibfnamefont{T.}~\bibnamefont{{Fischer}}},
  \bibnamefont{and}
  \bibinfo{author}{\bibfnamefont{G.}~\bibnamefont{{R{\"o}pke}}},
  \bibinfo{journal}{Universe} \textbf{\bibinfo{volume}{4}}, \bibinfo{pages}{67}
  (\bibinfo{year}{2018}).

\bibitem[{\citenamefont{Benic et~al.}(2015)\citenamefont{Benic, Blaschke,
  Alvarez-Castillo, Fischer, and Typel}}]{Benic2015}
\bibinfo{author}{\bibfnamefont{S.}~\bibnamefont{Benic}},
  \bibinfo{author}{\bibfnamefont{D.}~\bibnamefont{Blaschke}},
  \bibinfo{author}{\bibfnamefont{D.~E.} \bibnamefont{Alvarez-Castillo}},
  \bibinfo{author}{\bibfnamefont{T.}~\bibnamefont{Fischer}}, \bibnamefont{and}
  \bibinfo{author}{\bibfnamefont{S.}~\bibnamefont{Typel}},
  \bibinfo{journal}{Astron. Astrophys.} \textbf{\bibinfo{volume}{577}},
  \bibinfo{pages}{A40} (\bibinfo{year}{2015}).

\bibitem[{\citenamefont{{Kl{\"a}hn} and {Fischer}}(2015)}]{Klaehn2015}
\bibinfo{author}{\bibfnamefont{T.}~\bibnamefont{{Kl{\"a}hn}}} \bibnamefont{and}
  \bibinfo{author}{\bibfnamefont{T.}~\bibnamefont{{Fischer}}},
  \bibinfo{journal}{\apj} \textbf{\bibinfo{volume}{810}}, \bibinfo{eid}{134}
  (\bibinfo{year}{2015}).

\bibitem[{\citenamefont{Typel}(2016)}]{Typel2016}
\bibinfo{author}{\bibfnamefont{S.}~\bibnamefont{Typel}},
  \bibinfo{journal}{European Physical Journal A} \textbf{\bibinfo{volume}{52}},
  \bibinfo{eid}{16} (\bibinfo{year}{2016}).

\bibitem[{\citenamefont{Akmal et~al.}(1998)\citenamefont{Akmal, Pandharipande,
  and Ravenhall}}]{Akmal1998}
\bibinfo{author}{\bibfnamefont{A.}~\bibnamefont{Akmal}},
  \bibinfo{author}{\bibfnamefont{V.~R.} \bibnamefont{Pandharipande}},
  \bibnamefont{and} \bibinfo{author}{\bibfnamefont{D.~G.}
  \bibnamefont{Ravenhall}}, \bibinfo{journal}{\prc}
  \textbf{\bibinfo{volume}{58}}, \bibinfo{pages}{1804} (\bibinfo{year}{1998}).

\bibitem[{\citenamefont{Banik et~al.}(2014)\citenamefont{Banik, Hempel, and
  Bandyopadhyay}}]{Banik2014}
\bibinfo{author}{\bibfnamefont{S.}~\bibnamefont{Banik}},
  \bibinfo{author}{\bibfnamefont{M.}~\bibnamefont{Hempel}}, \bibnamefont{and}
  \bibinfo{author}{\bibfnamefont{D.}~\bibnamefont{Bandyopadhyay}},
  \bibinfo{journal}{\apjs} \textbf{\bibinfo{volume}{214}}, \bibinfo{eid}{22}
  (\bibinfo{year}{2014}).

\bibitem[{\citenamefont{{Goriely} et~al.}(2010)\citenamefont{{Goriely},
  {Chamel}, and {Pearson}}}]{Goriely2010}
\bibinfo{author}{\bibfnamefont{S.}~\bibnamefont{{Goriely}}},
  \bibinfo{author}{\bibfnamefont{N.}~\bibnamefont{{Chamel}}}, \bibnamefont{and}
  \bibinfo{author}{\bibfnamefont{J.~M.} \bibnamefont{{Pearson}}},
  \bibinfo{journal}{\prc} \textbf{\bibinfo{volume}{82}}, \bibinfo{eid}{035804}
  (\bibinfo{year}{2010}).

\bibitem[{\citenamefont{Wiringa et~al.}(1988)\citenamefont{Wiringa, Fiks, and
  Fabrocini}}]{Wiringa1988}
\bibinfo{author}{\bibfnamefont{R.~B.} \bibnamefont{Wiringa}},
  \bibinfo{author}{\bibfnamefont{V.}~\bibnamefont{Fiks}}, \bibnamefont{and}
  \bibinfo{author}{\bibfnamefont{A.}~\bibnamefont{Fabrocini}},
  \bibinfo{journal}{\prc} \textbf{\bibinfo{volume}{38}}, \bibinfo{pages}{1010}
  (\bibinfo{year}{1988}).

\bibitem[{\citenamefont{Shen et~al.}(2011)\citenamefont{Shen, Horowitz, and
  Teige}}]{Shen2011}
\bibinfo{author}{\bibfnamefont{G.}~\bibnamefont{Shen}},
  \bibinfo{author}{\bibfnamefont{C.~J.} \bibnamefont{Horowitz}},
  \bibnamefont{and} \bibinfo{author}{\bibfnamefont{S.}~\bibnamefont{Teige}},
  \bibinfo{journal}{\prc} \textbf{\bibinfo{volume}{83}}, \bibinfo{eid}{035802}
  (\bibinfo{year}{2011}).

\bibitem[{\citenamefont{Lattimer and Douglas~Swesty}(1991)}]{Lattimer1991}
\bibinfo{author}{\bibfnamefont{J.~M.} \bibnamefont{Lattimer}} \bibnamefont{and}
  \bibinfo{author}{\bibfnamefont{F.}~\bibnamefont{Douglas~Swesty}},
  \bibinfo{journal}{Nuclear Physics A} \textbf{\bibinfo{volume}{535}},
  \bibinfo{pages}{331} (\bibinfo{year}{1991}).

\bibitem[{\citenamefont{Lalazissis et~al.}(1997)\citenamefont{Lalazissis,
  K{\"o}nig, and Ring}}]{Lalazissis1997a}
\bibinfo{author}{\bibfnamefont{G.~A.} \bibnamefont{Lalazissis}},
  \bibinfo{author}{\bibfnamefont{J.}~\bibnamefont{K{\"o}nig}},
  \bibnamefont{and} \bibinfo{author}{\bibfnamefont{P.}~\bibnamefont{Ring}},
  \bibinfo{journal}{\prc} \textbf{\bibinfo{volume}{55}}, \bibinfo{pages}{540}
  (\bibinfo{year}{1997}).

\bibitem[{\citenamefont{Steiner et~al.}(2013)\citenamefont{Steiner, Hempel, and
  Fischer}}]{Steiner2013}
\bibinfo{author}{\bibfnamefont{A.~W.} \bibnamefont{Steiner}},
  \bibinfo{author}{\bibfnamefont{M.}~\bibnamefont{Hempel}}, \bibnamefont{and}
  \bibinfo{author}{\bibfnamefont{T.}~\bibnamefont{Fischer}},
  \bibinfo{journal}{\apj} \textbf{\bibinfo{volume}{774}}, \bibinfo{eid}{17}
  (\bibinfo{year}{2013}).

\bibitem[{\citenamefont{{Douchin} and {Haensel}}(2001)}]{Douchin2001}
\bibinfo{author}{\bibfnamefont{F.}~\bibnamefont{{Douchin}}} \bibnamefont{and}
  \bibinfo{author}{\bibfnamefont{P.}~\bibnamefont{{Haensel}}},
  \bibinfo{journal}{\aap} \textbf{\bibinfo{volume}{380}}, \bibinfo{pages}{151}
  (\bibinfo{year}{2001}).

\bibitem[{\citenamefont{Sugahara and Toki}(1994)}]{Sugahara1994a}
\bibinfo{author}{\bibfnamefont{Y.}~\bibnamefont{Sugahara}} \bibnamefont{and}
  \bibinfo{author}{\bibfnamefont{H.}~\bibnamefont{Toki}},
  \bibinfo{journal}{Nuclear Physics A} \textbf{\bibinfo{volume}{579}},
  \bibinfo{pages}{557} (\bibinfo{year}{1994}).

\bibitem[{\citenamefont{Toki et~al.}(1995)\citenamefont{Toki, Hirata, Sugahara,
  Sumiyoshi, and Tanihata}}]{Toki1995}
\bibinfo{author}{\bibfnamefont{H.}~\bibnamefont{Toki}},
  \bibinfo{author}{\bibfnamefont{D.}~\bibnamefont{Hirata}},
  \bibinfo{author}{\bibfnamefont{Y.}~\bibnamefont{Sugahara}},
  \bibinfo{author}{\bibfnamefont{K.}~\bibnamefont{Sumiyoshi}},
  \bibnamefont{and} \bibinfo{author}{\bibfnamefont{I.}~\bibnamefont{Tanihata}},
  \bibinfo{journal}{Nuclear Physics A} \textbf{\bibinfo{volume}{588}},
  \bibinfo{pages}{357} (\bibinfo{year}{1995}).

\bibitem[{\citenamefont{{Bauswein} et~al.}(2012)\citenamefont{{Bauswein},
  {Janka}, {Hebeler}, and {Schwenk}}}]{Bauswein2012a}
\bibinfo{author}{\bibfnamefont{A.}~\bibnamefont{{Bauswein}}},
  \bibinfo{author}{\bibfnamefont{H.-T.} \bibnamefont{{Janka}}},
  \bibinfo{author}{\bibfnamefont{K.}~\bibnamefont{{Hebeler}}},
  \bibnamefont{and}
  \bibinfo{author}{\bibfnamefont{A.}~\bibnamefont{{Schwenk}}},
  \bibinfo{journal}{\prd} \textbf{\bibinfo{volume}{86}}, \bibinfo{eid}{063001}
  (\bibinfo{year}{2012}).

\bibitem[{\citenamefont{{Bauswein} et~al.}(2013)\citenamefont{{Bauswein},
  {Goriely}, and {Janka}}}]{Bauswein2013a}
\bibinfo{author}{\bibfnamefont{A.}~\bibnamefont{{Bauswein}}},
  \bibinfo{author}{\bibfnamefont{S.}~\bibnamefont{{Goriely}}},
  \bibnamefont{and} \bibinfo{author}{\bibfnamefont{H.-T.}
  \bibnamefont{{Janka}}}, \bibinfo{journal}{\apj}
  \textbf{\bibinfo{volume}{773}}, \bibinfo{eid}{78} (\bibinfo{year}{2013}).

\bibitem[{\citenamefont{{Bauswein} et~al.}(2014)\citenamefont{{Bauswein},
  {Stergioulas}, and {Janka}}}]{Bauswein2014a}
\bibinfo{author}{\bibfnamefont{A.}~\bibnamefont{{Bauswein}}},
  \bibinfo{author}{\bibfnamefont{N.}~\bibnamefont{{Stergioulas}}},
  \bibnamefont{and} \bibinfo{author}{\bibfnamefont{H.-T.}
  \bibnamefont{{Janka}}}, \bibinfo{journal}{\prd}
  \textbf{\bibinfo{volume}{90}}, \bibinfo{eid}{023002} (\bibinfo{year}{2014}).

\bibitem[{\citenamefont{Fortin et~al.}(2018)\citenamefont{Fortin, Oertel, and
  Provid{\^e}ncia}}]{Fortin2018}
\bibinfo{author}{\bibfnamefont{M.}~\bibnamefont{Fortin}},
  \bibinfo{author}{\bibfnamefont{M.}~\bibnamefont{Oertel}}, \bibnamefont{and}
  \bibinfo{author}{\bibfnamefont{C.}~\bibnamefont{Provid{\^e}ncia}},
  \bibinfo{journal}{\pasa} \textbf{\bibinfo{volume}{35}}
  (\bibinfo{year}{2018}).

\bibitem[{\citenamefont{Marques et~al.}(2017)\citenamefont{Marques, Oertel,
  Hempel, and Novak}}]{Marques2017}
\bibinfo{author}{\bibfnamefont{M.}~\bibnamefont{Marques}},
  \bibinfo{author}{\bibfnamefont{M.}~\bibnamefont{Oertel}},
  \bibinfo{author}{\bibfnamefont{M.}~\bibnamefont{Hempel}}, \bibnamefont{and}
  \bibinfo{author}{\bibfnamefont{J.}~\bibnamefont{Novak}},
  \bibinfo{journal}{Phys. Rev.} \textbf{\bibinfo{volume}{C96}},
  \bibinfo{pages}{045806} (\bibinfo{year}{2017}).

\bibitem[{\citenamefont{Alford et~al.}(2005)\citenamefont{Alford, Braby, Paris,
  and Reddy}}]{Alford2005}
\bibinfo{author}{\bibfnamefont{M.}~\bibnamefont{Alford}},
  \bibinfo{author}{\bibfnamefont{M.}~\bibnamefont{Braby}},
  \bibinfo{author}{\bibfnamefont{M.}~\bibnamefont{Paris}}, \bibnamefont{and}
  \bibinfo{author}{\bibfnamefont{S.}~\bibnamefont{Reddy}},
  \bibinfo{journal}{\apj} \textbf{\bibinfo{volume}{629}}, \bibinfo{pages}{969}
  (\bibinfo{year}{2005}).

\bibitem[{\citenamefont{Read et~al.}(2009)\citenamefont{Read, Lackey, Owen, and
  Friedman}}]{Read2009a}
\bibinfo{author}{\bibfnamefont{J.~S.} \bibnamefont{Read}},
  \bibinfo{author}{\bibfnamefont{B.~D.} \bibnamefont{Lackey}},
  \bibinfo{author}{\bibfnamefont{B.~J.} \bibnamefont{Owen}}, \bibnamefont{and}
  \bibinfo{author}{\bibfnamefont{J.~L.} \bibnamefont{Friedman}},
  \bibinfo{journal}{\prd} \textbf{\bibinfo{volume}{79}}, \bibinfo{eid}{124032}
  (\bibinfo{year}{2009}).

\bibitem[{\citenamefont{{Bauswein} and {Janka}}(2012)}]{Bauswein2012}
\bibinfo{author}{\bibfnamefont{A.}~\bibnamefont{{Bauswein}}} \bibnamefont{and}
  \bibinfo{author}{\bibfnamefont{H.-T.} \bibnamefont{{Janka}}},
  \bibinfo{journal}{\prl} \textbf{\bibinfo{volume}{108}}, \bibinfo{eid}{011101}
  (\bibinfo{year}{2012}).

\bibitem[{\citenamefont{{Hotokezaka} et~al.}(2013)\citenamefont{{Hotokezaka},
  {Kiuchi}, {Kyutoku}, {Muranushi}, {Sekiguchi}, {Shibata}, and
  {Taniguchi}}}]{Hotokezaka2013a}
\bibinfo{author}{\bibfnamefont{K.}~\bibnamefont{{Hotokezaka}}},
  \bibinfo{author}{\bibfnamefont{K.}~\bibnamefont{{Kiuchi}}},
  \bibinfo{author}{\bibfnamefont{K.}~\bibnamefont{{Kyutoku}}},
  \bibinfo{author}{\bibfnamefont{T.}~\bibnamefont{{Muranushi}}},
  \bibinfo{author}{\bibfnamefont{Y.}~\bibnamefont{{Sekiguchi}}},
  \bibinfo{author}{\bibfnamefont{M.}~\bibnamefont{{Shibata}}},
  \bibnamefont{and}
  \bibinfo{author}{\bibfnamefont{K.}~\bibnamefont{{Taniguchi}}},
  \bibinfo{journal}{\prd} \textbf{\bibinfo{volume}{88}}, \bibinfo{eid}{044026}
  (\bibinfo{year}{2013}).

\bibitem[{\citenamefont{{Clark} et~al.}(2014)\citenamefont{{Clark}, {Bauswein},
  {Cadonati}, {Janka}, {Pankow}, and {Stergioulas}}}]{Clark2014}
\bibinfo{author}{\bibfnamefont{J.}~\bibnamefont{{Clark}}},
  \bibinfo{author}{\bibfnamefont{A.}~\bibnamefont{{Bauswein}}},
  \bibinfo{author}{\bibfnamefont{L.}~\bibnamefont{{Cadonati}}},
  \bibinfo{author}{\bibfnamefont{H.-T.} \bibnamefont{{Janka}}},
  \bibinfo{author}{\bibfnamefont{C.}~\bibnamefont{{Pankow}}}, \bibnamefont{and}
  \bibinfo{author}{\bibfnamefont{N.}~\bibnamefont{{Stergioulas}}},
  \bibinfo{journal}{\prd} \textbf{\bibinfo{volume}{90}}, \bibinfo{eid}{062004}
  (\bibinfo{year}{2014}).

\bibitem[{\citenamefont{{Clark} et~al.}(2016)\citenamefont{{Clark}, {Bauswein},
  {Stergioulas}, and {Shoemaker}}}]{Clark2016}
\bibinfo{author}{\bibfnamefont{J.~A.} \bibnamefont{{Clark}}},
  \bibinfo{author}{\bibfnamefont{A.}~\bibnamefont{{Bauswein}}},
  \bibinfo{author}{\bibfnamefont{N.}~\bibnamefont{{Stergioulas}}},
  \bibnamefont{and}
  \bibinfo{author}{\bibfnamefont{D.}~\bibnamefont{{Shoemaker}}},
  \bibinfo{journal}{Classical and Quantum Gravity}
  \textbf{\bibinfo{volume}{33}}, \bibinfo{eid}{085003} (\bibinfo{year}{2016}).

\bibitem[{\citenamefont{Chatziioannou et~al.}(2017)\citenamefont{Chatziioannou,
  Clark, Bauswein, Millhouse, Littenberg, and Cornish}}]{Chatziioannou2017}
\bibinfo{author}{\bibfnamefont{K.}~\bibnamefont{Chatziioannou}},
  \bibinfo{author}{\bibfnamefont{J.~A.} \bibnamefont{Clark}},
  \bibinfo{author}{\bibfnamefont{A.}~\bibnamefont{Bauswein}},
  \bibinfo{author}{\bibfnamefont{M.}~\bibnamefont{Millhouse}},
  \bibinfo{author}{\bibfnamefont{T.~B.} \bibnamefont{Littenberg}},
  \bibnamefont{and} \bibinfo{author}{\bibfnamefont{N.}~\bibnamefont{Cornish}},
  \bibinfo{journal}{\prd} \textbf{\bibinfo{volume}{96}}, \bibinfo{eid}{124035}
  (\bibinfo{year}{2017}.

\end{thebibliography}

\end{bibunit}

\end{document}